\documentclass[twocolumn,superscriptaddress,prl]{revtex4}
\pdfoutput = 1

\usepackage{graphicx}
\usepackage{epsfig}
\usepackage{verbatim}
\usepackage{bm}

\def\up{\uparrow}
\def\down{\downarrow}
\newcommand{\gammabar}{\ensuremath\gamma\kern-0.53em-}

\usepackage{times}
\usepackage{amsmath}
\usepackage{amsfonts}
\usepackage{amssymb}
\usepackage{graphicx}
\usepackage{pst-all}
\usepackage{leftidx}

\usepackage{tikz}

\makeatother

\makeatletter
\def\l@subsubsection#1#2{}
\makeatother

\begin{document}

\title{Physical Architecture for a Universal Topological Quantum Computer
\\ based on a Network of Majorana Nanowires}
\author{Maissam Barkeshli}
\affiliation{Station Q, Microsoft Research, Santa Barbara, California 93106-6105, USA}
\author{Jay D. Sau}
\affiliation{Department of Physics, Condensed Matter Theory Center and Joint Quantum Institute,
University of Maryland, College Park, Maryland 20742-4111, USA}
\date{\today}

\begin{abstract}
The idea of topological quantum computation (TQC) is to store and manipulate quantum information in an intrinsically fault-tolerant
manner by utilizing the physics of topologically ordered phases of matter. Currently, one of the most promising platforms for a topological qubit
is in terms of Majorana fermion zero modes (MZMs) in spin-orbit coupled superconducting nanowires. However, the topologically robust operations that
are possible with MZMs can be efficiently simulated on a classical computer and are therefore not sufficient for
realizing a universal gate set for TQC. Here, we show that an array of coupled semiconductor-superconductor nanowires with MZM
edge states can be used to realize a more sophisticated type of non-Abelian defect: a genon in an Ising $\times$ Ising
topological state. This leads to a possible implementation of the missing topologically protected $\pi/8$ phase gate
and thus universal TQC based on semiconductor-superconductor nanowire technology. We provide detailed numerical estimates of the relevant
energy scales, which we show to lie within accessible ranges.
\end{abstract}

\maketitle

\section*{Introduction}

The promise of topological quantum computation (TQC) is to encode and manipulate quantum information using
topological qubits \cite{kitaev2003,freedman2002,nayak2008}. The quantum states of a topological qubit
do not couple to any local operators, forming a non-local Hilbert space, and are therefore intrinsically robust to decoherence.
Currently, one of the most promising avenues towards developing
a topological qubit is in terms of Majorana
fermion zero modes (MZMs) in spin-orbit coupled superconducting nanowires \cite{read2000,kitaev2001,sau2010,lutchyn2010,alicea2012review,beenakker2013,mourik12,das12,rokhinson12,deng12,churchill13,nadj-Perge14,chang2015}.


Topologically protected qubits are most useful if the information stored in them can be controlled in a
topologically protected way. A system of topological qubits that are in a rich enough topological phase to
allow complete manipulation of the state space would lead to the construction of a universal topological quantum computer.
Unfortunately, most of the topological phases that appear within experimental reach such as topological superconductors hosting
MZMs \cite{alicea2012review}, the Pfaffian fractional quantum Hall (FQH) states \cite{nayak2008}, surface codes \cite{fowler2012,ioffe2002a,ioffe2002b,doucot2003,albuquerque2008,you2010a,xu2010,terhal2012,vijay2015}
and even the recently proposed parafermion zero modes\cite{barkeshli2012a,you2012,clarke2013,lindner2012,cheng2012,barkeshli2013genon} are
not complex enough to span the entire topological state space in a topologically protected way. 
In fact, the topologically protected unitary operations that are possible with MZMs correspond to the Clifford group, which can be efficiently
simulated on a classical computer \cite{gottesman1999}. Consequently, proposals
for utilizing MZMs in quantum computation require non-topological operations, and perhaps
interfacing them with conventional, non-topological qubits \cite{bravyi2006mr,bonderson2010arb,sau2010universal,bonderson2011bus,jiang2011}.

Despite much previous work, it remains a major open problem to find a viable path towards universal TQC.
Ref. \onlinecite{bravyi200universal} showed that some 2D systems harboring MZMs could support universal TQC given the
possibility of dynamical topology changes. However, proposals to exploit these ideas using topological superconductors in semiconductor-superconductor
heterostructures \cite{bonderson2010}, or the Moore-Read Pfaffian FQH state with
tilted interferometry \cite{freedman2006universal}, were since found to be insufficient even in principle, as they lack a crucial
ingredient: the existence of a finite energy quasiparticle excitation with an appropriate value of topological spin \cite{bonderson2013twisted}.
Other previously proposed physical platforms require
either (1) exotic non-Abelian fractional quantum Hall (FQH) states, such as those which possess
Fibonacci quasiparticles, or the $Z_4$ Read-Rezayi FQH state \cite{read1999,freedman2002,bonesteel2005,mong2014,vaezi2014prx,cui2015b,levaillant2015},
or (2) complex designer Hamiltonians based on Josephson junction arrays, which effectively realize non-Abelian discrete
gauge theories \cite{kitaev2003,mochon2003,doucot2004,cui2015}.
However, the viability of these proposals, even in principle, is questionable due to the difficulty of establishing the existence
of non-Abelian FQH states and the impractically low energy scales of previously proposed designer Hamiltonians \cite{albuquerque2008}.
Moreover, the scientific and technological advances required to realize such proposals are not directly relevant to those being
developed in the pursuit of MZMs in spin-orbit coupled superconducting nanowires.

In this paper, we show that a network of coupled superconducting nanowires hosting MZMs
can be used to realize a more powerful type of non-Abelian defect: a genon \cite{barkeshli2013genon,barkeshli2012a,barkeshli2010}
in an Ising $\times$ Ising topological state. The braiding of such genons can be shown to mathematically map onto the
required dynamical topology changes of Ref. \onlinecite{bravyi200universal} and therefore provides the missing topological single-qubit
$\pi/8$ phase gate \cite{barkeshli2013genon}. Combined with joint fermion parity measurements of MZMs, these operations provide a way to realize
universal TQC \cite{bravyi2002}.

Our proposal consists of the following basic building blocks.
First, we show that an array of suitably coupled MZMs in nanowire systems can realize a
two-dimensional phase of matter with Ising topological order \footnote{The Ising topologically ordered phase is
topologically distinct from a topological (e.g. p + ip) superconductor, as the latter does not have intrinsic topological order \cite{bonderson2013}.
The proposal of \cite{kells2014} is thus insufficient for this purpose.}.
We do this by showing how to engineer an effective Kitaev honeycomb spin model in a realistic physical system of coupled Majorana nanowires,
where each effective spin degree of freedom corresponds to a pair of Majorana nanowires.
We present two approaches to doing this, corresponding to whether capacitive charging energies or Josephson couplings are the dominant energy scales.
An analysis of the energy scales of a physically realistic system indicates that the Ising topological order could have energy gaps on the
order of a few percent of the charging energy of the Josephson junctions of the system; given present-day materials and technology,
and the constraints on the required parameter regimes, we estimate the possibility of energy gaps of up to several Kelvin.

Second, we show how short overpasses between neighboring chains, which are feasible with current nanofabrication technology,
can be used to create two effectively independent Ising phases, referred to as an Ising $\times$ Ising state. Changing the connectivity of the network
by creating a lattice dislocation allows the creation of a genon; this effectively realizes a twist defect
that couples the two layers together. Finally, the genons can be effectively braided with minimal to no physical movement of them,
by tuning the effective interactions between them.

\section*{Realizing the Kitaev Model}

\begin{figure}
	\centering
	\includegraphics[width=3.4in]{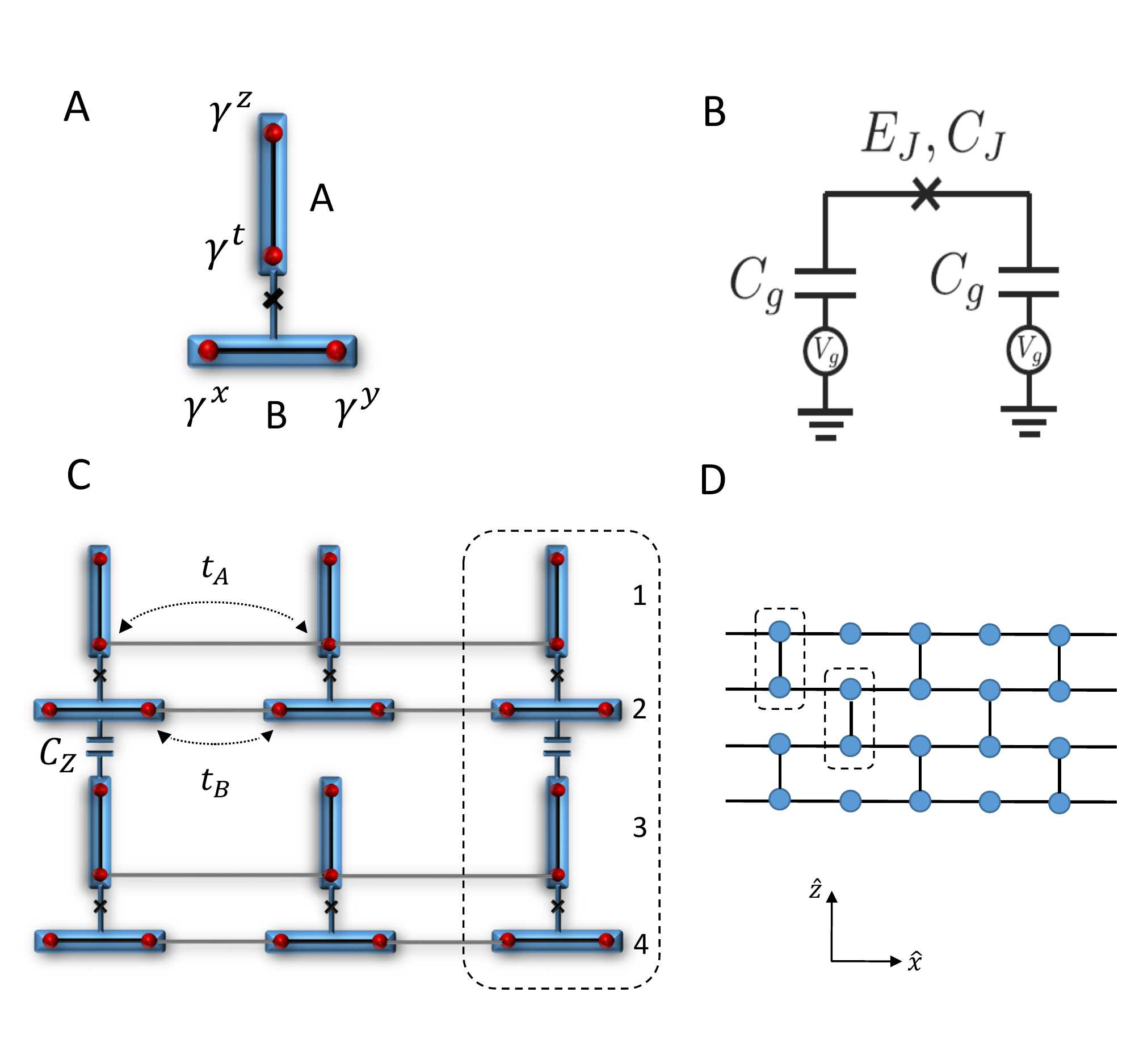}
	\caption{\label{fig1} (A) A pair of semiconductor-superconductor nanowire structures, labelled as $A$ and $B$.
The blue regions represent the superconductor, the black line represents the semiconducting nanowire in the
topological superconducting phase, and the red dots represent the MZM edge states.
(B) Circuit model for (a).
(C) A plaquette of a brick lattice consisting of six effective sites.
The semiconductor nanowires fully extend in the horizontal direction; the gray regions indicate normal (non-topological)
regions of the semiconductor nanowires. The required applied magnetic field can be taken to be normal to the plane.
$t_A$ and $t_B$ parameterize the electron tunneling as shown. Dashed rectangle
indicates a unit cell of the brick lattice, each containing four superconducting islands, or two pairs of $A$ and $B$ -type
structures, labelled $1,\cdots, 4$ as shown. (D) Depiction of the effective brick lattice. Dashed rectangles outline a unit cell of the lattice.
}
\end{figure}

We begin by providing a physical realization of a Kitaev model \cite{kitaev2006}, which can be described
by the following Hamiltonian with spin-1/2 degrees of freedom on each site $\vec{r}$ of a brick lattice
\footnote{A $\pi$ rotation about $S^z$ on every other site takes (\ref{2dspinH}) to the more conventional Kitaev form}:
\begin{align}
\label{2dspinH}
H_{\text{K}} = \sum_{\vec{r}} J_{yx} S^y_{\vec{r}} S^x_{\vec{r}+\hat{x}} + \sum_{\vec{R}} J_z S^z_{\vec{R}} S^z_{\vec{R}-\hat{z}},
\end{align}
where $S^\alpha_{\vec{r}}$ for $\alpha = x,y,z$ are taken to be the Pauli matrices.
We take the unit cell of the brick lattice to be two vertically separated neighboring spins; $\sum_{\vec{R}}$
is a sum over each two-spin unit cell and $\vec{R}$ refers to the top-most spin within the unit cell.

$H_{\text{K}}$ is most naturally solved by expressing the spins in terms of Majorana operators
$\tilde{\gamma}^j$ as $S^\alpha=i\tilde{\gamma}^\alpha\tilde{\gamma}^{t}$, together with
a gauge constraint $\prod_{\alpha=x,y,z,t}\tilde{\gamma}^j=1$ \cite{kitaev2006}. $H_{\text{K}}$ can thus be rewritten as
\begin{align}
\label{2dMajH}
H_{\text{K}} = \sum_{\vec{r}} J_{yx} \tilde{\gamma}^y_{\vec{r}} \tilde{\gamma}^x_{\vec{r}+\hat{x}}\tilde{\gamma}^t_{\vec{r}} \tilde{\gamma}^t_{\vec{r}+\hat{x}}
+ \sum_{\vec{R}} J_z \tilde{\gamma}^t_{\vec{R}}\tilde{\gamma}^z_{\vec{R}} \tilde{\gamma}^y_{\vec{R}-\hat{z}} \tilde{\gamma}^x_{\vec{R}-\hat{z}},
\end{align}
where we have made use of the gauge constraint in writing the $J_z$ term.
The gauge constraint allows us to separate the Majoranas into a set of $Z_2$ gauge fields
$u_{xy}(\vec{R}+\hat{x}/2)=i\tilde{\gamma}^x_{\vec{R}}\tilde{\gamma}^y_{\vec{R}+\hat{x}}$ and
$u_{zz}(\vec{R}+\hat{z}/2)=i\tilde{\gamma}^z_{\vec{R}}\tilde{\gamma}^z_{\vec{R}+\hat{z}}$, which commute with
$H_{\text{K}}$, together with Majorana modes $\tilde{\gamma}^t$ that couple to these $Z_2$ gauge fields.
The non-Abelian Ising phase corresponds to the regime where $\tilde{\gamma}^t$ forms a topological superconductor,
such that the $Z_2$ vortices of $u$ localize MZMs. Since the $Z_2$ vortices are deconfined finite-energy excitations
in the Ising phase, their topological twist $e^{\pi i/8}$ is well-defined and can be exploited for a
topologically protected $\pi/8$ phase gate.

The Majorana solution of $H_{\text{K}}$ suggests that it might be physically realizable in a system where each spin is represented using a pair of
proximity-induced superconducting nanowires with four MZMs $\gamma^{t,x,y,z}$, as shown in Fig.~\ref{fig1}A.
The semiconducting wires (such as InAs or InSb wires) can be either grown \cite{krogstrup2015} or lithographically
defined on 2D systems \cite{shabani2015}, and the superconductor thin films (such as Al, and perhaps other
superconductors such as Nb and NbTiN) can now be epitaxially grown with exceptional interface
qualities \cite{krogstrup2015,chang2015}\cite{shabani2015}.”
As we show below, such a physical realization of $H_{\text{K}}$ is indeed possible; we first provide a
proposal that physically implements the effective spins using the physics of charging energies, and subsequently
we provide a second proposal which utilizes quantum phase slips.

In our first proposal (see Fig. \ref{fig1}), an overall charging energy for the pair of islands, each of which is controlled by a capacitance
$C_g$ to a gate placed at a voltage $V_g$, is used to effectively generate the gauge constraint by constraining the total charge
of the pair of islands. For the four MZMs from each $A,B$ nanowire pair to be coherent with each other (in a sense which will be made
more precise below), it is necessary to retain some phase coherence between the neighboring islands that comprise a
single effective spin. This is obtained by connecting the islands $A$ and $B$ with a Josephson junction, with Josephson
coupling $E_J$ and capacitance $C_J$.

The effective Hamiltonian describing the Majorana and phase degrees of freedom for the system in Fig.~\ref{fig1}(A) can be written as
\begin{align}
\label{singleSpinH}
H_{\text{ss}} = &\sum_{j = A,B} H_{\text{BdG}} [ \Delta_{0j} e^{i \varphi_j}, \psi_j^\dagger, \psi_j]
\nonumber \\
&- E_J \cos(\varphi_A - \phi_B) + \frac{1}{2} \sum_{i,j = A,B} Q_i C_{ij}^{-1} Q_j.
\end{align}
Here $H_{\text{BdG}} [ \Delta_{0j} e^{i \varphi_j}, \psi_j^\dagger, \psi_j]$ is the BdG Hamiltonian for the
nanowire on the $j$th island, where $|\Delta_{0j}|$ is the proximity-induced superconducting gap
on the $j$th nanowire (at zero magnetic field). $Q_j = e (-2 i \partial_{\varphi_j} + N_j - n_{\text{off}j})$ 
is the excess charge on the $j$th superconducting island - nanowire
combination. $-i \partial_{\varphi_j}$ represents the number of Cooper pairs on the $j$th superconducting island,
$N_j = \int \psi_{j}^\dagger \psi_{j}$ is the total number of electrons on the $j$th nanowire,
and $n_{\text{off}j}$ is the remaining offset charge on the $j$th island, which can be tuned continuously with the gate voltage $V_{gj}$.
The capacitance matrix is given by $C_{ij} = (C_g + 2 C_J) \delta_{ij} - C_J$, for $j = A,B$.

In order to decouple the fermions $\psi_j$ in $H_{\text{BdG}}$ from the phase fluctuations $\varphi_j$, we
perform a unitary transformation $U = e^{-i \sum_{j =A,B} (N_j/2 - n_{Mj}/2 )\varphi_j}$:
$H_{ss} \rightarrow U H_{ss} U^\dagger = \sum_j H_{\text{BdG}}[ \Delta_{0j}, \psi_j^\dagger, \psi_j] + H_+ + H_-$, with
\begin{align}
H_- &= E_{C-} \left(- i \partial_{\varphi-} + \frac{n_{M-} - n_{\text{off},-}}{4} \right)^2 - E_J \cos( \varphi_-) ,
\nonumber \\
H_+ &= E_{C+} \left(N_+' + \frac{n_{M+} - n_{\text{off},+}}{4} \right)^2,
\end{align}
where we have defined $\varphi_- = \varphi_A - \varphi_B$, $E_{C-} = 4 e^2/(C_g + 2 C_J)$, $E_{C+} = 4 e^2/C_g$, $n_{\text{off}\pm} = n_{\text{off}A} \pm n_{\text{off}B}$,
$n_{M\pm} = n_{MA} \pm n_{MB}$, and $N_+' = -i\partial_{\varphi_+}/2$. For wires $A,B$ in the topological
superconducting phase, at energies below the single particle gap $\Delta_j$, $H_{\text{BdG}}$ creates  essentially decoupled Majorana
zero modes $\gamma^{j}$ which affect the phase dynamics only through the occupation numbers
$n_{MA} = (1+ i \gamma^z \gamma^t)/2$, $n_{MB} = (1+ i \gamma^x \gamma^y)/2$.
In order to allow the MZMs to remain free except for a constraint on the total fermion parity,
we consider tuning the gate voltages so that $n_{\text{off}+} = 2 m + 1$, where $m$ is an integer, and $n_{\text{off}-} = 0$.
The ground state of the system is then  two-fold degenerate, with
$N_+' = m/2$, $n_{M+} = 1$, and $n_{M-} = \pm 1$. There is an energy gap on the order of $E_{C+}$ to
violating the gauge constraint by changing the total charge of the system,
and, for $E_J < E_{C-}$, a gap of order $E_{C-}$ to excited states of $H_-$ that are related to fluctuations of
the relative phase $\varphi_-$.

For energy scales below $E_{C\pm}$, the system can therefore be described as an effective spin-1/2 system, with $S^z = n_{M-} = \pm 1$.
Tuning $n_{\text{off}-}$ slightly away from zero acts
like a Zeeman field for this effective spin degree of freedom, giving an effective Hamiltonian
$H_{\text{eff, ss}} = h_z S^z$, with $h_z \propto n_{\text{off}-}$. This is equivalent to tunneling terms between the
Majorana modes $\gamma^t$ and $\gamma^z$. Effective Zeeman fields $h_x S^x$, $h_y S^y$
can also be induced by allowing electrons to tunnel between the MZMs $\gamma^x$, $\gamma^t$ and
$\gamma^y$, $\gamma^t$ (see Supplemental Materials). We emphasize that the MZMs $\gamma^a$ are not exactly
equivalent to the $\tilde{\gamma}^a$ used in Eq. \ref{2dMajH}, because the two effective spin states
in this setup differ not only in the Majorana occupation numbers $n_{M-}$, but also in the wave function
of $\varphi_-$ (see Supplemental Material). We can think of $\tilde{\gamma}$ as corresponding to $\gamma$,
dressed with the $\varphi_-$ degrees of freedom.

The next step is to generate the quartic Majorana couplings in Eq.~\ref{2dMajH} by coupling
the different $A-B$ island pairs together. For example, the $J_z$ term in Eq.~\ref{2dMajH}
essentially represents a coupling between the occupation numbers
$n_{M,A,\vec{r}}$ and $n_{M,B,\vec{r}+\hat{z}}$ of neighboring SC islands in the lattice and can be realized
using a capacitor $C_Z$. This is shown in Fig. \ref{fig1}(C), where we have also introduced the labelling $1,...,4$
for the vertically coupled islands. For numerical optimization of energy scales, it is useful to also consider a capacitance $C_Z'$ between
islands $1$ and $4$, which is not shown explicitly in Fig. \ref{fig1}.

To estimate the resulting $J_z$ coupling, we consider a detailed model for two vertically coupled effective spins, consisting of four vertically
separated islands (see Fig. \ref{fig1}C), which is described by a Hamiltonian
\begin{align}
\label{H2s}
H_{2s} =& H_{12} + H_{34} + H_{1234}.
\end{align}
$H_{12}$ and $H_{34}$ are the Hamiltonians for the isolated units of the form of Eq.~\ref{singleSpinH},
while $H_{1234} = \sum_{\sigma_1, \sigma_2 = \pm} Q_{12,\sigma_1}Q_{34,\sigma_2} A_{\sigma_1\sigma_2}$
capacitively couples the two effective spins. As expected $H_{1234}$ couples the differences  $Q_{ij,\pm} = Q_i \pm Q_j$ of
the charges $Q_j$ on the islands. $A_{\sigma_1\sigma_2} = \frac{1}{4} (C_{13}^{-1} + \sigma_2 C_{14}^{-1}
+ \sigma_1 C_{23}^{-1} + C_{24}^{-1})$ are related to the four-island capacitance matrix (see Supplemental Materials).

The term $H_{1234}$ in Eq.~\ref{H2s} generates a coupling $J_z$ between the effective spins $1-2$ and $3-4$ by
coupling the charges $Q_j$ on the various islands. For small coupling capacitances $C_Z,C_Z'$, this can be estimated perturbatively.
The limits of validity of the perturbative estimate can be checked by a direct numerical calculation of the spectrum of $H_{2s}$,
which we have performed and presented in the Supplemental Material. An example of a suitable parameter regime is for
$C_Z = C_Z' = 0.5 C_g$, $C_J = 1.5 C_g$, and $E_J = 0.45 e^2/C_J$. In this case, we find
$J_z \approx 0.02 e^2/C_J$, while the gap to all other states in the system is $E_{\text{gauge}} \approx 10 J_z$.
Thus the gauge constraint is implemented effectively through a large energy penalty $E_{\text{gauge}}$, and
the system is well-described at low energies by the effective spin model
(or, equivalently, the constrained Majorana model).

The $J_{xy}$ terms in Eq.~\ref{2dMajH} involve coupling MZMs
in the horizontal direction. 
This quartic Majorana coupling can be obtained from single electron tunneling processes between the MZMs through
(normal) semiconductor wires that run horizontally, as shown in Fig. \ref{fig1}C.
The electron tunneling amplitudes $t_A$ and $t_B$ can also be controlled with a gate voltage.
The resulting Hamiltonian for the full 2D system shown in Fig. \ref{fig1}C,D is then
\begin{align}
\label{2dH}
H_{2D} &= \sum_{\vec{R}} H_{2s, \vec{R}} + H_{\text{tun}},
\nonumber \\
H_{\text{tun}} &= \sum_{\vec{r}} [t_A \psi_{t,\vec{r}}^\dagger \psi_{t,\vec{r}+\hat{x}} + t_B \psi_{y,\vec{r}}^\dagger \psi_{x,\vec{r}+\hat{x}} + H.c.],
\end{align}
where $H_{2s,\vec{R}}$ is the Hamiltonian for the two-spin unit cell at $\vec{R}$, given by (\ref{H2s}) above.
The single electron tunnelings $t_A$, $t_B$ violate the gauge constraint, which is related to fermion parity of the single
effective spin, and thus induce an energy penalty on the order of $E_{\text{gauge}}$.
We consider the limit where $t_A, t_B \ll E_{\text{gauge}}$, so that $H_{\text{tun}}$ can be treated perturbatively around the decoupled unit cell limit.
Assuming further that $t_A \ll \Delta_A$, $t_B \ll \Delta_B$, where $\Delta_j$ is the single-particle gap
on the $j$th superconducting nanowire, we can replace $\psi_{\alpha,\vec{r}}$, after the unitary transformation $U$, by the MZMs:
\begin{align}
U^\dagger \psi_{\alpha,\vec{r}} U &= e^{i  \varphi_{j\vec{r}}(1 - F_{pj,\vec{r}})/2} u^\alpha_{\vec{r}} \gamma^\alpha_{\vec{r}},
\end{align}
where we have set $\psi_{\alpha,\vec{r}} = u^\alpha_{\vec{r}} \gamma^\alpha_{\vec{r}}$ after the unitary transformation $U$, the c-number
$u^\alpha_{\vec{r}}$ is the wave function of the MZM, and
$F_{pA,\vec{r}} = i \gamma^z_{\vec{r}} \gamma^t_{\vec{r}}$, $F_{pB,\vec{r}} = i \gamma^x_{\vec{r}} \gamma^y_{\vec{r}}$
are the fermion parities of the $A$ and $B$ islands of site $\vec{r}$. It is useful to define
$\tilde{t}_A = t_A u_{t,\vec{r}}^* u_{t,\vec{r}+\hat{x}}$, $\tilde{t}_B = t_B u_{y,\vec{r}}^* u_{x,\vec{r}+\hat{x}} $, so that
after the unitary transformation by $U$, $H_{\text{tun}} = \sum_{\vec{r}} [\tilde{t}_A e^{-i (1+F_{p,A,\vec{r}})\varphi_{A\vec{r}}/2 + i  (1-F_{p,A,\vec{r}+\hat{x}})\varphi_{A\vec{r}+\hat{x}}/2} \gamma_{t,\vec{r}} \gamma_{t,\vec{r}+\hat{x}}
+ \tilde{t}_B e^{-i (1+F_{p,B,\vec{r}})\varphi_{B\vec{r}}/2 + i  (1-F_{p,B,\vec{r}+\hat{x}})\varphi_{B\vec{r}+\hat{x}}/2}  \gamma_{y,\vec{r}} \gamma_{x,\vec{r}+\hat{x}} + H.c.]$.
Treating $H_{\text{tun}}$ perturbatively around the decoupled unit cell limit, we obtain an effective Hamiltonian
\begin{align}
H_{\text{eff}} = &
\left|\frac{\tilde{t}_A \tilde{t}_B}{E_{\text{gauge}}}\right| \sum_{\vec{r}} \sum_{a,b = x,y} c_{ab} S^a_{\vec{r}} S^b_{\vec{r}+\hat{x}}
+ \sum_{\vec{R}} J_z S^z_{\vec{R}} S^z_{\vec{R}-\hat{z}} .
\end{align}
$c_{ab}$ are constants that depend on parameters of the model, in particular the angle $\theta \equiv Arg(\tilde{t}_A \tilde{t}_B^*)$.
For $\theta \approx 0, \pi$ and $E_J \approx E_{C_J} = e^2/C_J$, we find $c_{yx} \gg c_{xx}, c_{xy}, c_{yy}$.
Therefore, up to negligible corrections, we arrive at the effective Hamiltonian (\ref{2dspinH}) above, with
$J_{yx} = \left|\frac{\tilde{t}_A \tilde{t}_B}{E_{\text{gauge}}}\right| c_{yx}$.
Physically, the angle $\theta$ can be tuned by the applied magnetic flux piercing the loop defined by
the tunneling paths $t_A, t_B$ and also the angle between the Zeeman field and the Rashba splitting of the wire.

In the Supplemental Materials we provide detailed numerical estimates for the parameters of the effective spin model. For the
parameter choice $C_Z = C_Z' = 0.5 C_g$, $C_J = 1.5 C_g$, $E_J = 0.45 e^2/C_J$ described above, and $|\tilde{t}_A|, |\tilde{t}_B| \approx 0.2 E_{\text{gauge}}$
$\theta \approx 0, \pi$, we find that $|c_{yx}| \approx 1.75$, $|c_{xy}| = 0.3$, $c_{xx}, c_{yy} \approx 0$, and thus $J_{yx} \approx 0.016 e^2/C_J$.
Combined with the above estimate $J_z \approx 0.02 e^2/C_J$, we see that the energy scales $J_{yx}, J_z$ are on the order of a few percent
of the charging energy of the Josephson junctions. For Al-InAs-Al Josephson junctions, one can achieve\cite{doh2005}
$E_{J} \approx 3-4 \text{ K} \approx E_{C_J}/2$, implying that $J_{yx},J_z$ would have energy scales on the order of several
hundred milli-Kelvin. Nb-InAs-Nb Josephson junctions would in principle be able to support energy scales on the order of 7-8 times
larger, with $J_{z,xy}$ on the order of several Kelvin, due the correspondingly larger superconducting gap of Nb.

The phase diagram of Eq. (\ref{2dspinH})-(\ref{2dMajH}) contains two phases \cite{kitaev2006}:
a phase where the fermions $\tilde{\gamma}_t$ form a trivial insulating phase when $J_z \gg J_{yx}$, and a phase
where the fermions $\tilde{\gamma}^t$ are gapless with a Dirac-like node when
$J_z \approx J_{yx}$. Both phases have an Abelian topological sector associated
with the $Z_2$ gauge fields $u_{xy,z}$. It is possible to open a topological gap for the Dirac node,
and thus realize the non-Abelian Ising phase, by breaking the effective time-reversal symmetry of (\ref{2dspinH}) with a Zeeman field
$\sum_{\vec{r}} \sum_{\mu =x,y,z} h_a S_{\vec{r}}^a$, the implementation of which was described above.

\begin{figure}
	\centering
	\includegraphics[width=3.5in]{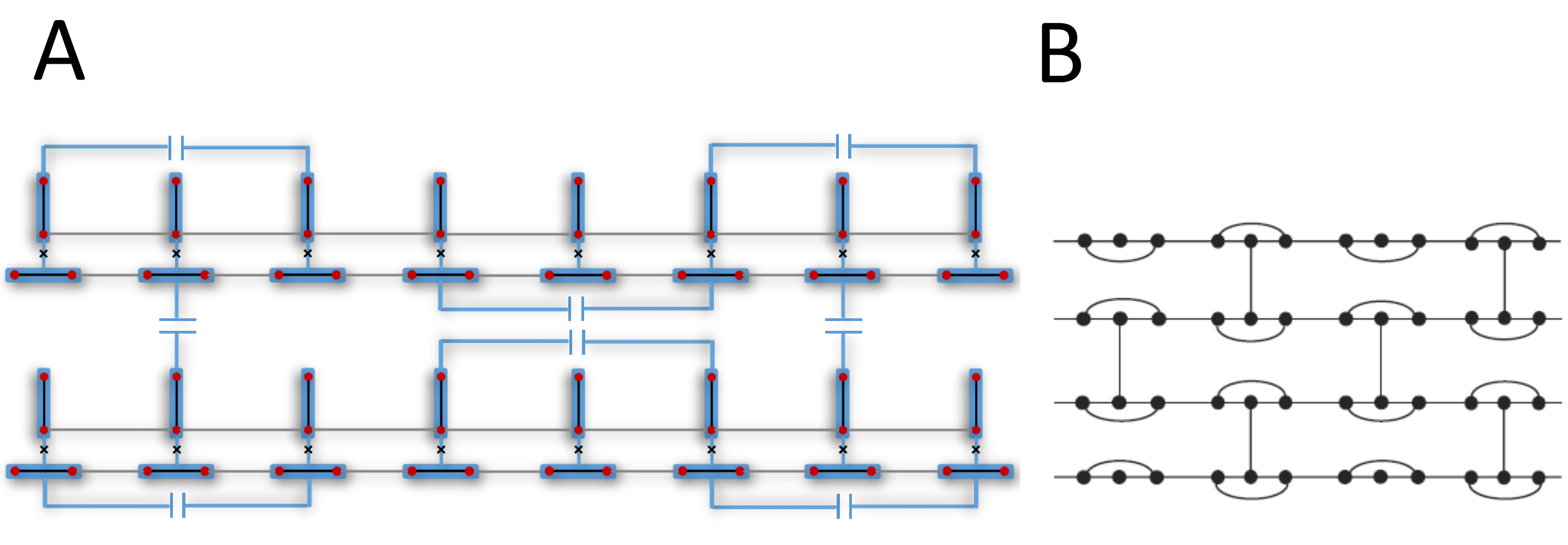}
	\caption{\label{fig2} Decorated brick lattice, to realize variant of the Kitaev model \cite{yao2007}. (A) A single plaquette shown made of
the superconductor-semiconductor system. (B) Effective lattice model. Each unit cell now consists of 6 effective sites of the lattice.
}
\end{figure}

A potentially more optimal approach to inducing the non-Abelian Ising phase is to use a modified structure where each point
of the brick lattice of Fig. \ref{fig1} is expanded into three points, with the couplings as
shown in Fig. \ref{fig2} \cite{yao2007}. The ground state on this lattice spontaneously breaks the effective
time-reversal symmetry of (\ref{2dspinH}) and gaps out the Dirac nodes in the regime where $J_z\sim J_{xy}$ to open a topological gap
on the order of $J_z \sim J_{xy}$.


In our system, the small perturbations $h_z$, $c_{xy}, c_{xx}, c_{yy}$
can be used to controllably tune the sign of the spontaneous time-reversal symmetry breaking and thus control whether
the system enters the Ising phase or its time-reversed conjugate, denoted $\overline{\text{Ising}}$.

\section*{Ising $\times$ Ising phase and genons}

\begin{figure}
	\centering
	\includegraphics[width=2.5in]{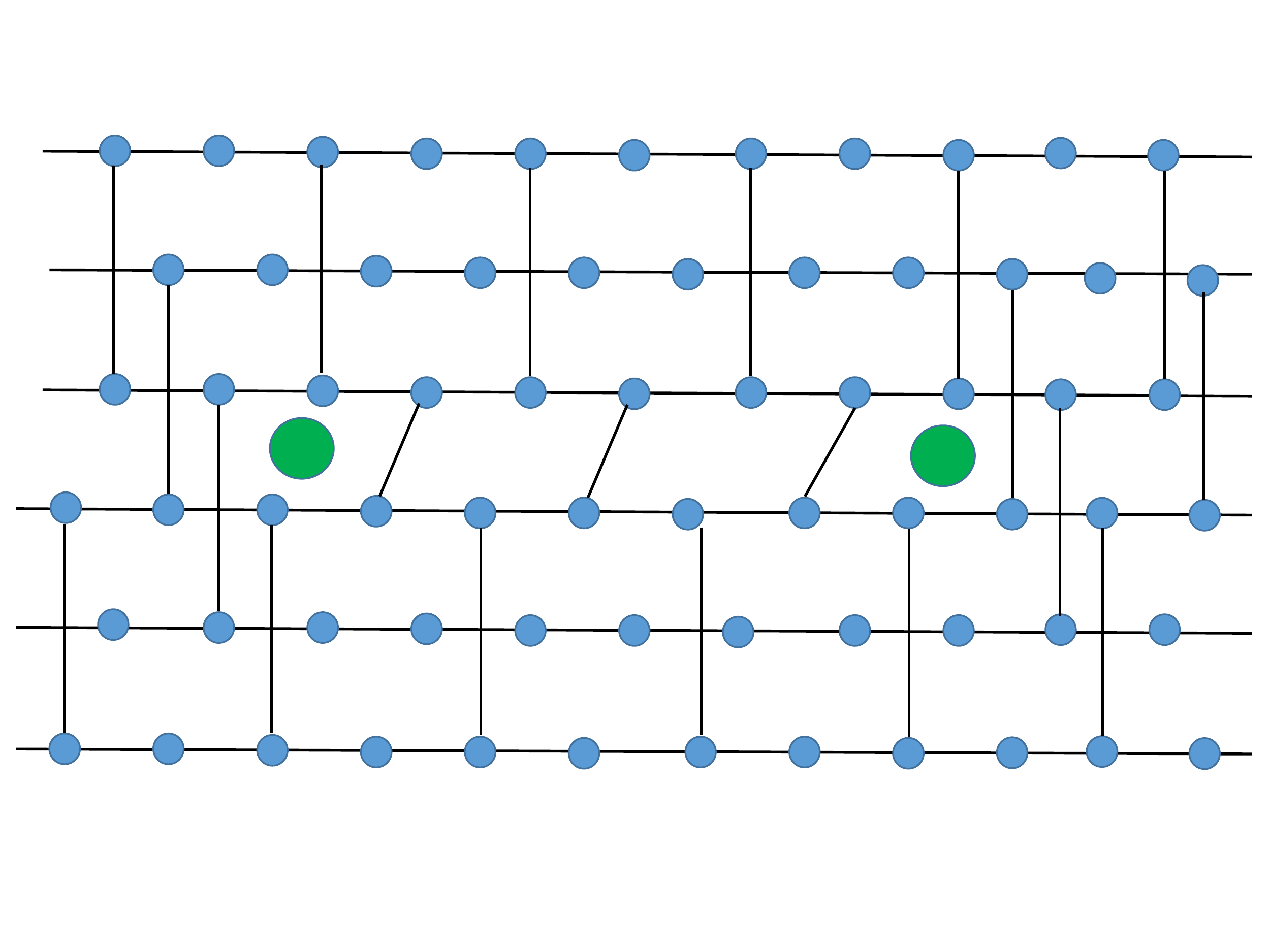}
	\caption{\label{fig3} Two effectively decoupled copies of the Kitaev model can be created with short overpasses where the vertical
superconducting wires skip over one chain and couple to the next chain. Genons (green circles) can be created
at the endpoint of a branch cut along which the vertical couplings connect the two copies of the model together.
}
\end{figure}

A crucial feature of the physical setup that we have proposed is that the vertical couplings between neighboring spins
only involve capacitances (or Josephson junctions as described in the quantum phase slip based implementation below).
This means that once a single copy of the model is realized, it is straightforward to realize two effectively independent copies of the model by creating
short overpasses. Specifically, this can be done as shown in Fig. \ref{fig3} by fabricating the superconducting wires that run
in the vertical direction to pass over one pair of nanowires and to couple capacitively to the next chain over in the vertical direction.

The Ising phase contains three topologically distinct classes of quasiparticle excitations, labelled as $\mathbb{I}$, $\psi$, and $\sigma$.
The Ising $\times$ Ising phase contains nine topologically distinct classes of quasiparticles, which we label as $(a,b)$, for $a,b = \mathbb{I}, \psi, \sigma$.

A genon, which we label $X_{\mathbb{I}}$ in the Ising $\times$ Ising phase is a defect in the capacitive couplings between
vertically separated chains, associated with the endpoint of a branch cut that effectively glues the two copies to
each other (see Fig. \ref{fig3})  \footnote{Note that only half of a branch cut is sufficient, following \cite{barkeshli2014exp}. A full branch cut
would also contain vertical couplings that skip over two chains.}.
This defect in the lattice configuration of the superconducting islands is not a
quasiparticle excitation of the system, but rather an \it extrinsically \rm imposed defect with projective
non-Abelian statistics \cite{barkeshli2013genon,barkeshli2014sdg} \footnote{See \cite{barkeshli2010,bombin2010,kitaev2012,barkeshli2012a,
you2012,clarke2013,lindner2012,cheng2012,vaezi2013}for other examples of twist defects in other topological states}.
$X_{\mathbb{I}}$ has quantum dimension $2$, and possesses the following fusion rules \footnote{Technically,
there are three topologically distinct types of genons \cite{barkeshli2014sdg}, as $X_{\mathbb{I}}$  can be bound
to the quasiparticles, although this additional complication will be ignored in the present discussion.}:
\begin{align}
X_\mathbb{I} \times X_{\mathbb{I}} &= (\mathbb{I},\mathbb{I}) + (\psi,\psi) + (\sigma,\sigma).
\end{align}

In \cite{barkeshli2013genon}, it was shown that the braiding of genons maps to Dehn twists of the Ising state on a high genus surface,
which is known \cite{bravyi200universal,freedman2006universal} to provide a topologically protected $\pi/8$ phase gate. The protocol
for implementing the $\pi/8$ phase gate using genon braiding was described in \cite{barkeshli2013genon}.
In the present system, the braiding of genons is complicated by the fact that it is difficult to continuously modify the physical
location of the genons to execute a braid loop in real space. Fortunately, this is not necessary, as the braiding of the genons can be implemented through a
different interaction-based approach, without moving the genons, as described for general anyon systems in \cite{bonderson2013braiding}.
To do this, we require that it be possible to project the joint fusion channel of any pair of genons into either the $(\mathbb{I},\mathbb{I})$ channel
or the $(\psi,\psi)$ channel. This can be done by adiabatically tuning the effective interactions between the genons,
similar to proposals for braiding MZMs in nanowire networks \cite{alicea2010b,sau2011}.

In order to implement the $\pi/8$ phase gate, we wish to start with two pairs of genons, labelled $1,..., 4$, and have
the ability to braid genons $2$ and $3$. To do this, we require an ancillary pair of genons, labelled $5$ and $6$.
The braiding process is then established by adiabatically changing the Hamiltonian of the system to effectively
project the genons $5$ and $6$ onto the fusion channel $b_{56}$,
then the genons $5$ and $3$ onto the fusion channel $b_{35}$, the genons $5$ and $2$ onto the fusion channel $b_{25}$,
and finally again the genons $5$ and $6$ onto the fusion channel $b_{56}'$ (see Fig. \ref{genonBraiding}).
\begin{figure}
	\centering
	\includegraphics[width=3.0in]{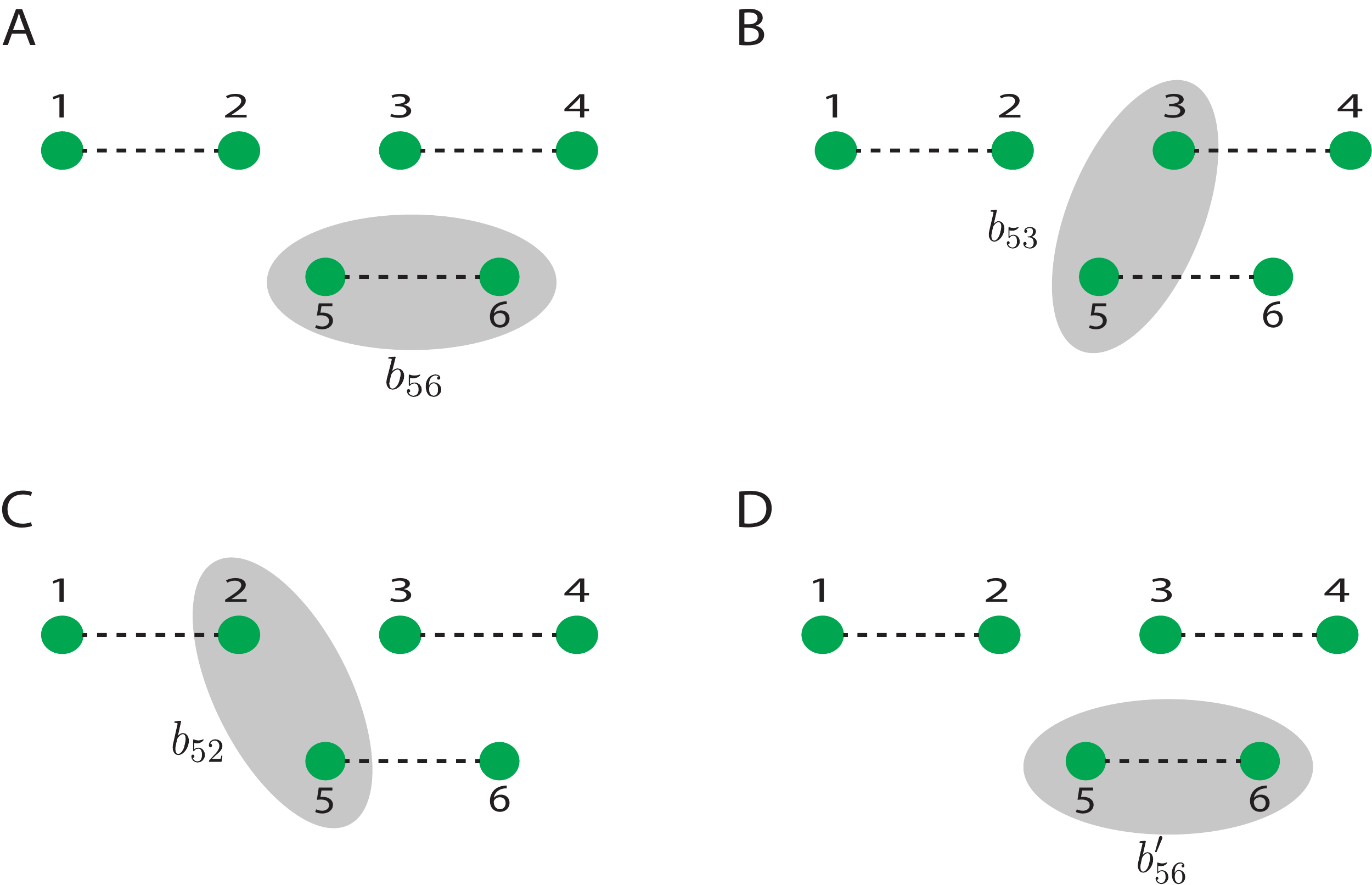}
	\caption{\label{genonBraiding} Genons can be braided without moving them by sequentially projecting different pairs of genons onto specific fusion channels, as depicted by the gray ellipses in A - D. This can be implemented by tuning the interactions between the genons by decreasing the quasiparticle gap along various paths.
}
\end{figure}
We will asssume here for simplicity that $b_{56} = b_{56}'$. If the genons $5$ and $6$ are created out of the vacuum, then
it will in fact be natural to have $b_{56} = b_{56}' = (\mathbb{I}, \mathbb{I})$. As long as $b_{56}, b_{35}$, $b_{25}$ are Abelian,
i.e. equal to either $(\mathbb{I},\mathbb{I})$ or $(\psi, \psi)$, then the results of \cite{barkeshli2013genon,bonderson2013braiding}
imply that the matrix obtained for a double braid (i.e. a full $2\pi$ exchange), is given by
\begin{align}
(R_{23})^2 = e^{i\phi} \left(\begin{matrix} 1 & 0 & 0 \\ 0 & 1 & 0 \\ 0 & 0 & e^{i\pi/4}\end{matrix} \right),
\end{align}
where $e^{i\phi}$ is an undetermined, non-topological phase.
In other words, the state obtains a relative phase of $e^{i\pi/8}$ if the fusion channel is
$(\sigma, \sigma)$ as compared with $(\mathbb{I}, \mathbb{I})$, or $(\psi, \psi)$.

When two genons are separated by a finite distance $L$, the effective Hamiltonian in the degenerate subspace spanned
by the genons obtains non-local Wilson loop operators:
\begin{align}
H_{\text{genon}} = t_\psi W_{(\psi,\psi)} + t_\sigma W_{(\sigma,\sigma)} + H.c.
\end{align}
$W_{(a,a)}$ describes the exchange of a $(a,a)$ particle between the two genons, which equivalently corresponds to a $(\mathbb{I},a)$ or $(a, \mathbb{I})$ particle
encircling the pair of genons. $t_a \propto e^{-L \epsilon_a/v_a}$, for $a = \psi, \sigma$, are the tunneling amplitudes, with $\epsilon_a$ being the energy gap for the $a$
quasiparticles, and $v_a$ their velocity.
When a $(1,a)$ quasiparticle encircles a topological charge $(b,b)$, it acquires a phase $S_{ab}/S_{\mathbb{I}b}$, where $S$ is
the modular $S$ matrix of the Ising phase \cite{nayak2008}. As we show in the Supplemental Materials, this implies that for
our purposes, we only need to ensure that $t_\sigma \neq 0$ and, in the case where $|t_\sigma| < |t_\psi|$, we must have $t_\psi < 0$.

The tunneling amplitudes $t_\psi$, $t_\sigma$ can be tuned physically by tuning the parameters of the model, such as the
electron tunneling amplitudes $t_A$, $t_B$, the capacitances $C_Z$, and the gate voltages $V_{gj}$.
Therefore, to tune the interactions between two desired genons, we tune the parameters of the model
in order to decrease the energy gap to the quasiparticle excitations along the path that connects them.
A more detailed study of this will be left for future work.

We note that it is also possible to implement effectively the same physics by using instead the Ising $\times$ $\overline{\text{Ising}}$ state.
In this case, the genons are replaced by holes with gapped boundaries, and the topological charge projections are implemented along various open lines
that connect the different gapped boundaries. A detailed discussion of this variation is presented in the Supplemental Materials.

\section*{Quantum Phase Slip Limit}


An alternative architecture is also possible, if we replace the purely capacitive coupling $C_Z$
by a Josephson junction, with Josephson coupling $E_{J_Z}$ and capacitance $C_Z$.
In this case we consider the limit where the Josephson energies $E_J$, $E_{J_Z}$ are much larger than the charging
energies $e^2 C^{-1}_{ij}$, leading to a state with long range phase coherence. In the limit where
the charging energies are ignored, the system has a large degeneracy due to the MZMs.
A small charging energy induces quantum phase slips of the superconducting phase of the islands; the amplitude
of the quantum phase slips depends on the occupation of the MZMs on the superconducting islands,
thus inducing an effective Hamiltonian in the space of states spanned by the MZMs. The effective
Hamiltonian takes the form: $H_{\text{2D}} = H_1 + H_2 + H_{\text{tun}}$, where $H_1$ consists of single-island phase
slips:
\begin{align}
H_{1} = \sum_{\vec{r}} ( \zeta_{\vec{r}}^A i \gamma_{\vec{r}}^z \gamma_{\vec{r}}^t + \zeta_{\vec{r}}^B i \gamma_{\vec{r}}^x \gamma_{\vec{r}}^y),
\end{align}
and $H_2$ consists of double island phase slips:
\begin{align}
H_2 = -\sum_{\vec{r}} \zeta_{\vec{r}}^{AB} \gamma_{\vec{r}}^z \gamma_{\vec{r}}^t \gamma_{\vec{r}}^x \gamma_{\vec{r}}^y
-\sum_{\vec{R}} \zeta_{Z}^{AB} \gamma_{\vec{R}}^x \gamma_{\vec{R}}^y \gamma_{\vec{R}-\hat{z}}^z \gamma_{\vec{R}-\hat{z}}^t.
\end{align}
$H_{\text{tun}}$ is the same as in Eqn. (\ref{2dH}) and describes electron tunneling in the horizontal direction.
The single island phase slips are modulated by the offset charge: $\zeta^j_{\vec{r}} \propto \cos(\pi n_{\text{off}, \vec{r}, j})$
and can therefore be tuned to zero using the gate voltages. Double island phase slips that are not included in $H_2$ can
be ignored in this limit, as can phase slips that involve more than two islands, as they are exponentially suppressed.
We wish to choose parameters to operate in the limit
$\tilde{t}_A, \tilde{t}_B, \zeta_Z^{AB} \ll \zeta^{AB}_{\vec{r}}$.
In this case, $\zeta^{AB}_{\vec{r}}$ effectively imposes the gauge constraint $\gamma_{\vec{r}}^z \gamma_{\vec{r}}^t \gamma_{\vec{r}}^x \gamma_{\vec{r}}^y = 1$
for states with energies much less than $\zeta^{AB}_{\vec{r}}$. Each site can therefore be described by a spin-1/2 degree of freedom.
The term involving $\zeta_Z^{AB}$ acts like a coupling $\zeta_Z^{AB} S^z_{\vec{R}} S^z_{\vec{R} - \hat{z}}$. As before, in this limit
$H_{\text{tun}}$ can be treated perturbatively, and gives rise to the desired coupling
$\frac{|\tilde{t}_A \tilde{t}_B|}{\zeta^{AB}} S^y_{\vec{r}} S^x_{\vec{r}+\hat{x}}$.



\section*{Acknowledgments}

We thank Michael Freedman, Charlie Marcus, Chetan Nayak, Zhenghan Wang, Javad Shabani, Vladimir Manucharyan,
Matthias Troyer, Roman Lutchyn, Meng Cheng, Parsa Bonderson for discussions. J. S. was supported by Microsoft Station Q,
startup funds from the University of Maryland and NSF-JQI-PFC During the preparation of
this manuscript we became aware of ongoing work that is related to the analysis of interaction-based braiding
of genons in the Ising $\times$ Ising state \cite{lindner2015}.

\bibliography{TI}

\newpage
\appendix
\newpage

\begin{widetext}

\section{Supplemental Material}

\section{Charging Energy Based Implementation}
\label{chargingSec}

\begin{figure}
	\centering
	\includegraphics[width=4.5in]{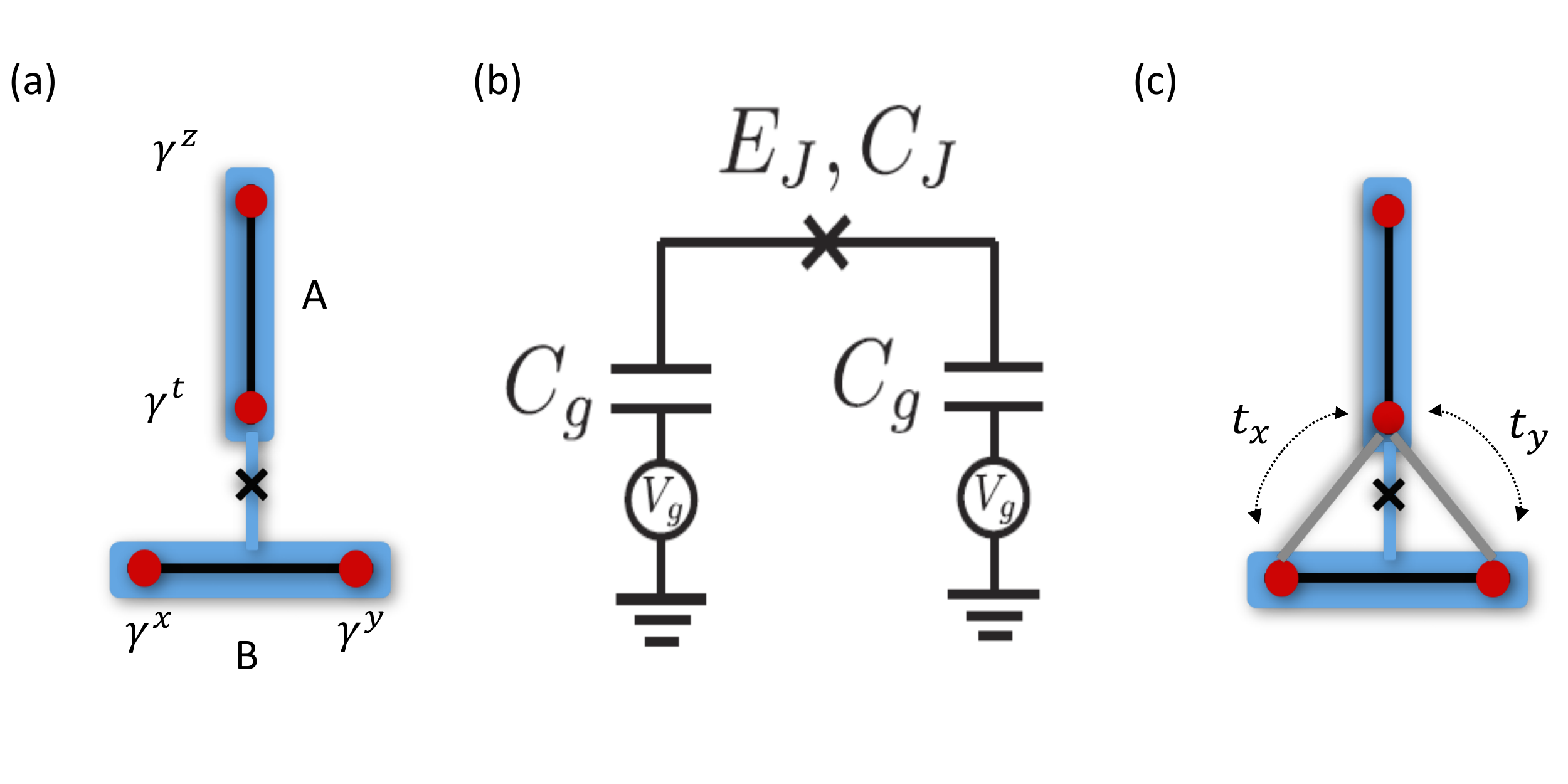}
	\caption{\label{capSpin} (a) A pair of $A$, $B$ superconducting islands, coupled together with a Josephson junction, denoted by the $\times$.
The red dots represent Majorana zero modes, labelled $\gamma^x, \gamma^y,\gamma^z,\gamma^t$. The Majorana zero modes are
localized at the end of the semiconducting nanowires, which are shown in black. Each $A$ and $B$ island
has a capacitance $C_g$ to a gate voltage $V_g$, which is then grounded. (b) An effective circuit for the pair of $A$, $B$ islands, indicating the
capacitances and Josephson junctions. (c) Grey wires indicate normal (non-superconducting) semiconductor wire.
Tuning the voltage on the semiconductor wires allows tuning the electron tunneling amplitudes $t_x$ and $t_y$ between the Majorana zero modes.
}
\end{figure}

We begin by describing an implementation where charging energies are the dominant energies in the system. We proceed by
incrementally increasing the complexity of our analysis, by first describing the physics of a single effective spin,
then two vertically coupled spins that will form the unit cell of the Kitaev model, and subsequently the horizontal
couplings that will link all of the two-spin unit cells into the full effective spin model.

\subsection{Single spin}
\label{chargingSecSS}

Let us consider the configuration shown in Fig. \ref{capSpin}(a), which consists of two superconducting islands, labelled $A$ and $B$,
each of which is proximity coupled to a Majorana nanowire. Each superconducting island is separated by a capacitance $C_g$ to a
gate voltage $V_g$. The $A$ and $B$ islands are coupled together through a Josephson junction, with Josephson coupling $E_J$ and
junction capacitance $C_J$. The effective Hamiltonian for this system is
\begin{align}
\label{singleSpinH}
H_{\text{ss}} = \sum_{j = A,B} H_{\text{BdG}} [ \Delta_{0j} e^{i \varphi_j}, \psi_j^\dagger, \psi_j] - E_J \cos(\varphi_A - \varphi_B)
+ \frac{1}{2} \sum_{i,j = A,B} Q_i C_{ij}^{-1} Q_j.
\end{align}
Here, $\varphi_j$ for $j = A,B$ is the superconducting phase on the $A$ and $B$ islands,
$H_{\text{BdG}} [ \Delta_{0j} e^{i \varphi_j}, \psi_j^\dagger, \psi_j]$ is the BdG Hamiltonian for the
nanowire on the $j$th island, where $|\Delta_{0j}|$ is the proximity-induced superconducting gap
on the $j$th nanowire at zero magnetic field. $Q_j$ is the excess charge on the $j$th superconducting island - nanowire
combination; it can be written as:
\begin{align}
Q_j = e (-2 i \partial_{\varphi_j} + N_j - n_{\text{off}j}),
\end{align}
where $-i \partial_{\varphi_j}$ represents the number of Cooper pairs on the $j$th superconducting island,
$N_j = \int \psi_{j}^\dagger \psi_{j}$ is the total number of electrons on the $j$th nanowire,
and $n_{\text{off}j}$ is the remaining offset charge on the $j$th island, which can be tuned continuously with the gate voltage $V_g$.

The capacitance matrix is given by
\begin{align}
C = \left( \begin{matrix}
C_g + C_J & - C_J \\
- C_J & C_g + C_J
\end{matrix} \right)
\end{align}
The charging energy term can be rewritten in terms of the total and relative charges on
the $A$ and $B$ islands:
\begin{align}
\frac{1}{2} \sum_{i,j = A,B} Q_i C_{ij}^{-1} Q_j &= \frac{1}{4 C_g} (Q_A + Q_B)^2  + \frac{1}{4} \frac{1}{C_g + 2C_J} (Q_A - Q_B)^2
\nonumber \\
&= \frac{e^2}{C_g} (-i \partial_{\varphi_A} - i \partial_{\varphi_B} + (N_+ - n_{\text{off}+})/2)^2 + \frac{e^2}{C_g + 2C_J}
(-i \partial_{\varphi_A} + i \partial_{\varphi_B} + (N_- - n_{\text{off}-})/2)^2,
\end{align}
where we have defined
\begin{align}
N_{\pm} &= N_{A} \pm N_B,
\nonumber \\
n_{\text{off}\pm} &= n_{\text{off}A} \pm n_{\text{off} B} .
\end{align}

The BdG Hamiltonian for the nanowire is given by
\begin{align}
H_{\text{BdG}}[ \Delta e^{i\varphi}, \psi^\dagger, \psi] = &
\int_0^L dx [ \psi^\dagger(x) \left( -
  \frac{1}{2m^*} \partial_x^2 - \mu + i \alpha \sigma_y \partial_x +
  g \mu_B \vec{B} \cdot \vec{\sigma} \right) \psi(x)
\nonumber \\
&+ \left( \Delta e^{i \varphi} \psi^\dagger_\up
  \psi^\dagger_\down + H.c. \right) ],
\end{align}
where $\psi = \left( \begin{matrix} \psi_\uparrow \\
    \psi_\downarrow \end{matrix} \right)$. Here we have taken $x$ to be the coordinate along the wire and $L$ is the length of the wire.
$\alpha$ is the Rasha spin-orbit coupling, $\mu$ is the chemical potential, and $m^*$ is the
effective mass of the electrons in the nanowire, $\vec{B}$ is the magnetic field and $g \mu_B |B|$ is the
Zeeman energy.

It is now useful to perform a unitary transformation $U = e^{-i \sum_{j =A,B} (N_j/2 - n_{Mj}/2 )\varphi_j}$ in order to
decouple the phase $\varphi_j$ from the fermions $\psi_j$ in $H_{\text{BdG}}$. Here,
$n_{Mj} = 0,1$ is the occupation number of the pair of Majorana zero modes on wire $j$. It is given in terms of the
Majorana zero modes as
\begin{align}
n_{MA} &= (1+ i \gamma^z \gamma^t)/2,
\nonumber \\
n_{MB} &= (1+ i \gamma^x \gamma^y)/2.
\end{align}
Under this transformation, the charge $Q_j$ transforms as:
\begin{align}
Q_j' = U^\dagger Q_j U = e(-2i \partial_{\varphi_j} + n_{Mj} - n_{\text{off}j}) .
\end{align}
Thus, taking
$H_{ss} \rightarrow U^\dagger H_{ss} U$, we obtain
\begin{align}
\label{hssprime}
H_{ss}' = &U^\dagger H_{ss} U = \sum_j H_{\text{BdG}}[ \Delta_{0j}, \psi_j^\dagger, \psi_j]
+ H_+ + H_-,
\nonumber \\
H_+ &= + \frac{4e^2}{C_g}  \left(- i \partial_{\varphi_+} + \frac{n_{M+} - n_{\text{off},+}}{4} \right)^2
\nonumber \\
H_- &= \frac{4e^2}{C_g + 2C_J} \left(- i \partial_{\varphi-} + \frac{n_{M-} - n_{\text{off},-}}{4} \right)^2 - E_J \cos( \varphi_-) .
\end{align}
Here, we have defined the combinations:
\begin{align}
n_{M\pm} &= n_{MA} \pm n_{MB} ,
\nonumber \\
\varphi_\pm &=  \varphi_A \pm \varphi_B  ,
\nonumber \\
\partial_{\varphi_{\pm}} &= \frac{1}{2} ( \partial_{\varphi_A} \pm \partial_{\varphi_B}) .
\end{align}
With this definition,
\begin{align}
[i\partial_{\varphi_{\rho}} (x), \varphi_{\rho'} (x')] = i \delta_{\rho \rho'} \delta(x - x'), \;\;\; \rho, \rho' = \pm
\end{align}
It is also useful to define
\begin{align}
N_j' &= -i \partial \varphi_j, \;\; j = A,B
\nonumber \\
N_{+}' &= (N_A' + N_B')/2
\nonumber \\
Q_{\pm}' &= Q_A' \pm Q_B'.
\end{align}
With these definitions, we see that the compactification of $\varphi_+, \varphi_-$ is:
\begin{align}
(\varphi_+, \varphi_-) \sim (\varphi_+ + 2\pi, \varphi_- + 2\pi) \sim (\varphi_+ +2\pi, \varphi_- - 2\pi).
\end{align}
While $\varphi_+$ and $\varphi_-$ are formally decoupled in the Hamiltonian, they are coupled through
their boundary conditions. The ground state of $H_{ss}'$ can now be written as
\begin{align}
\frac{1}{2} \int_{-2\pi}^{2\pi} d\varphi_+ d \varphi_- \psi_{n_{MA}, n_{MB}}(\varphi_+, \varphi_-) |\varphi_+, \varphi_- , n_{MA}, n_{MB} \rangle
\end{align}
Here, $|n_{MA}, n_{MB} \rangle$ is the state of the Majorana zero modes, and $|\varphi_+, \varphi_- \rangle$ is the state for the phase degrees of freedom.
Importantly, the wave function of $\varphi_+, \varphi_-$ itself does depend on the values of $n_{MA}$, $n_{MB}$.
Since the Hamiltonians for $H_+$ and $H_-$ are decoupled, we can immediately write the ground state wave function for $\varphi_+, \varphi_-$:
\begin{align}
\psi_{n_{MA}, n_{MB}}(\varphi_+, \varphi_-) = \frac{1}{\sqrt{2\pi}} e^{i N_+' \varphi_+} f_{n_{M-}}( \varphi_-),
\end{align}
where $f_{n_{M-}}(\varphi_-)$ is the ground state wave function for $H_-$, which is peaked at $\varphi_- = 0$.
Importantly, because of the compactification conditions on $\varphi_\pm$, we see that if $N'_+ = (N_A' + N_B')/2$ is
integer (half-integer), then $f_{n_{M-}}(\varphi_-)$ must be periodic (antiperiodic) in $\varphi_- \rightarrow \varphi_- + 2\pi$.

For energies much less than the single-particle gap $\Delta$ on the nanowire, we can ignore the excited single 
particle states associated with $H_{\text{BdG}}$, and we can describe the system by the following effective Hamiltonian:
\begin{align}
H_{\text{eff}} = \frac{4e^2}{C_g} \left( N_+' + \frac{n_{M+} - n_{\text{off}+}}{4} \right)^2 +
\frac{4e^2}{C_g + 2C_J} \left( - i \partial_{\varphi-} + \frac{n_{M-} - n_{\text{off},-}}{4} \right)^2 - E_J \cos( \varphi_-)
\end{align}
Let us define
\begin{align}
E_{C+} &= \frac{4e^2}{C_g},
\nonumber \\
E_{C-} &= \frac{4e^2}{C_g + 2C_J}.
\end{align}
We see that the effect of $H_+$ is to fix the total charge $N_+'$ on the pair of $A$, $B$ islands to a fixed value, and gives an
energy cost of $E_{C+}$ to increase the charge by one unit.
If we set
\begin{align}
n_{\text{off}+} = 2 m + 1, \;\; m \in \mathbb{Z},
\end{align}
then the ground state of the system will be given by
\begin{align}
N_+' =  m/2, \;\;\;\;\; n_{M+} = 1 .
\end{align}
Therefore the system will have two lowest energy states, associated with $n_{M-} = \pm 1$.
Thus we can define an effective spin degree of freedom:
\begin{align}
S^z \equiv n_{M -} = \pm 1.
\end{align}
We will denote these two states as $|S^z \rangle$:
\begin{align}
\label{ssStateDef}
|S^z \rangle \propto \int_{0}^{2\pi} d \varphi_+ d \varphi_- f_{S^z}(\varphi_-)|\varphi_+, \varphi_-, n_{MA} = \frac{1+S^z}{2}, n_{MB} = \frac{1 - S^z}{2} \rangle ,
\end{align}
where we have chosen $N_+' = m/2 = 0$ for simplicity.
Note that $|S^z = \pm 1 \rangle$ differ both in the value of $n_{M-} = \pm$ and also the wave function $f_{n_{M-}}(\varphi_-)$.
For future reference, it will also be useful to define the state
\begin{align}
\label{tildeState}
\widetilde{|S^z \rangle} \propto \int_{0}^{2\pi} d \varphi_+ d \varphi_- f_{-S^z}(\varphi_-)|\varphi_+, \varphi_-, n_{MA} = \frac{1+S^z}{2}, n_{MB} = \frac{1 - S^z}{2} \rangle,
\end{align}
which has the opposite wave function $f_{-S^z}(\varphi_-)$ as compared with $|S^z \rangle$.

The effective Hamiltonian in the two-dimensional space $|S^z = \pm 1 \rangle$ is given (up to an overall constant) by
\begin{align}
H_{\text{eff}} = h_z S^z .
\end{align}
The value of $h_z$ depends on parameters in $H_{\text{eff}}$, as described below.

\subsubsection{Numerical Solution}

The Hamiltonian $H_{ss}' = H_{\text{BdG}} + H_+ + H_-$ (see Eqn. (\ref{hssprime})). The three terms, $H_{\text{BdG}}$, $H_+$, $H_-$ commute with
each other and can be separately solved. As discussed above, $H_{\text{BdG}}$ is gapped for energies below $\Delta$, aside from the zero energy states
arising from the Majorana zero modes, while $H_+$ has a gap of $E_{C+}$. It is useful to solve $H_-$ numerically, for the different Majorana occupation
numbers.

In Fig. \ref{singleSpinPlot1} and Fig. \ref{singleSpinPlot2}, we plot the
energy spectra for the four lowest energy states of $H_{-}$, as a function of the offset charge $n_{\text{off},-}$, for the two different
values of the Majorana occupation numbers $S^z = n_{M-} = \pm 1$.  To connect with some standard notation in the literature for the well-known
Hamiltonian $H_-$, it will be useful to define
\begin{align}
E_C = E_{C-}/4 = \frac{e^2}{C_g+2C_J}.
\end{align}
We see that for energies much smaller than $E_C$, the system simply consists of the two states $|S^z = \pm 1\rangle$.
These are degenerate when $n_{\text{off}-} = 0$, and acquire a small splitting when $n_{\text{off}-} \neq 0$.

\begin{figure}
	\centering
	\includegraphics[width=4.0in]{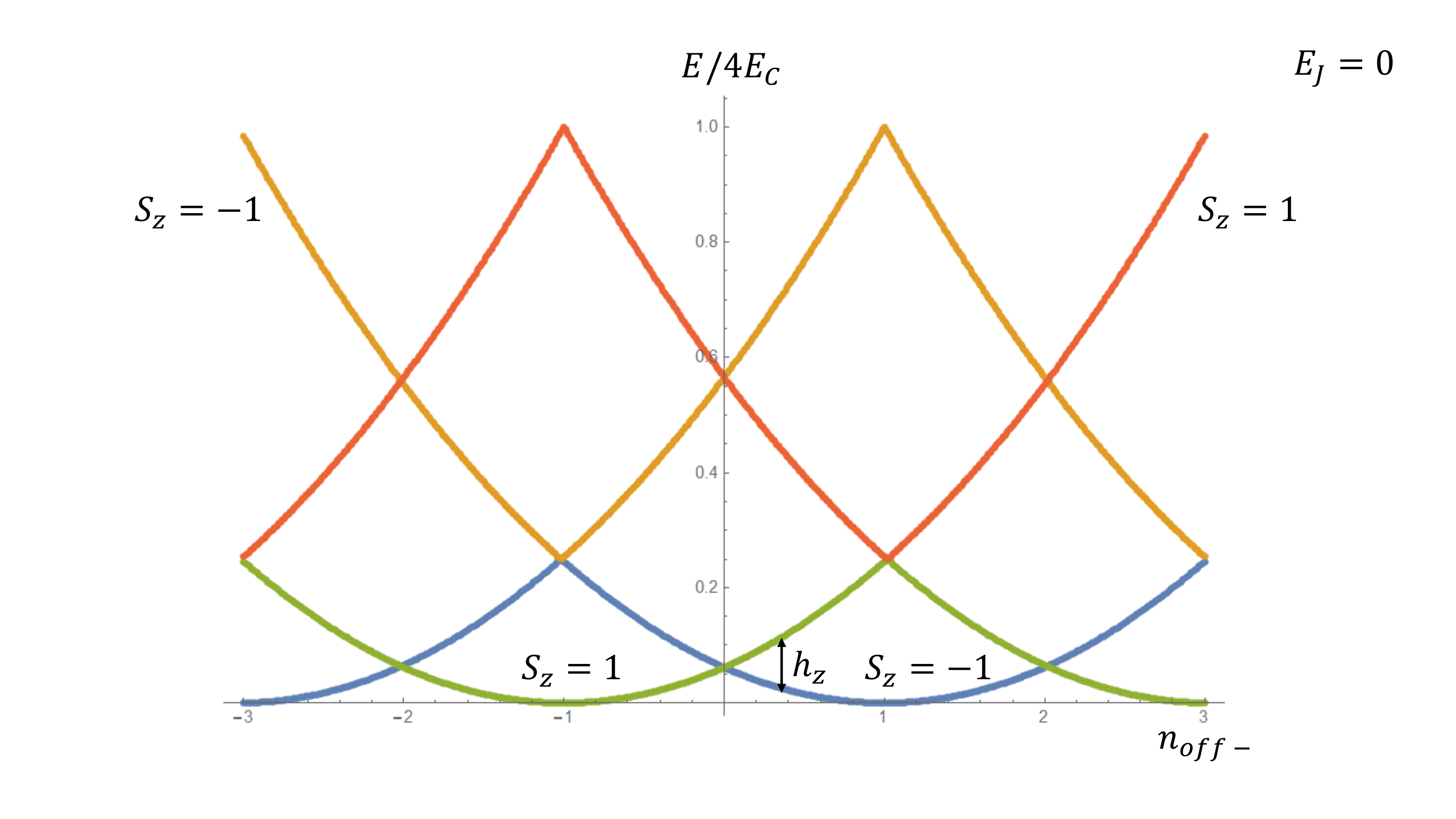}
	\caption{\label{singleSpinPlot1}Plot of energy spectrum of $H_-$, in units of $4 E_C$. We have set $E_J = 0$. Green and red curves are the ground state and the first excited state energies, respectively, for the case $S^z = n_{M-} = 1$, as a function of the offset charge $n_{\text{off}-}$. Blue and yellow curves are the ground state and first excited energies, respectively, for the case $S^z  = n_{M-} = -1$. We can see that at $n_{\text{off}-}=0$, there is a degeneracy between $S^z = \pm 1$, with a gap of order $E_C$ to all other states. Non-zero $n_{\text{off}-}$ acts like a Zeeman field that splits the energies of the two spin states.
}
\end{figure}

\begin{figure}
	\centering
	\includegraphics[width=4.0in]{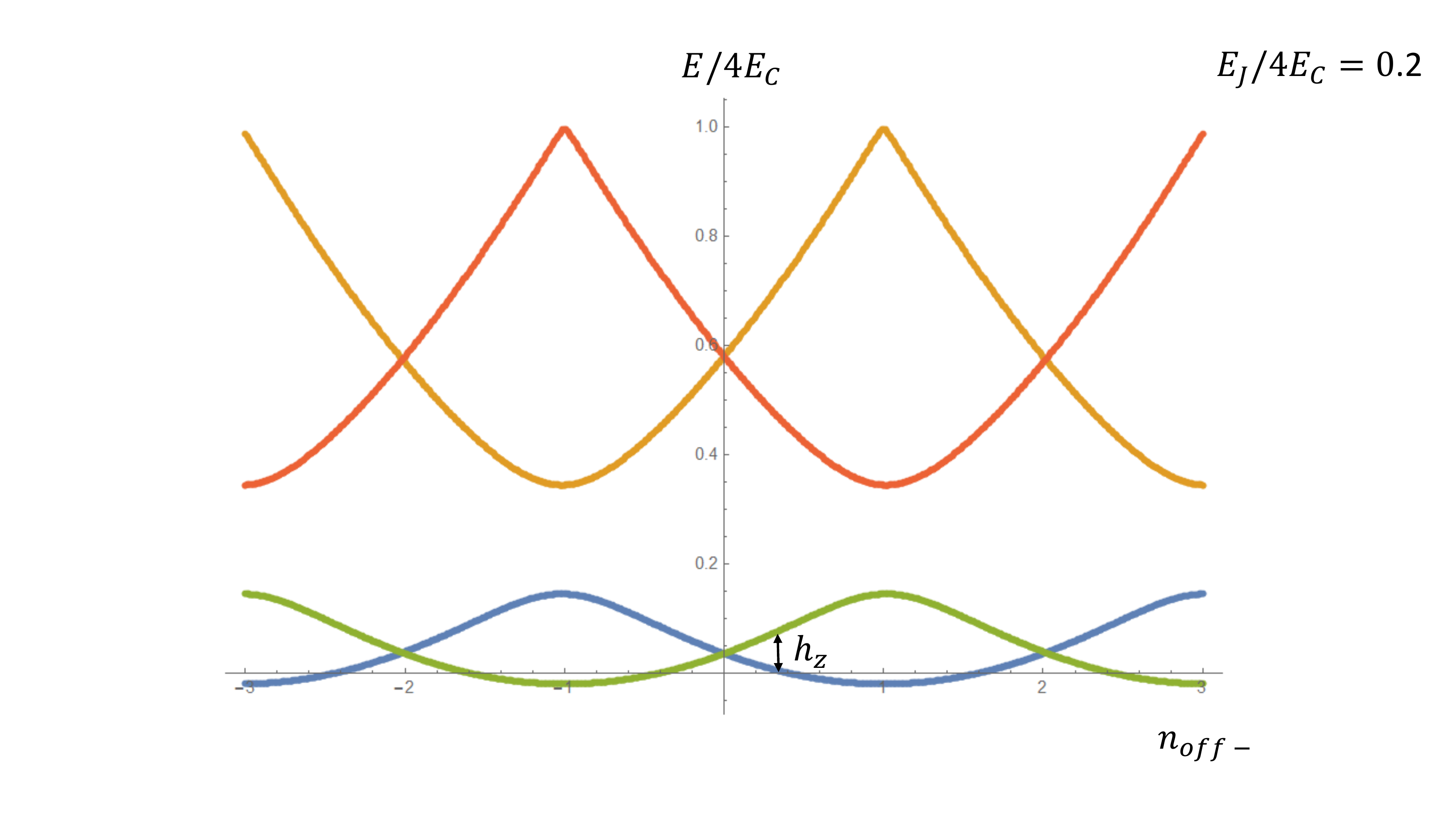}
	\caption{\label{singleSpinPlot2} Same plot as in Fig. \ref{singleSpinPlot1}, but for $E_J = 0.2 (4 E_C)$.
}
\end{figure}

\subsubsection{Analytical Solution}

The Hamiltonian $H_{-}$ can also be fully solved analytically through the use of Mathieu functions (see, e.g., Ref. \onlinecite{koch2007}).
Here we will provide this solution for reference. We find that
\begin{align}
h_z = E_c (a_{\nu_+}(-E_J/2E_C) - a_{\nu_-} (-E_J/2E_C))/2
\end{align}
where $a_\nu(q)$ is Mathieu's characteristic value, and
\begin{align}
\nu_{\pm} = 2[- n_{g\pm} + k(0,n_g\pm)],
\end{align}
\begin{align}
n_{g\pm} &= (\pm 1 - n_{\text{off}-})/4 + \frac{m \text{ mod } 2}{2},
\nonumber \\
k(0,n_{g\pm}) &= \text{int}(n_{g\pm})\sum_{l=\pm 1} (\text{int}(2n_{g\pm} +l/2) \text{ mod } 2 ).
\end{align}
int$(x)$ rounds $x$ to the nearest integer.
The average charge $Q_B'$ on the $B$ island is given by:
\begin{align}
\label{singleSzav}
\langle S^z| Q_B'| S^z \rangle &
= -\frac{1}{2} \langle S^z|  Q_-' |S^z \rangle
= - e \langle S^z| - 2 i\partial_{\varphi_-} + (S^z - n_{\text{off},-})/2 |S^z \rangle
\nonumber \\
&= \frac{e}{E_C}\frac{\langle S^z| \partial H_{ss}' |S^z \rangle}{\partial n_{\text{off}-}} = S^z \frac{e}{E_C} \frac{\partial h_z}{\partial n_{\text{off}-}}.
\end{align}
Evaluating the partial derivative:
\begin{align}
\frac{\partial h_z}{\partial n_{\text{off}-}} = \frac{E_c}{2}
(a'_{\nu_+}(-E_J/2E_C) - a'_{\nu_-}(-E_J/2E_C)) ,
\end{align}
where
\begin{align}
a'_\nu(x) \equiv \frac{\partial a_\nu(x)}{\partial \nu}.
\end{align}
Note we have assumed that $\partial k /\partial n_{g\pm} = 0$, which is true except for certain fine-tuned values of $n_{g\pm}$.

\subsubsection{Effective $S^x$ and $S^y$ terms}

Above we showed that tuning $n_{\text{off}-}$ away from zero effectively acts like a Zeeman field in the $S^z$ direction. Zeeman fields in the
$S^x$ and $S^y$ direction can also be generated, by allowing electron tunneling, with amplitude $t_x$ and $t_y$,
through the semiconducting wires as shown in Fig. \ref{capSpin}c.
Consider the following electron tunneling perturbations to $H_{ss}$:
\begin{align}
\delta H = t_x \psi_x^\dagger \psi_t + t_y \psi_y^\dagger \psi_t + H.c.
\end{align}
After the unitary transformation $U$, $\delta H$ changes:
\begin{align}
\delta H' = U^\dagger \delta H U = t_x (\psi_x')^\dagger \psi_t' + t_y (\psi_y')^\dagger \psi_t'  + H.c.,
\end{align}
where
\begin{align}
\psi_x' = U^\dagger \psi_x U = e^{i\varphi_B (1 - F_{pB})/2} \psi_x,
\nonumber \\
\psi_y' = U^\dagger \psi_y U = e^{i\varphi_B (1 - F_{pB})/2} \psi_y,
\nonumber \\
\psi_t' = U^\dagger \psi_t U = e^{i\varphi_A (1 - F_{pA})/2} \psi_t,
\end{align}
where
\begin{align}
F_{pA} &= i \gamma^z \gamma^t,
\nonumber \\
F_{pB} &= i \gamma^x \gamma^y
\end{align}
are the fermion parities of the $A$ and $B$ islands, respectively.
Assuming the regime
\begin{align}
t_x, t_y \ll \Delta,
\end{align}
where $\Delta$ is the single-particle gap in the semiconducting nanowire,
we can write the electron operators at low energies in terms of the Majorana zero modes:
\begin{align}
\psi_\alpha = u_\alpha \gamma^\alpha,
\end{align}
where $\alpha = x, y, z, t$, and $u_\alpha$ are complex numbers (whose magnitude is order unity) that depend on microscopic details. Thus, we obtain:
\begin{align}
\delta H' = t_x u_x^* u_t  \gamma^x e^{-i \varphi_B (1 - F_{pB})/2 + i \varphi_A (1-F_{pA})/2} \gamma^t
+ t_y u_y^* u_t \gamma^y e^{-i \varphi_B (1 - F_{pB})/2 + i \varphi_A (1-F_{pA})/2}\gamma^t + H.c.
\end{align}
Recall that $n_{M+} = 1$, and that $n_{MA} = (1+F_{pA})/2$, $n_{MB} = (1+F_{pB})/2$, which implies htat
\begin{align}
F_{pA} + F_{pB} = 0.
\end{align}
It is useful to define
\begin{align}
\tilde{t}_x &= t_x u_x^* u_t
\nonumber \\
\tilde{t}_y &= t_y u_y^* u_t
\end{align}
Thus, we get
\begin{align}
\delta H' &= \tilde{t}_x  e^{-i \varphi_B (1 - F_{pA})/2 + i \varphi_A (1-F_{pA})/2}\gamma^x \gamma^t
+ \tilde{t}_y e^{-i \varphi_B (1 - F_{pA})/2 + i \varphi_A (1-F_{pA})/2}\gamma^y \gamma^t + H.c.,
\nonumber \\
&= \tilde{t}_x  e^{i \varphi_- (1-F_{pA})/2}\gamma^x \gamma^t
+ \tilde{t}_y  e^{i \varphi_- (1 - F_{pA})/2}\gamma^y \gamma^t + H.c.,
\end{align}
where we have also commuted $\gamma^x$, $\gamma^y$ through the exponential term. Note further that
$S^z = n_{M-} = n_{MA} - n_{MB} = (F_{pA} - F_{pB})/2 = F_{pA}$. The above can then be rewritten as
\begin{align}
\delta H' =&\left(\frac{1+S^z}{2} + \frac{1-S^z}{2} e^{i\varphi_-}\right) (\tilde{t}_x \gamma^x \gamma^t
+ \tilde{t}_y \gamma^y \gamma^t) + H.c.,
\nonumber \\
=&\left( \frac{1+S^z}{2} + \frac{1-S^z}{2} e^{i\varphi_-}\right) ( \tilde{t}_x \gamma^x \gamma^t
+ \tilde{t}_y \gamma^y \gamma^t) + H.c.,
\nonumber \\
= &i \gamma^x \gamma^t \text{Im}\left( \tilde{t}_x (1+e^{i\varphi_-}) - \tilde{t}_y (1 - e^{i\varphi_-}) \right)
\nonumber \\
&+ i \gamma^y \gamma^t \text{Im}\left( \tilde{t}_y (1 + e^{i\varphi_-}) - \tilde{t}_x (1 - e^{i\varphi_-}) \right)
\end{align}
Thus, we have
\begin{align}
\delta H' &= a_x i\gamma^x \gamma^t + a_y i \gamma^y \gamma^t,
\end{align}
with
\begin{align}
a_x &= \text{Im}\left( \tilde{t}_x (1+e^{i\varphi_-}) - \tilde{t}_y (1 - e^{i\varphi_-}) \right)
\nonumber \\
a_y &= \text{Im}\left( \tilde{t}_y (1 + e^{i\varphi_-}) - \tilde{t}_x (1 - e^{i\varphi_-}) \right)
\end{align}
For $t_x, t_y \ll E_C$, we can treat $\delta H'$ perturbatively around $H_{ss}$. Thus, we get an
effective Hamiltonian $H_{\text{eff}}$, such that
\begin{align}
\langle m | H_{\text{eff}} | n \rangle &= \langle m | H_{ss} + \delta H' | n \rangle
\nonumber \\
&= \delta_{mn} E_m + \langle m | \delta H' |n \rangle,
\end{align}
where $|m\rangle$ are the normalized eigenstates of the unperturbed Hamiltonian $H_{ss}$. Then we can write the effective Hamiltonian in the low energy spin space as
\begin{align}
H_{\text{eff}} &=  \sum_m E_m |m\rangle \langle m | + \sum_{m,n} \langle m | \delta H' |n \rangle |m\rangle\langle n|
\nonumber \\
&= h_z S^z + h_x S^x + h_y S^y,
\end{align}
where
\begin{align}
h_z &= E_{S^z = 1} - E_{S^z = -1} + \langle S^z = 1 | \delta H ' | S^z = 1 \rangle - \langle S^z = -1 | \delta H' | S^z = -1 \rangle,
\nonumber \\
h_x &= \text{Re}[ \langle S^z = -1 | \delta H' |S^z = 1 \rangle ],
\nonumber \\
h_y &= \text{Im}[ \langle S^z = -1 | \delta H' |S^z = 1 \rangle ].
\end{align}

Recall that the two spin states of interest, $|S^z\rangle$ are defined as in Eq. (\ref{ssStateDef}).
Thus:
\begin{align}
\langle S^z | \delta H' | S^z \rangle &= 0,
\nonumber \\
\langle S^z | \delta H' | -S^z \rangle
&= \langle S^z | a_{x} i\gamma^x \gamma^t + a_{y} i\gamma^y \gamma^t | -S^z \rangle
\nonumber \\
&= \langle S^z | a_{x} \widetilde{| S^z \rangle} - i S^z \langle S^z | a_{y} \widetilde{| S^z \rangle},
\end{align}
where $\widetilde{|S^z \rangle }$ is defined in Eq. (\ref{tildeState}).
Therefore, we find:
\begin{align}
h_x = \text{Re}[ \langle -1 | a_{x} \widetilde{| -1 \rangle} + i \langle -1 | a_{y} \widetilde{| -1 \rangle} ]
\nonumber \\
h_y = \text{Im}[\langle -1 | a_{x} \widetilde{| -1 \rangle} + i \langle -1 | a_{y} \widetilde{| -1 \rangle}  ]
\end{align}

\subsection{Two spin unit cell}
\label{chargingSec2S}

\begin{figure}
	\centering
	\includegraphics[width=1.5in]{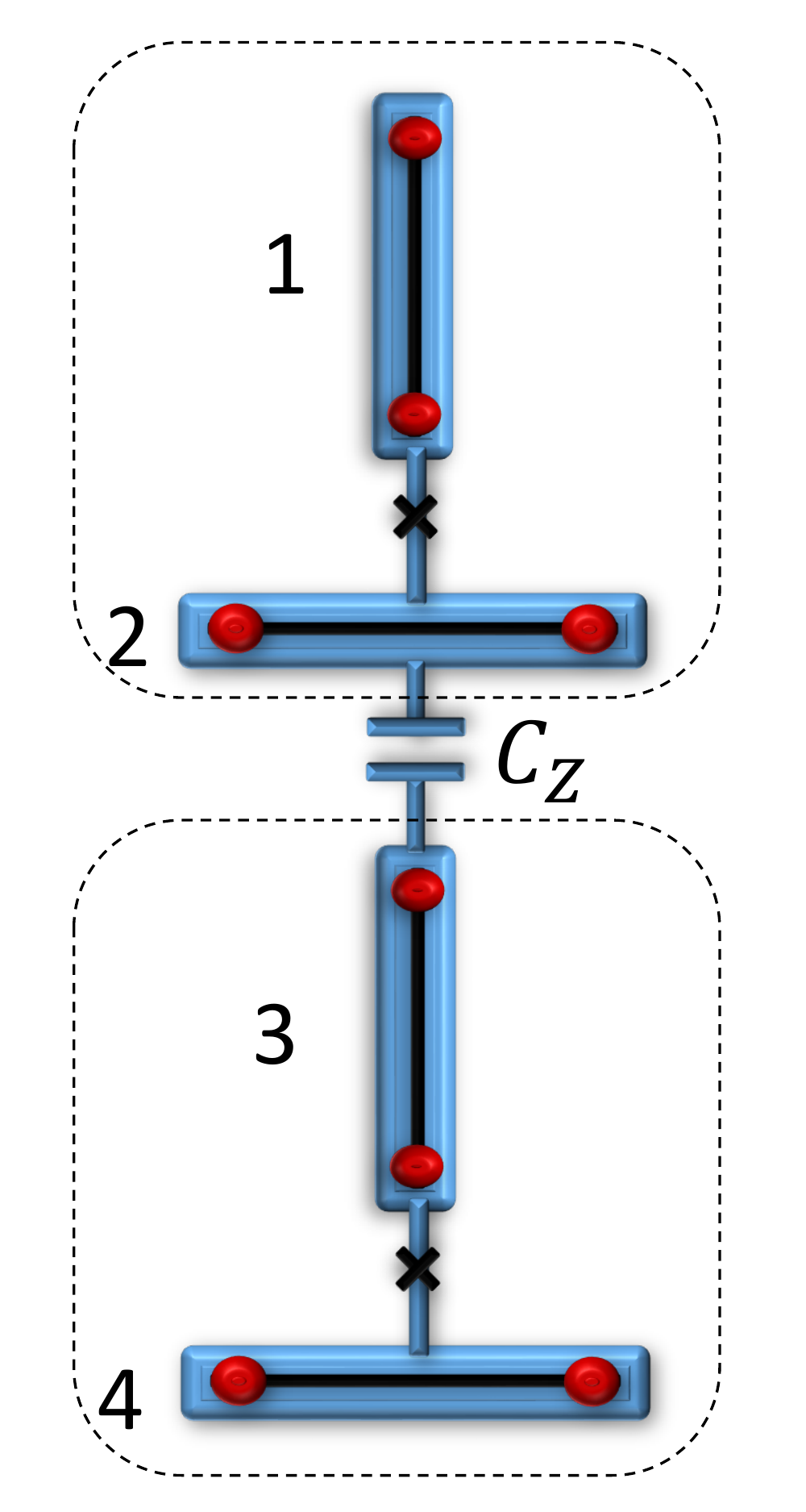}
	\caption{\label{twospinFig} Two vertically coupled spins, consisting of 4 superconducting islands. Each island is labelled 1,...,4 as shown. $C_z$ is a capacitor connecting islands $2$ and $3$. In order to optimize energy scales, we will also consider a capacitance $C_z'$ connecting islands $1$ and $4$, which we have not explicitly shown.
}
\end{figure}

Let us now consider a pair of vertically separated effective spins, which will form the unit cell for our
brick lattice Kitaev model. This consists of two pairs of $A$ and $B$ islands, as shown in Fig. \ref{twospinFig}. We wish
to consider a capacitive coupling $C_Z$, as shown in Fig. \ref{twospinFig}. In this analysis we will label the islands
$1,.., 4$ as shown. We will also consider a capacitance $C_Z'$, purely for subsequent numerical optimization of energy scales,
between islands $1$ and $4$, though this is not explicitly shown in Fig. \ref{twospinFig}.
The Hamiltonian for such a two-spin system is given by
\begin{align}
H_{2s} = \sum_{j=1}^4 H_{\text{BdG}}[ \Delta_{0j} , \psi_j^\dagger, \psi_j]
+ \frac{1}{2} \sum_{ij} Q_i C_{ij}^{-1} Q_j
- E_J (\cos(\varphi_1 - \varphi_2) + \cos(\varphi_3 - \varphi_4)).
\end{align}
The charges $Q_j$ (after the unitary tranformation discussed in the previous subsection) are
\begin{align}
Q_j = 2e \left( -i \partial_{\varphi_j} + \frac{n_{Mj} - n_{\text{off}j}}{2} \right).
\end{align}
Note we omit the primed superscripts in the preceding equation and throughout the rest of the discussion.
The capacitance matrix $C$ is now a $4\times 4$ matrix:
\begin{align}
C = \left( \begin{matrix}
C_g + C_J +C_Z' & - C_J & 0 & -C_Z' \\
-C_J & C_g + C_J + C_Z & -C_Z & 0 \\
0 & -C_Z & C_g + C_J + C_Z & - C_J \\
-C_Z' & 0 & -C_J & C_g + C_J + C_Z'
\end{matrix} \right)
\end{align}

It is useful to write $H_{2s}$ as
\begin{align}
H_{2s} = H_{12} + H_{34} + H_{1234},
\end{align}
where
\begin{align}
H_{12} &= \sum_{j=1}^2 H_{\text{BdG}}[ \Delta_{0j} , \psi_j^\dagger, \psi_j]
+ \frac{1}{2} \sum_{i,j =1}^2 Q_i C_{ij}^{-1} Q_j
- E_J \cos(\varphi_1 - \varphi_2)
\nonumber \\
&= \sum_{j=1}^2 H_{\text{BdG}}[ \Delta_{0j} , \psi_j^\dagger, \psi_j] - E_J \cos(\varphi_{12,-})
\nonumber \\
&+ Q_{12,+}^2 \frac{1}{4}( C_{11}^{-1}/2 + C_{22}^{-1}/2 + C_{12}^{-1} ) + Q_{12,-}^2 \frac{1}{4}(C_{11}^{-1}/2 + C_{22}^{-1}/2 - C_{12}^{-1} ) + Q_{12,+} Q_{12,-} \frac{1}{4}(C^{-1}_{11} - C^{-1}_{22})
\nonumber \\
H_{34} &= \sum_{j=3}^4 H_{\text{BdG}}[ \Delta_{0j} , \psi_j^\dagger, \psi_j]
+ \frac{1}{2} \sum_{i,j =3}^4 Q_i C_{ij}^{-1} Q_j
- E_J \cos(\varphi_3 - \varphi_4)
\nonumber \\
&= \sum_{j=3}^4 H_{\text{BdG}}[ \Delta_{0j} , \psi_j^\dagger, \psi_j] - E_J \cos(\varphi_{34,-})
\nonumber \\
&+ Q_{34,+}^2 \frac{1}{4}( C_{33}^{-1}/2 + C_{44}^{-1}/2 + C_{34}^{-1} ) + Q_{34,-}^2 \frac{1}{4}(C_{33}^{-1}/2 + C_{44}^{-1}/2 - C_{34}^{-1} ) + Q_{34,+} Q_{34,-} \frac{1}{4}(C^{-1}_{33} - C^{-1}_{44})
\nonumber \\
H_{1234} &= \sum_{i = 1}^2 \sum_{j = 3}^4 Q_i C_{ij}^{-1} Q_j
\nonumber \\
&= \sum_{\sigma_1, \sigma_2 = \pm} Q_{12,\sigma_1}Q_{34,\sigma_2} A_{\sigma_1\sigma_2}
\end{align}
where we have defined $Q_{ij,\pm} = Q_i \pm Q_j$ and
\begin{align}
A_{\sigma_1\sigma_2} &= \frac{1}{4} (C_{13}^{-1} + \sigma_2 C_{14}^{-1} + \sigma_1 C_{23}^{-1} + C_{24}^{-1})
\end{align}
The terms $H_{12}$ and $H_{34}$ are just the Hamiltonians for a single effective spin, which was analyzed in the previous section.
We label these spins by $S^z_{\vec{r}}$ and $S^z_{\vec{r}-\hat{z}}$. $\vec{r}$ and $\vec{r}-\hat{z}$ label the two different
effective sites, as shown in Fig. \ref{twospinFig}. $H_{1234}$, then, couples the two effective spins.

\subsubsection{Analytical treatment}
We now wish to treat $H_{1234}$ perturbatively around the decoupled limit $H_{12} +
H_{34}$. This is valid if
\begin{align}
A_{--}, A_{-+}, A_{+-} \ll E_C.
\end{align}

To lowest order in perturbation theory, we can replace $H_{1234}$ with the effective Hamiltonian $H_{\text{eff},z}$:
\begin{align}
\label{Jzanalysis}
\langle S^z_{\vec{r}} S^z_{\vec{r}-\hat{z}} | H_{\text{eff},z} |S^z_{\vec{r}} S^z_{\vec{r}-\hat{z}} \rangle &=
\langle S^z_{\vec{r}} S^z_{\vec{r}-\hat{z}} | H_{1234} |S^z_{\vec{r}} S^z_{\vec{r}-\hat{z}} \rangle
\nonumber \\
&= A_{--} \langle S^z_{\vec{r}} S^z_{\vec{r}-\hat{z}} | Q_{12,-}Q_{34,-} |S^z_{\vec{r}} S^z_{\vec{r}-\hat{z}} \rangle
\nonumber \\
&= A_{--} \langle S^z_{\vec{r}}  | Q_{12,-} |S^z_{\vec{r}} \rangle  \langle S^z_{\vec{r}-\hat{z}}| Q_{34,-} |S^z_{\vec{r}-\hat{z}} \rangle
\nonumber \\
&=J_z S^z_{\vec{r}} S^z_{\vec{r}-\hat{z}},
\end{align}
where we have defined (see eq. \ref{singleSzav})
\begin{align}
J_z = A_{--}  \left( \frac{e}{E_C} \frac{\partial h_z}{\partial n_{\text{off}-}}\right)^2 .
\end{align}
The second equality in eq. (\ref{Jzanalysis}) follows because $\langle Q_{12,-} \rangle = \langle Q_{34,-}\rangle = 0$, so only the $\langle Q_{12,-}Q_{34,-}\rangle$ term remains non-zero.  We have assumed for simplicity that the two $A$,$B$ island pairs have the same parameters
$E_J$, $E_C$, $n_{\text{off}\pm}$.

Therefore, to first order in perturbation theory, the effect of the vertical capacitances $C_Z$, $C_Z'$, which couple the two spins at $\vec{r}$ and $\vec{r} - \hat{z}$,
is to induce an $S^z_{\vec{r}} S^z_{\vec{r}-\hat{z}}$ coupling.

\subsubsection{Numerical Solution}

We can also more comprehensively analyze the two-spin model by employing a numerical solution.
In Fig. \ref{twoSpinPlot1} - \ref{twoSpinPlot3}, we present results of such a numerical solution
for certain choices of parameters.

\begin{figure}
	\centering
	\includegraphics[width=7in]{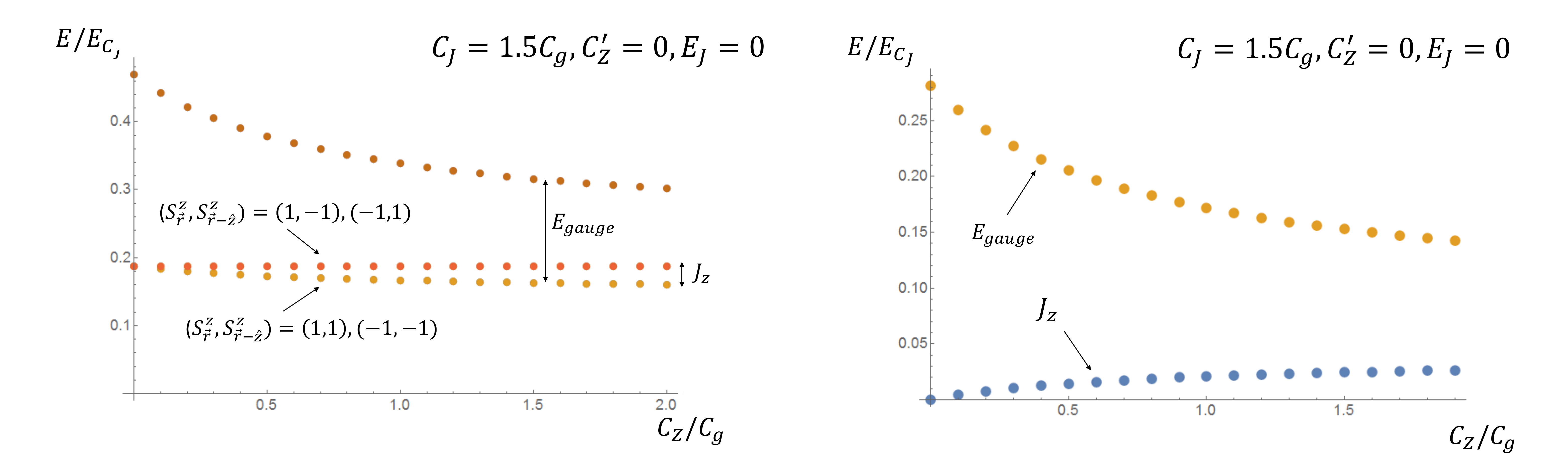}
	\caption{\label{twoSpinPlot1} Left panel: Plot of energy spectrum of $H_{2s}$ (ignoring the excited state spectrum of $H_{\text{BdG}}$), for the lowest energy states,
as a function of the capacitance $C_Z$. The other parameters are set to $C_J = 1.5 C_g$, $C_{Z}' = 0$, $E_J = 0$. The lowest energy curve is doubly
degenerate, and is associated with the states $(S^z_{\vec{r}}, S^z_{\vec{r}-\hat{z}}) = (1,1), (-1,-1)$. The next excited state, whose energy difference
with the lowest energy curve defines $J_z$, is also doubly degenerate and associated with the spin states
$(S^z_{\vec{r}}, S^z_{\vec{r}-\hat{z}}) = (1,-1), (-1,1)$. The next excited state lies outside of the effective "spin" subspace that we are interested in,
and we define the gap to these excited states as $E_{\text{gauge}}$. The notation $E_{\text{gauge}}$ is used because
states with energies $E > E_{\text{gauge}}$ can violate the "gauge" constraint $\gamma^x \gamma^y \gamma^z \gamma^t = 1$ that is required for the Kitaev spin model.
Right panel: Plot of $J_z$ and $E_{\text{gauge}}$.
}
\end{figure}

\begin{figure}
	\centering
	\includegraphics[width=7in]{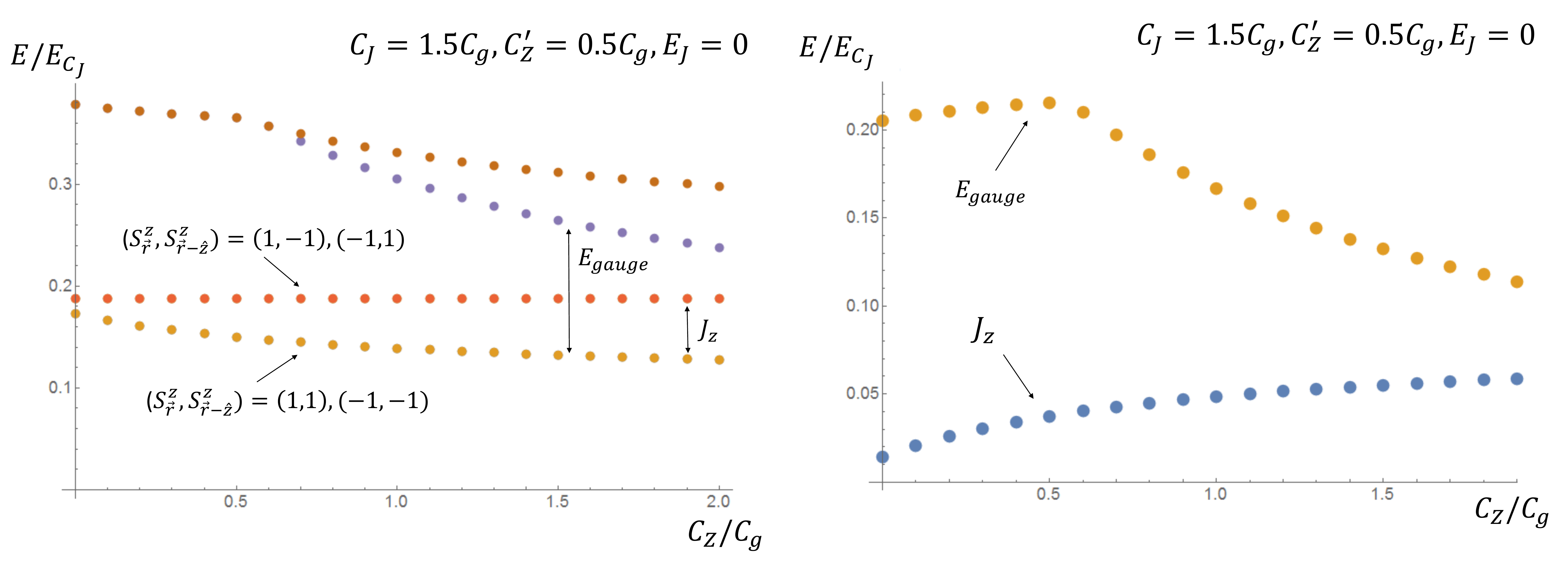}
	\caption{\label{twoSpinPlot2} Same as in Fig. \ref{twoSpinPlot1}, with different parameters as indicated.
}
\end{figure}

\begin{figure}
	\centering
	\includegraphics[width=7in]{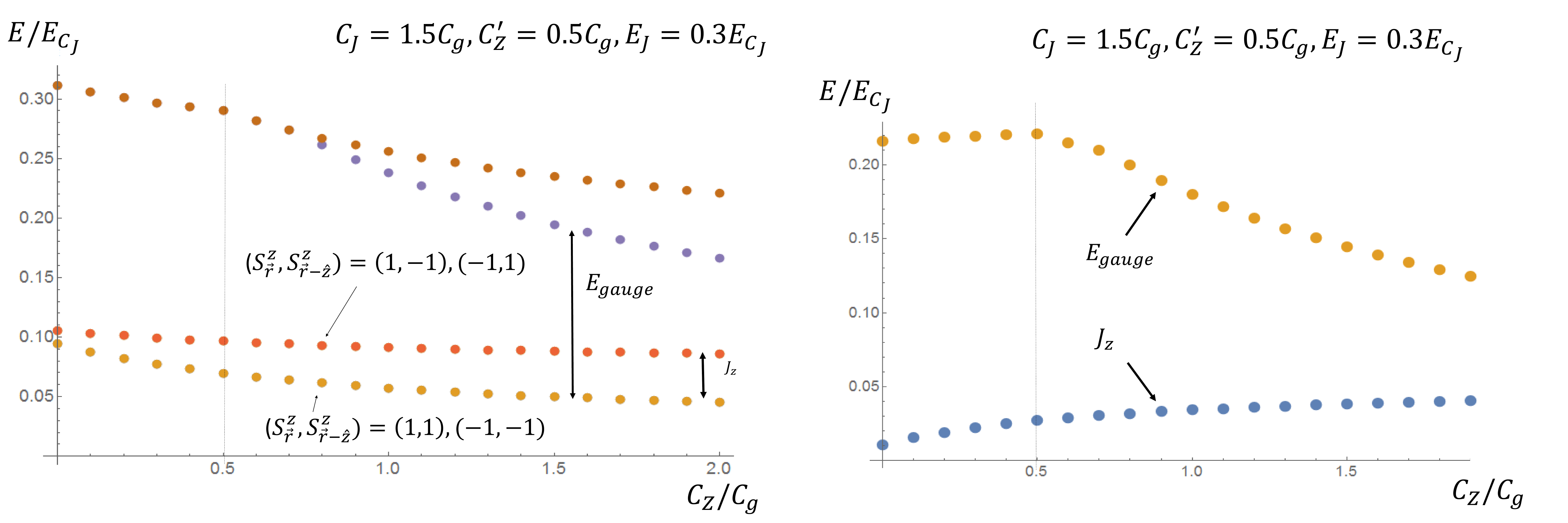}
	\caption{\label{twoSpinPlot3} Same as in Fig. \ref{twoSpinPlot1}, with different parameters as indicated.
}
\end{figure}

\begin{figure}
	\centering
	\includegraphics[width=7in]{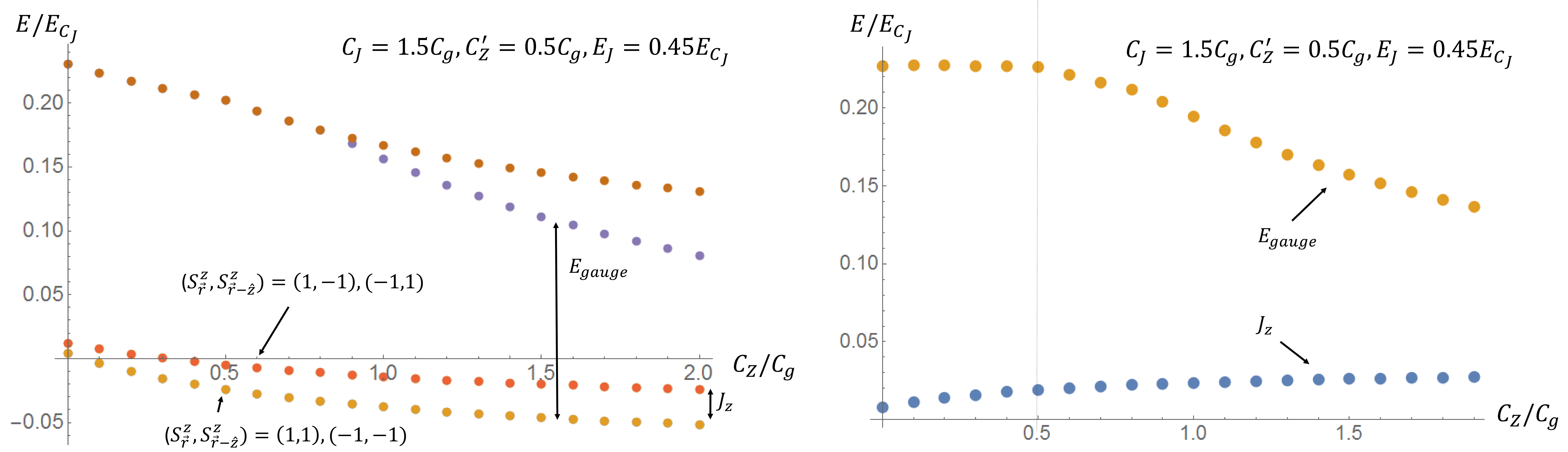}
	\caption{\label{twoSpinPlot4} Same as in Fig. \ref{twoSpinPlot1}, with different parameters as indicated. We see that a particularly
optimal point occurs when $C_Z = 0.5 C_g$.
}
\end{figure}

\subsection{Horizontally coupled unit cells: four spins}
\label{fourSpinCapSec}

\begin{figure}
	\centering
	\includegraphics[width=4in]{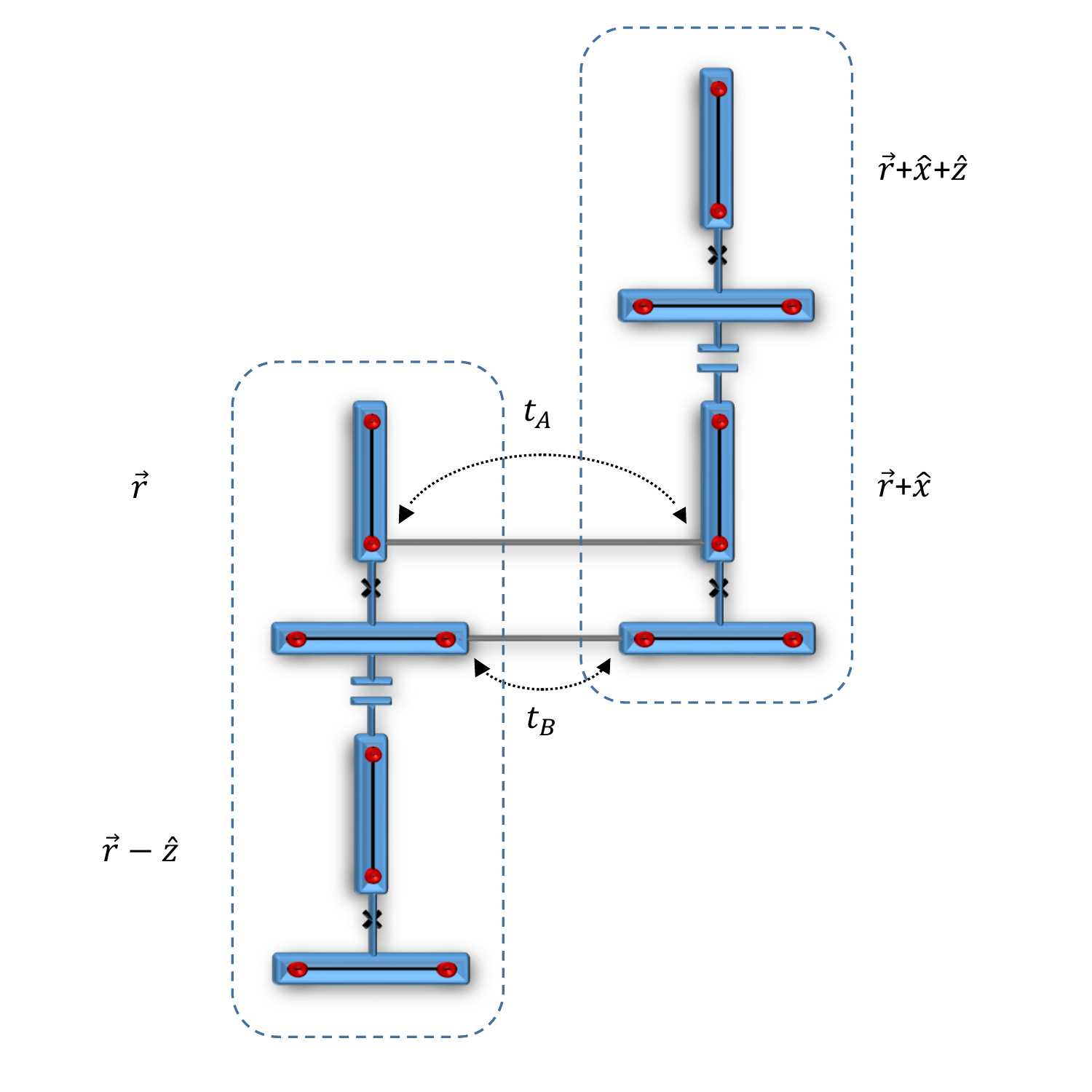}
	\caption{\label{fourspinFig} Two horizontally coupled unit cells, consisting of 4 spins total. Dashed lines encircle each unit cell. The locations of the spins are
 $\vec{r}$, $\vec{r} - \hat{z}$, $\vec{r} + \hat{x}$, $\vec{r}+\hat{x} + \hat{z}$. $t_A$ and $t_B$ indicate electron tunneling, as shown.
}
\end{figure}

Let us now consider horizontally coupling two unit cells, as shown in Fig. \ref{fourspinFig}. We connect two horizontally separated unit cells with semiconductor wire, as shown, which allows electrons to tunnel between the end points of the wires, with tunneling amplitudes $t_A$ and $t_B$, as shown. The effective Hamiltonian for this system is now
\begin{align}
H_{4s} = \sum_I H_{2s,I} + H_{\text{tun}},
\end{align}
where $H_{2s,I}$ is the Hamiltonian for the $I$th unit cell, which is given by $H_{2s}$ above.
$H_{\text{tun}}$ contains the horizontal couplings, as we explain below.

We wish to show that in a suitable parameter regime, at low energies the effective Hamiltonian can be described by the following spin model:
\begin{align}
H_{\text{eff},4s} = J_z S^z_{\vec{r}} S^z_{\vec{r}-\hat{z}} + J_z S^z_{\vec{r}+\hat{x}} S^z_{\vec{r}+\hat{x} + \hat{z}}
+ J_{yx} S^y_{\vec{r}} S^x_{\vec{r}+\hat{x}} ,
\end{align}
with corrections to this effective Hamiltonian being much smaller in energy scale than $J_z, J_{yx}$.

We consider the electron tunneling terms $t_A$ and $t_B$, as shown in Fig \ref{fourspinFig}.
This gives rise to an effective tunneling Hamiltonian (written in the basis before the unitary transformation $U$):
\begin{align}
H_{\text{tun}} &=  [t_A \psi_{t,\vec{r}}^\dagger \psi_{t,\vec{r}+\hat{x}} + H.c.]
+ [t_B \psi_{y,\vec{r}}^\dagger \psi_{x,\vec{r}+\hat{x}} + H.c.]
\end{align}
Here, $\vec{r}$ labels the effective spins, each of which consists of an $A$ and a $B$ island. After the
unitary transformation by $U$, we have
\begin{align}
H_{\text{tun}}' = &U^\dagger H_{\text{tun}} U =  [t_A (\psi'_{t,\vec{r}})^\dagger \psi'_{t,\vec{r}+\hat{x}} + H.c.]
\nonumber \\
&+ [t_B (\psi'_{y,\vec{r}})^\dagger \psi'_{x,\vec{r}+\hat{x}} + H.c.],
\end{align}
where
\begin{align}
\psi'_{\alpha,\vec{r}} = U^\dagger \psi_{\alpha,\vec{r}} U &= e^{i  \varphi_{j\vec{r}}(1 - F_{pj,\vec{r}})/2} \psi_{\alpha,\vec{r}},
\nonumber \\
(\psi'_{\alpha,\vec{r}})^\dagger &= \psi_{\alpha,\vec{r}}^\dagger e^{-i \varphi_{j\vec{r}}(1 - F_{pj,\vec{r}})/2 }
\end{align}
where $j = A,B$ depending on whether $\alpha=z,t$ or $x,y$. Therefore, the tunneling Hamiltonian is, after the unitary transformation:
\begin{align}
H_{\text{tun}}' = & [t_A \psi_{t,\vec{r}}^\dagger e^{-i (1-F_{p,A,\vec{r}})\varphi_{A\vec{r}}/2 + i  (1-F_{p,A,\vec{r}+\hat{x}})\varphi_{A\vec{r}+\hat{x}}/2} \psi_{t,\vec{r}+\hat{x}} + H.c.]
\nonumber \\
&+ [t_B \psi_{y,\vec{r}}^\dagger e^{-i (1-F_{p,B,\vec{r}})\varphi_{B\vec{r}}/2 + i  (1-F_{p,B,\vec{r}+\hat{x}})\varphi_{B\vec{r}+\hat{x}}/2}  \psi_{x,\vec{r}+\hat{x}} + H.c.],
\end{align}

We consider the limit where
\begin{align}
t_A, t_B \ll \Delta,
\end{align}
where $\Delta$ is the single-particle gap in the nanowire. In this limit,
the electron operator $\psi$ can be replaced by
\begin{align}
\psi_{\alpha \vec{r}} = u_{\alpha,\vec{r}} \gamma^\alpha_{\vec{r}} ,
\end{align}
where $u_{\alpha,\vec{r}}$ are complex numbers that depend sensitively on microscopic details.
Let us also define
\begin{align}
\tilde{t}_A &= t_A (u^t_{\vec{r}})^* u^t_{\vec{r}+\hat{x}},
\nonumber \\
\tilde{t}_B &= t_B (u^y_{\vec{r}})^* u^x_{\vec{r}+\hat{x}} ,
\end{align}
We therefore write $H_{\text{tun}}'$ as
\begin{align}
H_{\text{tun}}' &= \Lambda_{\vec{r}},
\nonumber \\
\Lambda_{\vec{r}} &= [\tilde{t}_A \gamma^t_{\vec{r}} e^{-i (1-F_{p,A,\vec{r}})\varphi_{A\vec{r}}/2 + i  (1-F_{p,A,\vec{r}+\hat{x}})\varphi_{A\vec{r}+\hat{x}}/2} \gamma^t_{\vec{r}+\hat{x}}
\nonumber \\
&+ \tilde{t}_B \gamma^y_{\vec{r}} e^{-i (1-F_{p,B,\vec{r}})\varphi_{B\vec{r}}/2 + i  (1-F_{p,B,\vec{r}+\hat{x}})\varphi_{B\vec{r}+\hat{x}}/2} \gamma^x_{\vec{r}+\hat{x}} + H.c.]
\nonumber \\
&= [\tilde{t}_A e^{-i (1+F_{p,A,\vec{r}})\varphi_{A\vec{r}}/2 + i  (1-F_{p,A,\vec{r}+\hat{x}})\varphi_{A\vec{r}+\hat{x}}/2} \gamma^t_{\vec{r}} \gamma^t_{\vec{r}+\hat{x}}
\nonumber \\
&+ \tilde{t}_B e^{-i (1+F_{p,B,\vec{r}})\varphi_{B\vec{r}}/2 + i  (1-F_{p,B,\vec{r}+\hat{x}})\varphi_{B\vec{r}+\hat{x}}/2}  \gamma^y_{\vec{r}} \gamma^x_{\vec{r}+\hat{x}} + H.c.]
\end{align}
These single electron tunneling processes violate the charging energy constraint and are therefore suppressed in the limit
\begin{align}
\tilde{t}_j \ll E_{\text{gauge}},
\end{align}
where $E_{\text{gauge}}$ is the energy cost to adding a single electron to the two-spin unit cell.
Perturbing in $\tilde{t}_j/E_{\text{gauge}}$, we obtain an effective Hamiltonian:
\begin{align}
H_{\text{eff}} &= -\frac{1}{E_{\text{gauge}}} \Lambda_{\vec{r}}^\dagger \Lambda_{\vec{r}} + \mathcal{O}( \tilde{t}^4/E_{\text{gauge}}^3)
\end{align}

Expanding, we obtain, up to a constant term,
\begin{align}
\label{Heffxy}
H_{t,\text{eff}} = -\frac{1}{E_{\text{gauge}}} (h_{t;1} + h_{t;2} + h_{t;3})
\end{align}
\begin{align}
h_{t;1} &=  -\tilde{t}_A^2 e^{-i (\varphi_{A,\vec{r}} - \varphi_{A,\vec{r}+\hat{x}})}
- \tilde{t}_B^2  e^{-i (\varphi_{B,\vec{r}} - \varphi_{B,\vec{r}+\hat{x}})} + H.c.
\end{align}
\begin{align}
h_{t;2}= &2 \tilde{t}_A \tilde{t}_B e^{-i (1+F_{p,A,\vec{r}})\varphi_{A\vec{r}}/2 + i  (1-F_{p,A,\vec{r}+\hat{x}})\varphi_{A\vec{r}+\hat{x}}/2} e^{-i (1+F_{p,B,\vec{r}})\varphi_{B\vec{r}}/2 + i  (1-F_{p,B,\vec{r}+\hat{x}})\varphi_{B\vec{r}+\hat{x}}/2}
\nonumber \\
&\gamma^t_{\vec{r}} \gamma^t_{\vec{r}+\hat{x}} \gamma^y_{\vec{r}} \gamma^x_{\vec{r}+\hat{x}} + H.c.
\nonumber \\
= &2 \tilde{t}_A \tilde{t}_B e^{-i\varphi_{+,\vec{r}}/2 - i (F_{p+,\vec{r}} \varphi_{+,\vec{r}} + F_{p-\vec{r}}\varphi_{-\vec{r}})/4} e^{i \varphi_{+,\vec{r}+\hat{x}}/2 - i (F_{p+,\vec{r}} \varphi_{+,\vec{r}+\hat{x}} + F_{p-\vec{r}}\varphi_{-\vec{r}+\hat{x}})/4}
\nonumber \\
&\gamma^t_{\vec{r}} \gamma^t_{\vec{r}+\hat{x}} \gamma^y_{\vec{r}} \gamma^x_{\vec{r}+\hat{x}} + H.c.
\end{align}
\begin{align}
h_{t;3} = &2 \tilde{t}_A \tilde{t}_B^* e^{-i (1+F_{pA,\vec{r}}) \varphi_A/2 + i (1 - F_{pA,\vec{r}+\hat{x}}) \varphi_{A,\vec{r}+\hat{x}}/2} e^{i (1- F_{pB,\vec{r}})\varphi_{B,\vec{r}}/2 - i (1+F_{pB,\vec{r}+\hat{x}})\varphi_{B,\vec{r}+\hat{x}}/2}
\nonumber \\
& \gamma^t_{\vec{r}}\gamma^t_{\vec{r}+\hat{x}} \gamma^x_{\vec{r}+\hat{x}} \gamma^y_{\vec{r}} + H.c.
\nonumber \\
=& 2 \tilde{t}_A \tilde{t}_B^* e^{-i \varphi_{-,\vec{r}}/2 - i(F_{p+\vec{r}} \varphi_{+,\vec{r}} + F_{p-\vec{r}}\varphi_{-\vec{r}})/4 }e^{i \varphi_{-,\vec{r}+\hat{x}}/2 - i(F_{p+\vec{r}+\hat{x}} \varphi_{+,\vec{r}+\hat{x}} + F_{p-\vec{r}+\hat{x}}\varphi_{-\vec{r}+\hat{x}})/4 }
\nonumber \\
& \gamma^t_{\vec{r}}\gamma^t_{\vec{r}+\hat{x}}\gamma^x_{\vec{r}+\hat{x}} \gamma^y_{\vec{r}} + H.c
\end{align}

Note that in the limit within which we are working,
\begin{align}
\frac{2 |\tilde{t}_A \tilde{t}_B|}{E_{\text{gauge}}} , \frac{2|\tilde{t}_A|^2}{E_{\text{gauge}}} , \frac{2 |\tilde{t}_B|^2}{E_{\text{gauge}}} \ll E_{\text{gauge}}
\end{align}
Thus, due to the charging energy on each site $\vec{r}$,
$\varphi_+$ is highly fluctuating independently on each site. Treating (\ref{Heffxy}) perturbatively
around the decoupled limit $H_{2s}$, we see that we can set
\begin{align}
F_{p+,\vec{r}} = F_{pA,\vec{r}} + F_{pB,\vec{r}} = 0,
\end{align}
\begin{align}
F_{p-,\vec{r}} = 2 F_{pA,\vec{r}} = -2 F_{pB,\vec{r}} = 2 S^z_{\vec{r}} .
\end{align}
Moreover, we can replace $h_{t;2}$,$h_{t;1}$,$h_{t;3}$ by their expectation values in the ground state manifold of $H_{2s}$:
\begin{align}
\langle m | h_{t;1,\text{eff}} |n \rangle = \langle m| h_{t;1} |n\rangle = 0
\end{align}
\begin{align}
\langle m | h_{t;2,\text{eff}} |n \rangle = \langle m | h_{t;2} | n \rangle = &2 t_A t_B u_{t,\vec{r}}^* u_{t,\vec{r}+\hat{x}} u_{y,\vec{r}}^* u_{x,\vec{r}+\hat{x}}
\langle m | e^{-i\varphi_{+,\vec{r}}/2 - i S^z_{\vec{r}}\varphi_{-\vec{r}}/2} e^{i \varphi_{+,\vec{r}+\hat{x}}/2 - i S^z_{\vec{r}+\hat{x}}\varphi_{-\vec{r}+\hat{x}}/2}
\nonumber \\
&\gamma^t_{\vec{r}} \gamma^t_{\vec{r}+\hat{x}} \gamma^y_{\vec{r}} \gamma^x_{\vec{r}+\hat{x}} + H.c. |n \rangle = 0.
\end{align}

We define
\begin{align}
\tilde{t}_A \tilde{t}_B^* = |\tilde{t}_A \tilde{t}_B| e^{i\theta}
\end{align}
Note that the phase $\theta$ depends on two quantities: the magnetic flux normal to the system, and the
angle between the Zeeman field and the Rashba spin-orbit field. These can both be tuned, and therefore $\theta$ can be viewed as a tunable quantity.
\begin{align}
\langle m | h_{t;3,\text{eff}} |n \rangle =& \langle m | h_{t;3} |n \rangle
\nonumber \\
= & \langle m | 2 \tilde{t}_A \tilde{t}_B^*
e^{-i \varphi_{-,\vec{r}}/2 - i S^z_{\vec{r}}\varphi_{-\vec{r}}/2 } e^{i \varphi_{-,\vec{r}+\hat{x}}/2 - i S^z_{\vec{r}+\hat{x}}\varphi_{-\vec{r}+\hat{x}}/2 }
\gamma^t_{\vec{r}}\gamma^t_{\vec{r}+\hat{x}} \gamma^x_{\vec{r}+\hat{x}} \gamma^y_{\vec{r}}  + H.c. | n \rangle
\nonumber \\
= \langle m |&\frac{\tilde{t}_A \tilde{t}_B^*}{2}
 \left( 1+ e^{-i\varphi_{-,\vec{r}}} + (e^{-i\varphi_{-,\vec{r}}} - 1) S^z_{\vec{r}} \right)
\left( (e^{i\varphi_{-,\vec{r}+\hat{x}}} + 1) + (1 - e^{i\varphi_{-,\vec{r}+\hat{x}}}) S^z_{\vec{r}+\hat{x}} \right)
\nonumber \\
&\times \gamma^t_{\vec{r}}\gamma^t_{\vec{r}+\hat{x}} \gamma^x_{\vec{r}+\hat{x}} \gamma^y_{\vec{r}}  + H.c. |n \rangle .
\end{align}
The eigenstates of interest can be labelled as
\begin{align}
|S^z_{\vec{r} - \hat{z}} S^z_{\vec{r}} S^z_{\vec{r}+\hat{x}} S^z_{\vec{r}+\hat{x} + \hat{z}} \rangle .
\end{align}
Thus, we are interested in the matrix elements:
\begin{align}
\langle S^z_{\vec{r} - \hat{z}} S^z_{\vec{r}} S^z_{\vec{r}+\hat{x}} S^z_{\vec{r}+\hat{x} + \hat{z}}  | h_{t;3} |(S^{z}_{\vec{r} - \hat{z}})' (S^{z}_{\vec{r}})' (S^{z}_{\vec{r}+\hat{x}})' (S^{z}_{\vec{r}+\hat{x} + \hat{z}})'  \rangle .
\end{align}
We see that the only non-zero matrix elements are those for which $(S^{z}_{\vec{r}-\hat{z}})' = S^z_{\vec{r} - \hat{z}}$, $(S^{z}_{\vec{r}})' = - S^z_{\vec{r}}$,
$(S^{z}_{\vec{r}+\hat{x}})' = -S^z_{\vec{r}+\hat{x}}$, and $(S^{z}_{\vec{r}+\hat{x}+\hat{z}})' = S^z_{\vec{r}+\hat{x}+\hat{z}}$. Thus, we need to compute
\begin{align}
\langle S^z_{\vec{r} - \hat{z}} S^z_{\vec{r}} S^z_{\vec{r}+\hat{x}} S^z_{\vec{r}+\hat{x} + \hat{z}}  | h_{t;3} |S^z_{\vec{r} - \hat{z}}, -S^z_{\vec{r}}, -S^z_{\vec{r}+\hat{x}}, S^z_{\vec{r}+\hat{x} + \hat{z}}  \rangle .
\end{align}
In terms of these matrix elements, we can then write the effective Hamiltonian:
\begin{align}
h_{t;3,\text{eff}} =& \sum_{\{S^z\}} \langle S^z_{\vec{r} - \hat{z}} S^z_{\vec{r}} S^z_{\vec{r}+\hat{x}} S^z_{\vec{r}+\hat{x} + \hat{z}}  | h_{t;3} |S^z_{\vec{r} - \hat{z}}, -S^z_{\vec{r}}, -S^z_{\vec{r}+\hat{x}}, S^z_{\vec{r}+\hat{x} + \hat{z}}  \rangle
\nonumber \\
&\times | S^z_{\vec{r} - \hat{z}} S^z_{\vec{r}} S^z_{\vec{r}+\hat{x}} S^z_{\vec{r}+\hat{x} + \hat{z}}\rangle  \langle S^z_{\vec{r} - \hat{z}}, -S^z_{\vec{r}}, -S^z_{\vec{r}+\hat{x}}, S^z_{\vec{r}+\hat{x} + \hat{z}}|
\nonumber \\
=& \sum_{s_1,s_2,s_3,s_4 = \pm 1} h_{t;3}^{s_1,s_2,s_3,s_4}\frac{(1+ s_1 S^z_{\hat{r}-\hat{z}})}{2}  \frac{S^x_{\vec{r}} + s_2 i S^y_{\vec{r}}}{2} \frac{S^x_{\vec{r}+\hat{x}} + s_3 i S^y_{\vec{r+\hat{x}}}}{2} \frac{(1+ s_4 S^z_{\hat{r}+\hat{x}+\hat{z}})}{2}
\nonumber \\
=& \sum_{a,d = 1,z} \sum_{b,c = x,y} c_{abcd} S^a_{\vec{r}-\hat{z}} S^b_{\vec{r}} S^c_{\vec{r}+\hat{x}} S^d_{\vec{r}+\hat{x}+\hat{z}},
\end{align}
where
\begin{align}
c_{zxx1} =& \frac{1}{16} \sum_{s}  h_{t;3}^{s_1,s_2,s_3,s_4} s_1, &c_{1xxz} &= \frac{1}{16} \sum_{s}  h_{t;3}^{s_1,s_2,s_3,s_4} s_4
\nonumber \\
c_{zxy1} =& \frac{1}{16} \sum_{s}  h_{t;3}^{s_1,s_2,s_3,s_4} i s_1 s_3, &c_{1xyz} &= \frac{1}{16} \sum_{s}  h_{t;3}^{s_1,s_2,s_3,s_4} i s_4 s_3
\nonumber \\
c_{zyx1} =& \frac{1}{16} \sum_{s}  h_{t;3}^{s_1,s_2,s_3,s_4} i s_1 s_2, &c_{1yxz} &= \frac{1}{16} \sum_{s}  h_{t;3}^{s_1,s_2,s_3,s_4} i s_4 s_2
\nonumber \\
c_{zyy1} =& -\frac{1}{16} \sum_{s}  h_{t;3}^{s_1,s_2,s_3,s_4} s_1 s_2 s_3, &c_{1yyz} &= -\frac{1}{16} \sum_{s}  h_{t;3}^{s_1,s_2,s_3,s_4} s_4 s_2 s_3
\end{align}

\begin{align}
c_{zxxz} =& \frac{1}{16} \sum_{s}  h_{t;3}^{s_1,s_2,s_3,s_4} s_1 s_4, & c_{1xx1} =& \frac{1}{16} \sum_{s}  h_{t;3}^{s_1,s_2,s_3,s_4}
\nonumber \\
c_{zxyz} =& \frac{1}{16} \sum_{s}  h_{t;3}^{s_1,s_2,s_3,s_4} i s_1 s_3 s_4, & c_{1xy1} =& \frac{1}{16} \sum_{s}  h_{t;3}^{s_1,s_2,s_3,s_4} i s_3
\nonumber \\
c_{zyxz} =& \frac{1}{16} \sum_{s}  h_{t;3}^{s_1,s_2,s_3,s_4} i s_1 s_2 s_4, & c_{1yx1} =& \frac{1}{16} \sum_{s}  h_{t;3}^{s_1,s_2,s_3,s_4} i s_2
\nonumber \\
c_{zyyz} =& -\frac{1}{16} \sum_{s}  h_{t;3}^{s_1,s_2,s_3,s_4} s_1 s_2 s_3 s_4, & c_{1yy1} =& -\frac{1}{16} \sum_{s}  h_{t;3}^{s_1,s_2,s_3,s_4} s_2 s_3
\end{align}

Now, simplifying $h_{t;3}^{s_1,s_2,s_3,s_4}$, we get:
\begin{align}
h_{t;3}^{s_1,s_2,s_3,s_4} &= i |\tilde{t}_A \tilde{t}_B |
\left( \langle s_1,s_2,s_3,s_4|
i \text{Im}\left( e^{i\theta} (e^{-i\varphi_{-,\vec{r}}}-1)(e^{i\varphi_{-,\vec{r}+\hat{x}}} + 1) \right) | s_1,\tilde{s}_2,\tilde{s}_3,s_4 \rangle \right.
\nonumber \\
&+s_2 \langle s_1,s_2,s_3,s_4|
\text{Re}\left( e^{i\theta} (e^{-i\varphi_{-,\vec{r}}} + 1)(e^{i\varphi_{-,\vec{r}+\hat{x}}} + 1) \right) | s_1,\tilde{s}_2,\tilde{s}_3,s_4 \rangle
\nonumber \\
&+s_3 \langle s_1,s_2,s_3,s_4|
\text{Re}\left( e^{i\theta} (e^{-i\varphi_{-,\vec{r}}} - 1)(-e^{i\varphi_{-,\vec{r}+\hat{x}}} + 1) \right) | s_1,\tilde{s}_2,\tilde{s}_3,s_4 \rangle
\nonumber \\
&\left. +s_2 s_3 \langle s_1,s_2,s_3,s_4|
i\text{Im}\left( e^{i\theta} (e^{-i\varphi_{-,\vec{r}}} + 1)(1 - e^{i\varphi_{-,\vec{r}+\hat{x}}} ) \right) | s_1,\tilde{s}_2,\tilde{s}_3,s_4 \rangle \right)
\end{align}
Here we have defined
\begin{align}
| s_1,s_2,s_3,s_4 \rangle = |S^z_{\vec{r} - \hat{z}} = s_1, S^z_{\vec{r}}=s_2, S^z_{\vec{r}+\hat{x}}=s_3, S^z_{\vec{r}+\hat{x} + \hat{z}} = s_4 \rangle .
\end{align}
The states with the tildes over the $s$'s indicate that the phase mode has the opposite wave function as compared with the spin degree of freedom,
as described for the single spin case in Eq. (\ref{tildeState}).
Let us define
\begin{align}
A_{s_1,s_2,s_3,s_4}^{\sigma_1,\sigma_2} &= \langle s_1,s_2,s_3,s_4|
\text{Re}\left( e^{i\theta} (e^{-i\varphi_{-,\vec{r}}} + \sigma_1)(\sigma_2 e^{i\varphi_{-,\vec{r}+\hat{x}}} + 1) \right) | s_1,\tilde{s}_2,\tilde{s}_3,s_4 \rangle.
\nonumber \\
B_{s_1,s_2,s_3,s_4}^{\sigma_1,\sigma_2} &= \langle s_1,s_2,s_3,s_4|
i\text{Im}\left( e^{i\theta} (e^{-i\varphi_{-,\vec{r}}} + \sigma_1)(\sigma_2 e^{i\varphi_{-,\vec{r}+\hat{x}}} + 1) \right) | s_1,\tilde{s}_2,\tilde{s}_3,s_4 \rangle.
\end{align}
In terms of $A_{s_1,s_2,s_3,s_4}^{\sigma_1,\sigma_2}$, we have
\begin{align}
h_{t;3}^{s_1,s_2,s_3,s_4} &= i |\tilde{t}_A \tilde{t}_B |\left(
B_{s_1,s_2,s_3,s_4}^{-1,1} + s_2 A_{s_1,s_2,s_3,s_4}^{1,1} + s_3 A_{s_1,s_2,s_3,s_4}^{-1,-1} + s_2 s_3 B_{s_1,s_2,s_3,s_4}^{1,-1} \right)
\end{align}
We see that we need to compute the following expectation values for the two-spin system:
\begin{align}
v_{s_1,s_2;\pm} &= \langle S^z_{\vec{r}-\hat{z}} = s_1, S^z_{\vec{r}} = s_2| e^{\pm i \varphi_{\vec{r}}} | S^z_{\vec{r}-\hat{z}} = s_1, S^z_{\vec{r}} = \tilde{s}_2\rangle
\nonumber \\
w_{s_1,s_2;\pm} &= \langle S^z_{\vec{r} - \hat{z}} = s_1, S^z_{\vec{r}} = s_2| e^{\pm i \varphi_{\vec{r}-\hat{z}}} | S^z_{\vec{r} - \hat{z}} = \tilde{s}_1, S^z_{\vec{r}} = s_2\rangle,
\nonumber \\
g_{1,s_1,s_2} &= \langle S^z_{\vec{r} - \hat{z}} = s_1, S^z_{\vec{r}} = s_2 | S^z_{\vec{r}-\hat{z}} = s_1, \widetilde{S^z_{\vec{r}} = s_2} \rangle
\nonumber \\
g_{2,s_1,s_2} &= \langle S^z_{\vec{r} - \hat{z}} = s_1, S^z_{\vec{r}} = s_2 | \widetilde{S^z_{\vec{r}-\hat{z}} = s_1}, S^z_{\vec{r}} = s_2 \rangle
\end{align}
In terms of these expectation values,
\begin{align}
A_{s_1,s_2,s_3,s_4}^{\sigma_1,\sigma_2} =& \frac{1}{2} e^{i\theta} ( v_{s_1s_2,-} + \sigma_1 g_{1,s_1,s_2})(\sigma_2 w_{s_3s_4,-} + g_{2,s_3,s_4})
\nonumber \\
&+  \frac{1}{2} e^{-i\theta} (v_{s_1s_2,+} + \sigma_1 g_{1,s_1,s_2})(\sigma_2 w_{s_3s_4,+} + g_{2,s_3,s_4})
\nonumber \\
B_{s_1,s_2,s_3,s_4}^{\sigma_1,\sigma_2} =& \frac{1}{2} e^{i\theta} (v_{s_1s_2,-} + \sigma_1 g_{1,s_1,s_2})(\sigma_2 w_{s_4s_3,-} + g_{2,s_3,s_4})
\nonumber \\
&-  \frac{1}{2} e^{-i\theta} (v_{s_1s_2,+} + \sigma_1 g_{1,s_1,s_2})(\sigma_2 w_{s_4s_3,+} + g_{2,s_3,s_4})
\end{align}
Note that the eigenstates $| S^z_{\vec{r}-\hat{z}} = s_1, S^z_{\vec{r}} = \tilde{s}_2\rangle$ can be changed by a phase, which will modify the
expressions above. It is useful to pick a choice of phase so that, when possible, $g,v,w > 0$.

In Figs. \ref{JxyPlot1}-\ref{JxyPlot3}, we display some results for the numerical calculation of $c_{abcd}$. We see that in all cases,
the only appreciable couplings are $c_{1ab1}$; $c_{zbcd}$ and $c_{abcz}$ are both quite small. This is because the amplitudes $h_{t;3}^{s_1s_2s_3s_4}$ have a very
weak dependence on $s_1$ and $s_4$ and thus essentially cancel each other in the sum.

When $E_J = 0$, then Fig. \ref{JxyPlot1} shows that at $\theta = 0$, only $c_{1xy1} = -c_{1yx1} = 1$; for all values of $\theta$, at least two of
the four couplings $c_{1xx1}$, $c_{1xy1}$, $c_{1yx1}$, and $c_{1yy1}$ are of the same order. As shown in Figs. \ref{JxyPlot2}-\ref{JxyPlot3},
when $E_J \neq 0 $, we can enter the regime where only $c_{1yx1}$ is appreciable while all others are much smaller. In fact,
we see from the equations above that if the matrix element
$\langle s_1,s_2,s_3,s_4 | e^{i\varphi_{-,\vec{r}}} | s_1, \tilde{s}_2, \tilde{s}_3,s_4 \rangle$ is close to one, which is the case for large $E_J$,
then the only appreciable term in $c_{1bc1}$ will be $c_{1yx1}$. This is the reason that we need to include $E_J \neq 0$. The optimal point would
be to take $E_J$ to be as large as possible; however this would dramatically reduce the $J_z$ coupling, which was calculated in the previous section.
Therefore, we need to find an optimal point where $E_J$ is non-zero so that only $c_{1yx1}$ is appreciable, while $J_z$ is still large enough.

\begin{figure}
	\centering
	\includegraphics[width=5.0in]{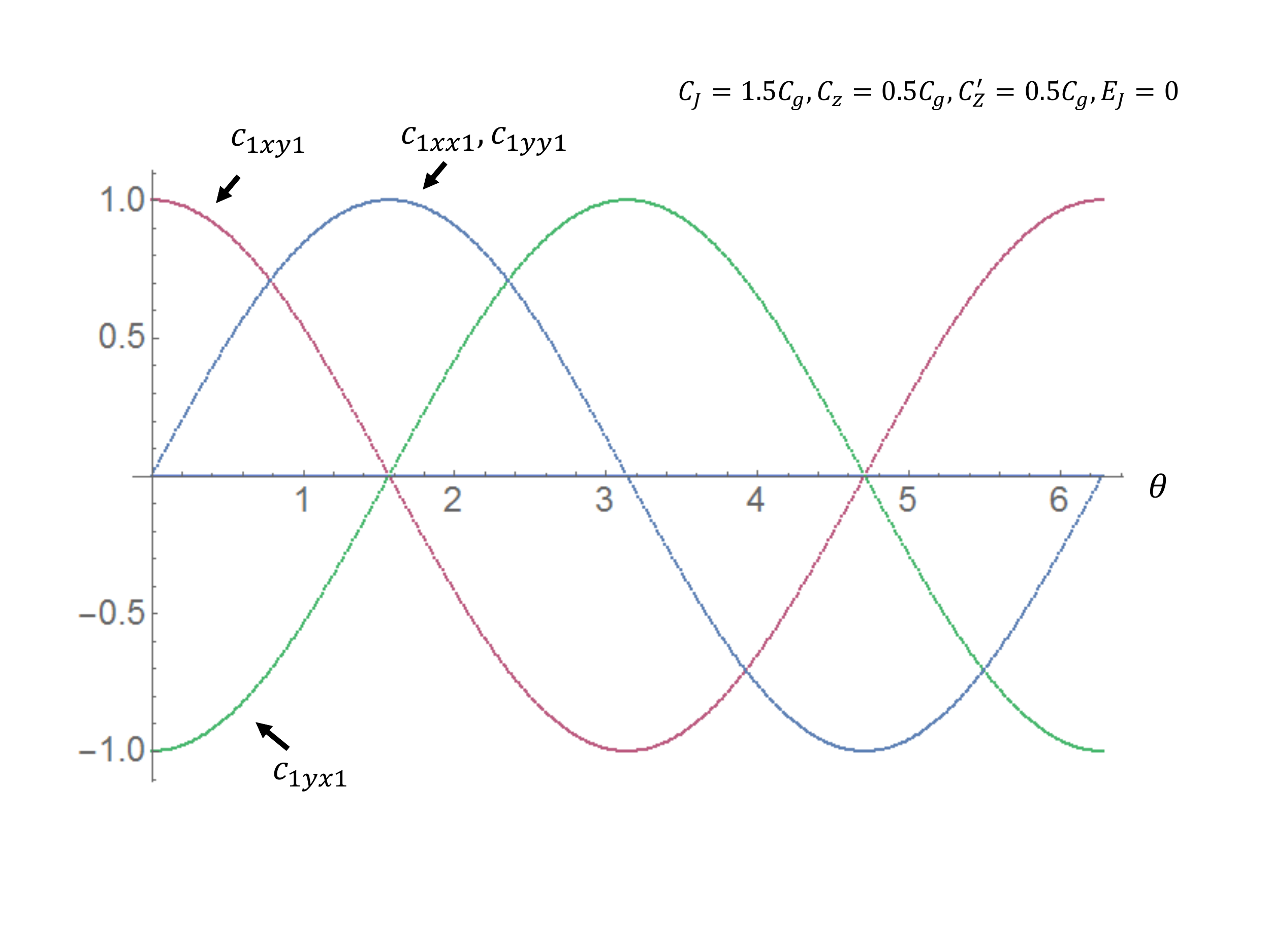}
	\caption{\label{JxyPlot1} Plot of the 16 parameters $c_{abcd}$, for $a,d = 1,z$ and $b,c = x,y$, as a function of $\theta$.
The only ones that differ appreciably from zero are $c_{1bc1}$. }
\end{figure}

\begin{figure}
	\centering
	\includegraphics[width=5.0in]{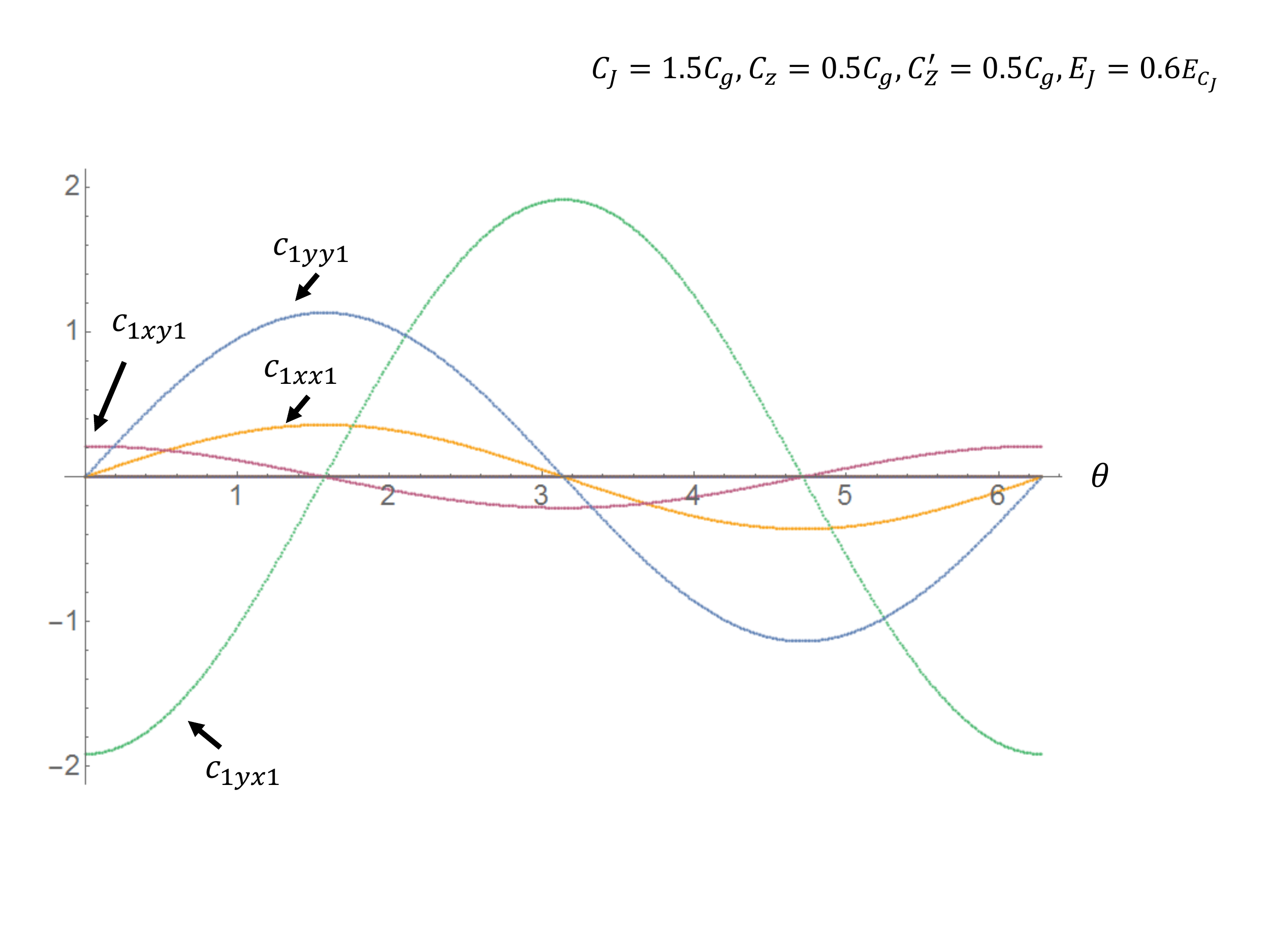}
	\caption{\label{JxyPlot2} Plot of the 16 parameters $c_{abcd}$, for $a,d = 1,z$ and $b,c = x,y$, as a function of $\theta$.
The only ones that differ appreciably from zero are $c_{1bc1}$. We see that for $\theta \approx 0, \pi$, $c_{1yx1}$ is much larger
than $c_{1xy1}$, $c_{1xx1}$, and $c_{1yy1}$.
}
\end{figure}

\begin{figure}
	\centering
	\includegraphics[width=5.0in]{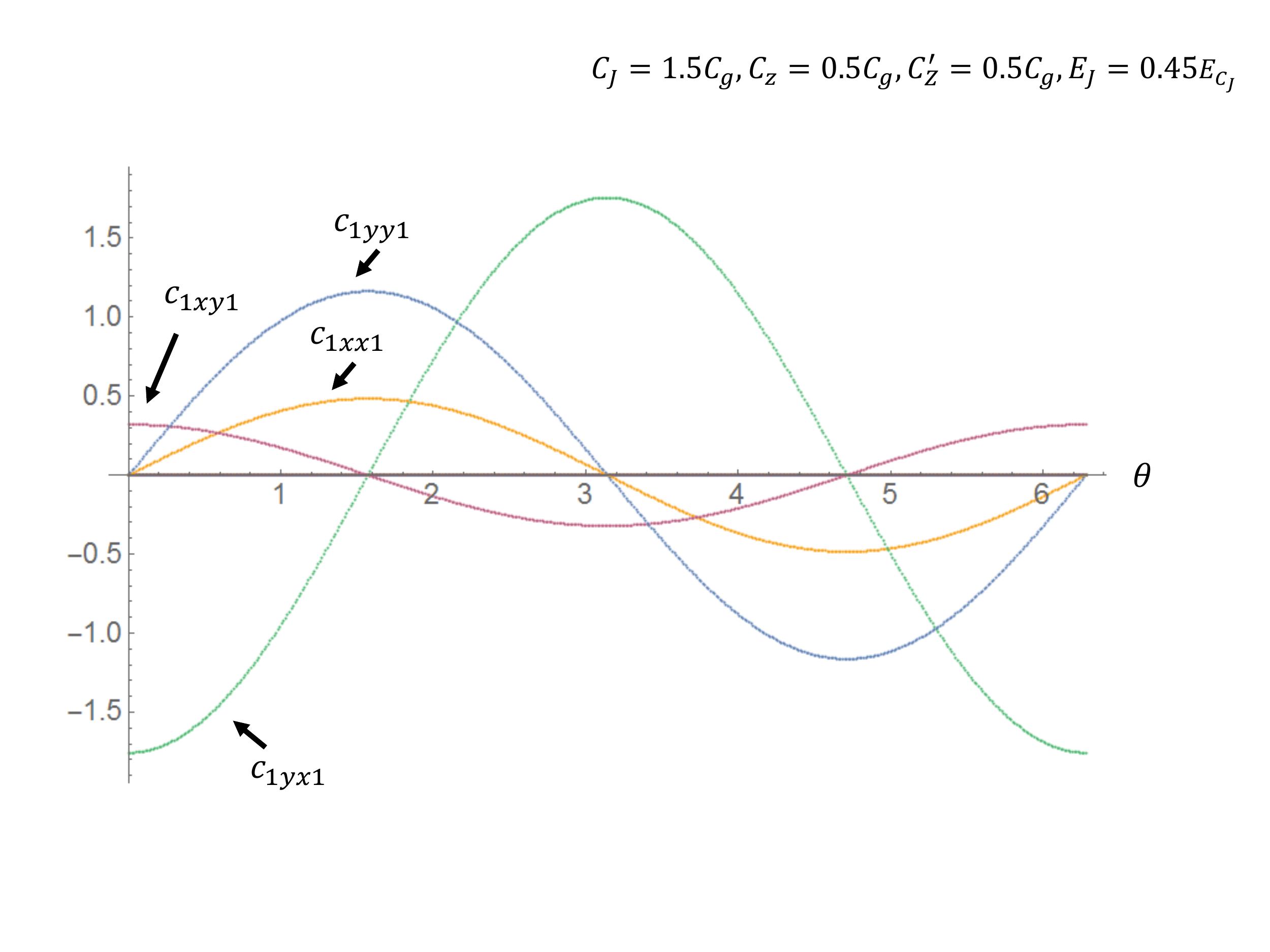}
	\caption{\label{JxyPlot3} Plot of the 16 parameters $c_{abcd}$, for $a,d = 1,z$ and $b,c = x,y$, as a function of $\theta$.
}
\end{figure}

\subsection{2D}

\begin{figure}
	\centering
	\includegraphics[width=4in]{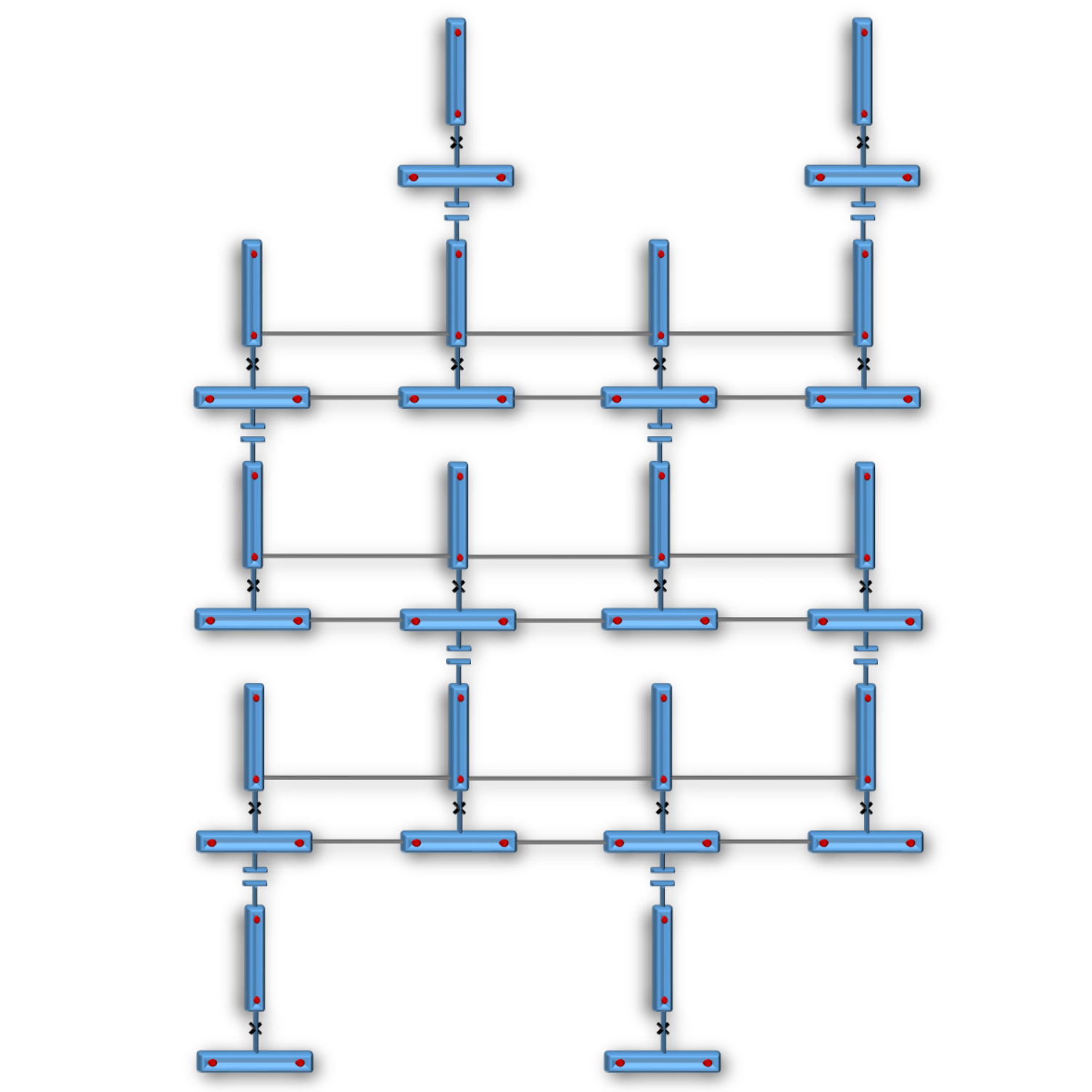}
	\caption{\label{2dfig} Two-dimensional array.
}
\end{figure}

Let us now consider the 2D network shown in Fig. \ref{2dfig} . The effective Hamiltonian for this is given by
\begin{align}
H_{2D} = \sum_I H_{2s,I} + H_{\text{tun}},
\end{align}
where $H_{2s,I}$ is the Hamiltonian for the $I$th unit cell; for each unit cell, this is given by $H_{2s}$ above.
$H_{\text{tun}}$ contains the horizontal tunneling terms, which were analyzed for the case of two horizontally coupled
unit cells in the preceding section. For the full 2D system, it is given by:
\begin{align}
H_{\text{tun}} &=  \sum_{\vec{r}} [t_A \psi_{t,\vec{r}}^\dagger \psi_{t,\vec{r}+\hat{x}} + t_B \psi_{y,\vec{r}}^\dagger \psi_{x,\vec{r}+\hat{x}} + H.c.].
\end{align}
Perturbing around the independent unit cell limit $t_A, t_B = 0$, a simple generalization of the analysis of the preceding section gives the following effective
Hamiltonian, which operates in the subspace spanned by the two effective spin states on each site:
\begin{align}
\label{Heff2d}
H_{2D;\text{eff}} = \sum_{\vec{R}} J_z S^z_{\vec{R}} S^z_{\vec{R}-\hat{z}} +  J_{yx} \sum_{\vec{r}} S^y_{\vec{r}} S^x_{\vec{r}+\hat{x}}
 + \delta H,
\end{align}
where $\sum_{\vec{R}}$ sums over all unit cells, and $\vec{R}$ refers to the top spin of each two-spin unit cell.
As we have shown in the preceding sections, there exist parameter regimes where the additional terms in $\delta H$ are negligible:
\begin{align}
||\delta H|| \ll J_z, J_{yx} .
\end{align}
$H_{2D;\text{eff}}$ can be recognized to be the Kitaev honeycomb spin model.\cite{kitaev2006} Specifically, one can perform a $\pi$ spin rotation
around $S^z$ on every other site, which brings the $H_{2D;\text{eff}}$ into the form
\begin{align}
\label{Heff2d}
H_{2D;\text{eff}} = \sum_{\vec{R}} \left( J_z S^z_{\vec{R}} S^z_{\vec{R}-\hat{z}} +  J_{y} S^y_{\vec{R}} S^y_{\vec{R}+\hat{x}} + J_x S^x_{\vec{R}} S^x_{\vec{R} - \hat{x}} \right)
 + \delta H,
\end{align}
which is the more familiar form of the Kitaev model. Here, $J_y = - J_x = J_{yx}$.

\subsection{Numerical Estimates of Energy Scales}
\label{chargingEstimates}

In Figs. \ref{singleSpinPlot1}-\ref{JxyPlot3}, we have presented the results of several numerical calculations of the
energy spectra of the single effective spin, the two spin unit cell,
and the couplings constants of the effective spin interaction terms. From Figs. \ref{twoSpinPlot1} - \ref{twoSpinPlot4}, we see
that the $J_z$ interactions, which couple the vertically separated spins via an interaction $S^z_{\vec{R}} S^z_{\vec{R}-\hat{z}}$,
must be on the order of a few percent of the Josephson charging energy $e^2/C_J$, in order for the
low energy spin manifold to be comfortably separated from the rest of the excitations of the system.
Figs. \ref{JxyPlot1} -- \ref{JxyPlot3} show that a finite Josephson coupling $E_J$ is
required, so that the horizontal couplings will be in the appropriate regime of the Kitaev honeycomb model.

While we have not performed an exhaustive optimization, our preliminary calculations suggest that the following parameter regime is a good one:
\begin{align}
C_J = 1.5 C_g \;\; C_Z = C_Z' = 0.5 C_g \;\; E_J = 0.45 E_{C_J} \;\; \theta = 0, \;\; |\tilde{t}_A| = |\tilde{t}_B| \approx 0.1 E_{\text{gauge}} .
\end{align}
With this choice of parameters, we find that $J_z \approx 0.02 E_{C_J}$, while the energy cost to the other excited states of the two-spin unit cell is approximately
ten times as large, $E_{\text{gauge}} \approx 0.23 E_{C_J}$. This gives a comfortable energy window that separates the low-lying effective spin states and the rest of
the states of the sytsem. Fig. \ref{JxyPlot3} shows that with this choice of parameters, $c_{1yx1} \approx -1.75$, while $c_{1xy1} \approx 0.3$, and $c_{1xx1}=c_{1yy1} = 0$, as are
all other horizontal coupling terms. This gives almost a factor of approximately 6 between the horizontal couplings that we want and the undesired ones. In terms of
absolute energy scales, we have:
\begin{align}
|J_{x}| = |J_{y}| = 1.75 |\tilde{t}_A \tilde{t}_B|/E_{\text{gauge}} .
\end{align}
If we set $|\tilde{t}_A| = |\tilde{t}_B| = 0.2 E_{\text{gauge}}$ to ensure the single electron tunneling processes are suppressed relative to the second order
process, we find $|J_{x}| = |J_{y}| = 1.75 \times 0.04 E_{\text{gauge}} = 0.016 E_{C_J}$, which is almost the same order as the $J_z$ estimate above.

Therefore, we see that to get an effective spin model whose dominant interactions are the Kitaev interactions, while all other interactions are suppressed, we can
get energy scales that are roughly in the range of a few percent of the Josephson charging energies $E_{C_J}$. To get a large energy scale, then, we wish to use
a physical setup with the largest possible Josephson charging energy $e^2/C_J$, which can also simultaneously accommodate a Josephson coupling $E_J \approx 0.5 e^2/C_J$.

Typical Al-Al$_{x}$O$_{1-x}$ Josephson junctions have Josephson charging energies on the order of $E_{C_J} \approx 1 \text{ K}$,
which can therefore give interaction strengths $J_z, J_{x}, J_{y} \approx 20 \text{ mK}$.

Josephson junctions made from gated semiconductor wires, such as Al-InAs-Al junctions, can yield much larger Josephson
charging energies, because the distance between the superconductors (ie the length of the nanowire junction) can be much larger.
For example, let us consider InAs wires with radius $r$ and a distance $d$ between the Al superconductors. For $r = 20 - 60$ nm and $d = 100 - 450$ nm,
critical supercurrents $I_c = 1 - 135$ nA have been measured,\cite{doh2005} which corresponds to $E_J = \hbar I_c/2e \approx 0.05 - 3$ K. If we consider $d = 100$ nm
and the superconductor consisting of a wire of Al epitaxially grown on the InAs nanowire with total radius $100$ nm, we can estimate
$C_J = \epsilon_r \epsilon_0 \pi (100 \text{nm})^2/(100 \text{nm})$. Taking $\epsilon_r = 15$ for InAs, this implies a charging energy $e^2/C_J \approx 40$ K.
This can be further reduced by increasing the radius or decreasing $d$.

Interestingly, devices with $d = 30$ nm have also been fabricated, and have been reported to yield critical currents as high as $I_c = 800$ nA,
for InAs nanowires with radius $r = 40$ nm.\cite{abay2012} This corresponds to $E_J \approx 40$ K. If we assume the parallel plate capacitor formula
for a radius $40$ nm and $d = 30$ nm, we would get $e^2/C_J \approx 75$ K. However, it is not clear whether such a high supercurrent is due to
unwanted parasitic effects that are introduced during the fabrication process.

To put this on a somewhat more theoretical footing, consider that the supercurrent is typically given by
\begin{align}
I_c R_N = \pi \Delta/2e,
\end{align}
where $R_N$ is the normal-state resistance of the junction, and $\Delta$ is the superconducting gap. For a semiconducting wire with $N_c$ channels,
this implies
\begin{align}
E_J = \hbar I_c/2e = \frac{\hbar}{e^2} \frac{\pi}{4} \Delta N_c \frac{e^2}{h} = \Delta N_c/8
\end{align}
where the conductance is $e^2/h$ per channel. For Al, with $\Delta = 1.2$ K, this implies that $E_J = 0.15 N_c$ K

The above considerations, and in particular the experimental measurements, suggest that it could be possible,
with Al-InAs-Al junctions, to get to a regime where $E_{C_J} = 2 E_J \approx 5-10$ K. This would then imply
\begin{align}
J_z, J_{x}, J_{y} \approx 0.1 - 0.2 \text{ K}.
\end{align}

Note that Nb is also a candidate material that can be used in these setups, instead of Al. Indeed, Nb-InAs-Nb Josephson junctions
have been fabricated and measured.\cite{gunel2012} While the use of Nb presents certain technical obstacles
for fabricating the required semiconductor-superconductor heterostructures, it has the advantage that the superconducting gap is much larger,
$\Delta_{\text{Nb}} \approx 9$ K. This implies that the energy scales considered above will be a factor of $\Delta_{\text{Nb}}/\Delta_{\text{Al}} = 7.5$ larger if Nb is
used instead and good contact can be made between the Nb and the InAs. So far such an enhancement in the critical supercurrent has not been
observed due to contact quality, but there are no fundamental obstacles to improving this contact quality and thus achieving this factor of $7-8$ enhancement.

This would then suggest the theoretical possibility
\begin{align}
J_z, J_{x}, J_{y} \approx 0.5 - 2 \text{ K}.
\end{align}

Superconductors with even larger gaps, such as NbTiN, could potentially yield even larger energy scales.

\section{Quantum Phase Slip Based Implementation}


\begin{figure}
	\centering
	\includegraphics[width=5.0in]{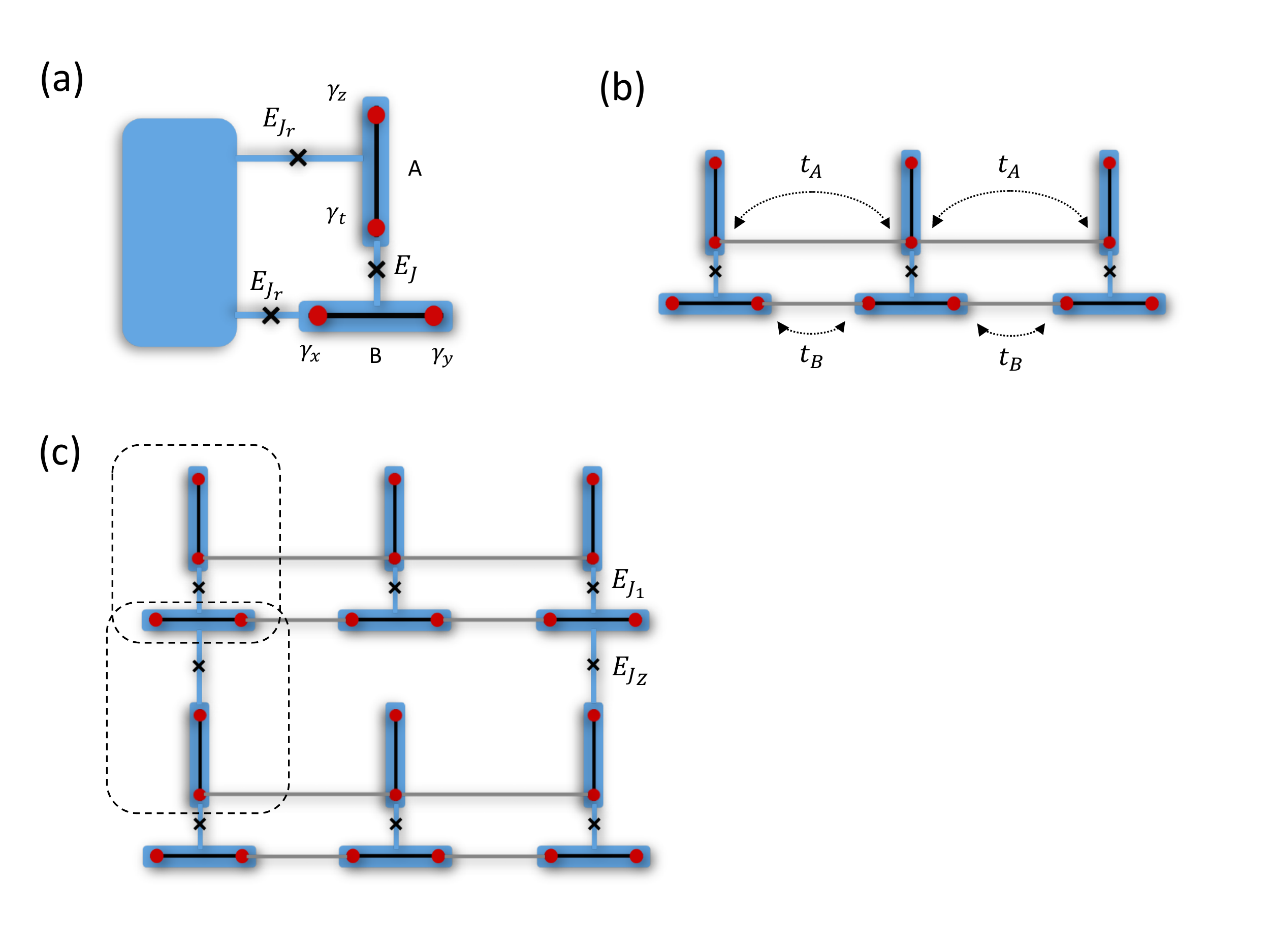}
	\caption{\label{microArc} (a) Single effective spin. The $A$ and $B$ islands are each coupled to a large superconductor with Josephson coupling $J_r$. This is to emulate the embedding of the single spin into a larger network. (b) 1D chain. $t_A$ and $t_B$ indicate electron tunneling. (c) A single plaquette of the 2D network. The Josephson coupling
between different spins is $E_{J_Z}$, while $E_{J_1}$ is the Josephson coupling between the $A$ and $B$ island of a single spin.
}
\end{figure}

\subsection{Single spin}

We now consider an alternative possible architecture, which utilizes the physics of quantum phase slips to engineer an effective Kitaev spin model.
The main building block of this architecture is a set of four Majorana fermion zero modes, as shown in Fig. \ref{microArc}A. Each pair of Majorana zero modes arises from
the endpoints of a spin-orbit coupled semiconductor nanowire, proximity coupled to a superconducting island. Each pair of Majorana fermion zero modes gives rise to two
degenerate states, associated with whether the fermion parity on the island is even or odd. The four Majorana zero modes will be labelled as $\gamma^x$, $\gamma^y$, $\gamma^z$, and $\gamma^t$, as shown in Fig. \ref{microArc}A.

The effective Hamiltonian for the system shown in Fig. \ref{microArc} A is
\begin{align}
\label{singleSpinHqps}
H_{\text{ss}} = \sum_{j = A,B} H_{\text{BdG}} [ \Delta_{0j} e^{i \varphi_j}, \psi_j^\dagger, \psi_j] - E_J \cos(\varphi_A - \varphi_B) - E_{J_r} \sum_{i=A,B} \cos(\varphi_i)
+ \frac{1}{2} \sum_{i,j = A,B} Q_i C_{ij}^{-1} Q_j
\end{align}
Here, $\varphi_j$ for $j = A,B$ is the superconducting phase on the $A$ and $B$ islands,
$H_{\text{BdG}} [ \Delta_{0j} e^{i \varphi_j}, \psi_j^\dagger, \psi_j]$ is the BdG Hamiltonian for the
nanowire on the $j$th island, where $|\Delta_{0j}|$ is the proximity-induced superconducting gap
on the $j$th nanowire at zero magnetic field. $Q_j$ is the excess charge on the $j$th superconducting island - nanowire
combination; it can be written as:
\begin{align}
Q_j = e (-2 i \partial_{\varphi_j} + N_j - n_{\text{off}j}),
\end{align}
where $-i \partial_{\varphi_j}$ represents the number of Cooper pairs on the $j$th superconducting island,
$N_j = \int \psi_{j}^\dagger \psi_{j}$ is the total number of electrons on the $j$th nanowire,
and $n_{\text{off}j}$ is the remaining offset charge on the $j$th island, which can be tuned continuously with the gate voltage $V_g$.
The capacitance matrix is
\begin{align}
C = \left(\begin{matrix}
C_g + C_J + C_r & - C_J \\
-C_J & C_g + C_J + C_r
\end{matrix} \right)
\end{align}

As in Sec. \ref{chargingSecSS}, we perform a unitary transformation $U = e^{-i \sum_{j =A,B} (N_j/2 - n_{Mj}/2 )\varphi_j}$ in order to
decouple the phase $\varphi_j$ from the fermions $\psi_j$ in $H_{\text{BdG}}$. Here,
$n_{Mj} = 0,1$ is the occupation number of the pair of Majorana zero modes on wire $j$. It is given in terms of the
Majorana zero modes as
\begin{align}
n_{MA} &= (1+ i \gamma^z \gamma^t)/2,
\nonumber \\
n_{MB} &= (1+ i \gamma^x \gamma^y)/2.
\end{align}
Under this transformation, the charge $Q_j$ transforms as:
\begin{align}
Q_j' = U^\dagger Q_j U = e(-2i \partial_{\varphi_j} + n_{Mj} - n_{\text{off}j}) .
\end{align}
Thus, taking
$H_{ss} \rightarrow U^\dagger H_{ss} U$, we obtain
\begin{align}
H_{ss}' = &U^\dagger H_{ss} U = \sum_j H_{\text{BdG}}[ \Delta_{0j}, \psi_j^\dagger, \psi_j]- E_J \cos(\varphi_A - \varphi_B) - E_{J_r} \sum_{i=A,B} \cos(\varphi_i)
+ \frac{1}{2} \sum_{i,j = A,B} Q'_i C_{ij}^{-1} Q'_j
\nonumber \\
=& \sum_j H_{\text{BdG}}[ \Delta_{0j}, \psi_j^\dagger, \psi_j]- E_J \cos(\varphi_-) - 2 E_{J_r} \cos(\varphi_+/2) \cos(\varphi_-/2)
\nonumber \\
&+ \frac{1}{4} (Q_+')^2 (C_{AA}^{-1} + C_{AB}^{-1}) +\frac{1}{4} (Q_-')^2 (C_{AA}^{-1} - C_{AB}^{-1}),
\end{align}
where we have used $C_{AA} = C_{BB}$,
and we have set
\begin{align}
Q_{\pm} = Q_A \pm Q_B.
\end{align}
In what follows we will drop the prime superscripts for convenience.

It is helpful to consider the Lagrangian for this system, which is given by
\begin{align}
L_{\varphi} = \frac{1}{2} \frac{1}{4 e^2} \dot{\phi}_i C_{ij}\dot{\phi}_j + \frac{1}{2}(n_{Mi} - n_{\text{off}}) \dot{\varphi}_i
+ E_J \cos(\varphi_A - \varphi_B) + E_{J_r} \sum_{i=A,B} \cos(\varphi_i)
\end{align}

We consider the limit where
\begin{align}
E_J, E_{J_r} \gg e^2 C^{-1}_{IJ},
\end{align}
in which case $\varphi_A$, $\varphi_B$ are pinned, while the conjugate variables
$\hat{N}_A$, $\hat{N}_B$ are highly fluctuating. For energy scales below $E_J, E_{J_r}$, the
effective Hamiltonian of this two island system takes the form
\begin{align}
\label{Heff0Da}
H_{\text{0D}} = \zeta_{A} i \gamma^x \gamma^y + \zeta_B i \gamma^z \gamma^t - \zeta_{AB} \gamma^x \gamma^y \gamma^z \gamma^t
\end{align}
The first two terms are due to quantum phase slip events where either $\varphi_A$ or $\varphi_B$ change by $2\pi$. The last term is
due to the quantum phase slip event where both islands $A$ and $B$ collectively change their phase by $2\pi$, relative to the phase
$\Phi$ of the reservoir. These phase slip processes effectively measure the fermion parity of the region undergoing the phase slip,
which can be expressed in terms of the Majorana fermion modes.

The quantum phase slip amplitudes, $\zeta_{A}, \zeta_B, \zeta_{AB}$, can be computed using the standard instanton calculation,
in the limit of dilute instantons. To leading order in $e^{- \sqrt{8 (E_J + E_{J_r})/E_{C_{AA}}}}$, we estimate this to be
\begin{align}
\zeta_{A} =& -\cos(\pi n_{\text{off}A}) 8 E_{C_{AA}} \sqrt{\frac{2}{\pi}} \left(\frac{E_J+ E_{J_r}}{2E_{C_{AA}}} \right)^{3/4} e^{- \sqrt{8 (E_J + E_{J_r})/E_{C_{AA}}}}
\nonumber \\
\zeta_{B} =& -\cos(\pi n_{\text{off}B}) 8 E_{C_{BB}} \sqrt{\frac{2}{\pi}} \left(\frac{E_J+ E_{J_r}}{2E_{C_{BB}}} \right)^{3/4} e^{- \sqrt{8 (E_J + E_{J_r})/E_{C_{BB}}}}
\nonumber \\
\zeta_{AB} =& - \cos(\pi n_{\text{off}+}) 8 E_{C_{AB}}  \sqrt{\frac{2}{\pi}} \left(\frac{E_{J_r}}{E_{C_{AB}}} \right)^{3/4} e^{- \sqrt{8 (2 E_{J_r})/E_{C_{AB}}}}
\end{align}
where $E_{C_{ii}} = \frac{e^2}{2C_{ii}}$, and $E_{C_{AB}} = \frac{e^2}{4} (C^{-1}_{AA} + C^{-1}_{AB})$.

We assume that the direct coupling between these different Majorana zero modes, which is generated by
electron tunneling between the two ends of the wires, is much smaller than all other energy scales in the
problem, and can therefore be ignored.

Observe now that if we set
\begin{align}
n_{\text{off}I} = 1/2, \;\; I = A, B,
\end{align}
then the effective Hamiltonian is simply
\begin{align}
H_{\text{eff},ss} = -\zeta_{AB} \gamma^x \gamma^y \gamma^z \gamma^t ,
\end{align}
which enforces the constraint
\begin{align}
\gamma^x \gamma^y \gamma^z \gamma^t = 1
\end{align}
for states with energies much less than $\zeta_{AB}$. The ground state subspace of this system is therefore doubly degenerate
and acts like a spin, with
\begin{align}
S^z = i \gamma^z \gamma^t.
\end{align}
Slightly tuning the offset $n_{\text{off}A}$ and/or $n_{\text{off}B}$ away from $1/2$ acts like a Zeeman field $h_z S^z$.

\subsection{1D Chain}

Next, let us consider a 1D chain of the $A,B$ superconducting island pairs introduced in the previous subsection. The coupling between the effective sites of the chain consists of the Majorana fermion tunneling terms shown in Fig. \ref{microArc}B, with tunneling amplitudes $t_A$ and $t_B$. In addition to the single particle tunneling terms between the Majorana zero modes, there are pair tunneling terms that induce Josephson couplings $J_A$, $J_B$ between the islands. Finally, there are cross-capacitances between the different superconducting islands; we will consider the regime where the charging energies due to these cross-capacitances are much smaller than the tunneling and Josephson couplings, and can therefore be ignored. The effective Hamiltonian of the chain is therefore
\begin{align}
\label{Heff1Da}
H_{1D} =& \sum_{\vec{r}} [ h_{\vec{r}} + h_{\vec{r},\vec{r}+\hat{x}} ],
\end{align}
where each $h_{\vec{r}}$ is given by the Hamiltonian $H_{ss}$ described in the previous section, and
\begin{align}
\label{Hnn}
h_{\vec{r},\vec{r}+\hat{x}} &=  t_A \psi_{t\vec{r}}^\dagger \psi_{t\vec{r}+\hat{x}} + t_B \psi_{y\vec{r}}^\dagger \psi_{x\vec{r}+\hat{x}}
\nonumber \\
&- J_A \cos(\varphi_{\vec{r},A} - \varphi_{\vec{r}+\hat{x},A}) -
J_B \cos(\varphi_{\vec{r},B} - \varphi_{\vec{r}+\hat{x},B}) .
\end{align}
Note that the pair tunnelings $J_A$ and $J_B$ in this setup are generated by pair tunneling between the Majorana zero modes, and thus
$J_A < t_A$, $J_B < t_B$.
Performing the unitary transformation $U$ in the previous section, and then setting
\begin{align}
\psi_{\alpha\vec{r}} = u^\alpha \gamma^\alpha_{\vec{r}} ,
\end{align}
as in the previous section, we get:
\begin{align}
h_{\vec{r},\vec{r}+\hat{x}} &=
[\tilde{t}_A e^{-i (1+F_{p,A,\vec{r}})\varphi_{A\vec{r}}/2 + i  (1-F_{p,A,\vec{r}+\hat{x}})\varphi_{A\vec{r}+\hat{x}}/2} \gamma^t_{\vec{r}} \gamma^t_{\vec{r}+\hat{x}}
\nonumber \\
&+ \tilde{t}_B e^{-i (1+F_{p,B,\vec{r}})\varphi_{B\vec{r}}/2 + i  (1-F_{p,B,\vec{r}+\hat{x}})\varphi_{B\vec{r}+\hat{x}}/2}  \gamma^y_{\vec{r}} \gamma^x_{\vec{r}+\hat{x}} + H.c.]
\nonumber \\
&- J_A \cos(\varphi_{\vec{r},A} - \varphi_{\vec{r}+\hat{x},A}) - J_B \cos(\varphi_{\vec{r},B} - \varphi_{\vec{r}+\hat{x},B}) .
\end{align}
Now, since we are in the limit of large Josephson couplings, all of the phases of the superconducting islands can be set equal to each other, which we can set
to zero without loss of generality:
\begin{align}
\varphi_{A\vec{r}} = \varphi_{B\vec{r}} = 0 ,
\end{align}
with corrections coming from instanton events. Thus, $h_{\vec{r},\vec{r}+\hat{x}}$ becomes
\begin{align}
h_{\vec{r},\vec{r}+\hat{x}} =
[2 i \text{Im}(\tilde{t}_A) \gamma^t_{\vec{r}} \gamma^t_{\vec{r}+\hat{x}} + 2 i \text{Im}(\tilde{t}_B) \gamma^y_{\vec{r}} \gamma^x_{\vec{r}+\hat{x}}] .
\end{align}

We wish to consider the limit where
\begin{align}
\tilde{t}_A, \tilde{t}_B \ll \zeta_{AB} .
\end{align}
In this case, the single tunneling events are suppressed; perturbing to second order in $h_{\vec{r},\vec{r}+\hat{x}}$,
we obtain an effective Hamiltonian:
\begin{align}
h_{\vec{r},\vec{r}+\hat{x}}^{\text{eff}} = -\frac{4 \text{Im}(\tilde{t}_A) \text{Im}(\tilde{t}_B)}{\zeta_{AB}}
\gamma^y_{\vec{r}} \gamma^t_{\vec{r}} \gamma^x_{\vec{r}+\hat{x}} \gamma^t_{\vec{r}+\hat{x}}
\end{align}

Therefore, the effective Hamiltonian of the 1D chain is given by:
\begin{align}
H_{1D} =\sum_{\vec{r}}  [ \zeta_{A} i \gamma_{\vec{r}}^x \gamma_{\vec{r}}^y + \zeta_B i \gamma_{\vec{r}}^z \gamma_{\vec{r}}^t
- \zeta_{AB} \gamma_{\vec{r}}^x \gamma_{\vec{r}}^y \gamma_{\vec{r}}^z \gamma_{\vec{r}}^t
-\frac{4 \text{Im}(\tilde{t}_A) \text{Im}(\tilde{t}_B)}{\zeta_{AB}}
\gamma^y_{\vec{r}} \gamma^t_{\vec{r}} \gamma^x_{\vec{r}+\hat{x}} \gamma^t_{\vec{r}+\hat{x}} ]
\end{align}
If we assume for simplicity that $J_A = J_B = J$, then the quantum phase slip amplitudes are
essentially as given in Eqn. \ref{ssQPS}, with $E_{J_r} = 2J$, and $C_r$ being twice the capacitance across
the semiconductor wires.

\subsection{2D Network}

Now we would like to assemble the 1D chains described above into a two-dimensional network. We consider the network shown in
Fig. \ref{microArc}C, which effectively forms a brick (honeycomb) lattice. Each unit cell of the lattice consists of two pairs of superconducting
islands: two $A$ islands and two $B$ islands, separated vertically from each other. The Hamiltonian for the system can be written as
\begin{align}
H_{2D} = \sum_{\vec{R}} H_{2s,\vec{R}} + H_{\text{tun}},
\end{align}
where $\vec{R}$ is the location of the top spin of each unit cell, $\sum_{\vec{R}}$ is thus a sum over unit cells, and
the Hamiltonian for each unit cell $H_{2s;\vec{R}}$:
\begin{align}
H_{2s;\vec{R}} = &\sum_{I=1}^4 H_{BdG} + \frac{1}{2} \sum_{I,J = 1}^4 Q_I C^{-1}_{IJ} Q_J
- E_{J_1} \cos(\varphi_{A\vec{R}} - \varphi_{B\vec{R}}) - E_{J_1}  \cos(\varphi_{A\vec{R}-\hat{z}} - \varphi_{B\vec{R}-\hat{z}})
\nonumber \\
& - E_{J_Z} \cos(\varphi_{B\vec{R}} - \varphi_{A\vec{R}-\hat{z}}) .
\end{align}
The $4\times 4$ capacitance matrix now is
\begin{align}
C = \left(\begin{matrix}
C_g + C_J & - C_J & 0 & 0 \\
-C_J & C_g + C_J + C_{J_Z} & - C_{J_Z} & 0 \\
0 & -C_{J_Z} & C_g + C_J + C_{J_Z} & -C_J \\
0 & 0 & -C_J & C_g + C_J
\end{matrix} \right)
\end{align}
The tunneling Hamiltonian $H_{\text{tun}}$ is:
\begin{align}
H_t = \sum_{\vec{r}} [t_A \psi_{t\vec{r}}^\dagger \psi_{t\vec{r}+\hat{x}} + t_B \psi_{y\vec{r}}^\dagger \psi_{x\vec{r}+\hat{x}} + H.c. ]
\end{align}
We have ignored the capacitance between horizontally separated islands, as they are assumed to be far enough apart that their capacitance
is negligible.

Based on the analysis of the single spin case and the 1D chain, we can now immediately see that the low energy effective Hamiltnian for this
system can be written as
\begin{align}
H_{2D;\text{eff}} = H_1 + H_2 + H_{t;\text{eff}},
\end{align}
where $H_1$ consists of single island phase slips:
\begin{align}
H_1 = \sum_{\vec{R}} ( \zeta_{\vec{R}}^A i \gamma_{\vec{R}}^z \gamma_{\vec{R}}^t + \zeta_{\vec{R}-\hat{z}}^A \gamma_{\vec{R}-\hat{z}}^z \gamma_{\vec{R}-\hat{z}}^t
+ \zeta_{\vec{R}}^B i \gamma_{\vec{R}}^x \gamma_{\vec{R}}^y + \zeta_{\vec{R}-\hat{z}}^B \gamma_{\vec{R}-\hat{z}}^x \gamma_{\vec{R}-\hat{z}}^y)
\end{align}
$H_2$ consists of two-island phase slips:
\begin{align}
\label{ssQPS}
H_2 = -&\sum_{\vec{r}} \zeta_{\vec{r}}^{AB} \gamma_{\vec{r}}^z \gamma_{\vec{r}}^t \gamma_{\vec{r}}^x \gamma_{\vec{r}}^y
-\sum_{\vec{R}} \zeta_{Z}^{AB} \gamma_{\vec{R}}^x \gamma_{\vec{R}}^y \gamma_{\vec{R}-\hat{z}}^z \gamma_{\vec{R}-\hat{z}}^t
\nonumber \\
&- \sum_{\vec{r}} \zeta_{X}^{AA} \gamma_{\vec{r}}^z \gamma_{\vec{r}}^t \gamma_{\vec{r}+\hat{x}}^z \gamma_{\vec{r}+\hat{x}}^t
- \sum_{\vec{r}} \zeta_{X}^{BB} \gamma_{\vec{r}}^x \gamma_{\vec{r}}^y \gamma_{\vec{r}+\hat{x}}^x \gamma_{\vec{r}+\hat{x}}^y
\end{align}
These phase slip amplitudes are given in terms of the Josephson couplings and charging energies:
\begin{align}
\zeta^{A}_{\vec{R}} =& -\cos(\pi n_{\text{off}A,\vec{R}}) 8 E_{C_{11}} \sqrt{\frac{2}{\pi}} \left(\frac{E_{J_1}}{2E_{C_{11}}} \right)^{3/4} e^{- \sqrt{8 (E_{J_1})/E_{C_{11}}}}
\nonumber \\
\zeta^{A}_{\vec{R}-\hat{z}} =& -\cos(\pi n_{\text{off}A,\vec{R}-\hat{z}}) 8 E_{C_{33}} \sqrt{\frac{2}{\pi}}
\left(\frac{E_{J_1} + E_{J_Z}}{2E_{C_{11}}} \right)^{3/4} e^{- \sqrt{8 (E_{J_1}+ E_{J_Z})/E_{C_{33}}}}
\nonumber \\
\zeta^{B}_{\vec{R}} =& -\cos(\pi n_{\text{off}B,\vec{R}}) 8 E_{C_{22}} \sqrt{\frac{2}{\pi}}
\left(\frac{E_{J_1} + E_{J_Z}}{2E_{C_{22}}} \right)^{3/4} e^{- \sqrt{8 (E_{J_1}+ E_{J_Z})/E_{C_{22}}}}
\nonumber \\
\zeta^{B}_{\vec{R}-\hat{z}} =& -\cos(\pi n_{\text{off}B,\vec{R}-\hat{z}})
8 E_{C_{44}} \sqrt{\frac{2}{\pi}} \left(\frac{E_{J_1}}{2E_{C_{44}}} \right)^{3/4} e^{- \sqrt{8 (E_{J_1})/E_{C_{44}}}}
\end{align}
\begin{align}
\zeta_{\vec{r}}^{AB} &= - \cos(\pi (n_{\text{off}A,\vec{r}} + n_{\text{off}B,\vec{r}} ))
8 E_{C_{AB}} \sqrt{\frac{2}{\pi}} \left(\frac{E_{J_Z}}{2E_{C_{AB}}} \right)^{3/4} e^{- \sqrt{8 (E_{J_Z})/E_{C_{AB}}}}
\nonumber \\
\zeta_{Z}^{AB} &= - \cos(\pi (n_{\text{off}B,\vec{r}} + n_{\text{off}A,\vec{r}-\hat{z}} ))
8 E_{C_{BA}} \sqrt{\frac{2}{\pi}} \left(\frac{2E_{J_1}}{2E_{C_{BA}}} \right)^{3/4} e^{- \sqrt{8 (2E_{J_1})/E_{C_{BA}}}}
\end{align}
where we have defined
\begin{align}
E_{C_{AB}} &= \frac{e^2}{2} \frac{1}{2C_g + C_Z}
\nonumber \\
E_{C_{BA}} &= \frac{e^2}{2} \frac{1}{2C_g + 2 C_J}
\end{align}

The phase slip amplitudes $\zeta_{X}^{AA}$, $\zeta_X^{BB}$ are approximately given by products of the single island phase slips,
\begin{align}
\zeta_X^{AA} \approx \zeta_{\vec{R}}^A \zeta_{\vec{R}-\hat{z}}^A,
\nonumber \\
\zeta_X^{BB} \approx \zeta_{\vec{R}}^B \zeta_{\vec{R}-\hat{z}}^B,
\end{align}
because the horizontal Josephson couplings are negligible.

All other two-island phase slips, and collective phase slips of more than two islands, can be ignored,
as their amplitudes are exponentially suppressed relative to the terms considered here.
Finally, $H_{\text{tun}}$ includes the Majorana tunneling terms $t_A$ and $t_B$:
\begin{align}
H_{\text{tun}} = \sum_{\vec{r}} [2 i \text{Im}(\tilde{t}_A) \gamma_{t,\vec{r}} \gamma_{t,\vec{r}+\hat{x}}
+ 2 i \text{Im}(\tilde{t}_B) \gamma_{y,\vec{r}} \gamma_{x,\vec{r}+\hat{x}}]
\end{align}

If we set $n_{\text{off}A,\vec{r}} = n_{\text{off}B,\vec{r}} = 1/2$, then the single island phase slips vanish.
The effective Hamiltonian becomes
\begin{align}
\label{H2Deff}
H_{2D;\text{eff}} = -&\sum_{\vec{r}} \zeta_{\vec{r}}^{AB} \gamma_{\vec{r}}^z \gamma_{\vec{r}}^t\gamma_{\vec{r}}^x \gamma_{\vec{r}}^y
-\sum_{\vec{R}} \zeta_{Z}^{AB} \gamma_{\vec{R}}^x \gamma_{\vec{R}}^y \gamma_{\vec{R}-\hat{z}}^z \gamma_{\vec{R}-\hat{z}}^t
-\frac{4 \text{Im}(\tilde{t}_A) \text{Im}(\tilde{t}_B)}{\zeta_{AB}}
\gamma^y_{\vec{r}} \gamma^t_{\vec{r}} \gamma^x_{\vec{r}+\hat{x}} \gamma^t_{\vec{r}+\hat{x}}
\end{align}
If we further consider the regime where
\begin{align}
\zeta^{AB}_{\vec{r}} \gg \zeta_Z^{AB}, 2\text{Im} (\tilde{t}_A), 2 \text{Im}(\tilde{t}_B),
\end{align}
we see that the system can be described by an effective spin model:
\begin{align}
H_{2D;\text{eff}} = \sum_{\vec{R}} J_z S^z_{\vec{R}} S^z_{\vec{R}-\hat{z}} + J_{yx} \sum_{\vec{r}} S^y_{\vec{r}} S^x_{\vec{r}+\hat{x}},
\end{align}
with
\begin{align}
J_z &= \zeta_Z^{AB},
\nonumber \\
J_{yx} &= \frac{4 \text{Im}(\tilde{t}_A) \text{Im}(\tilde{t}_B)}{\zeta_{AB}}
\end{align}

Upon rotating every other spin around the $z$ axis by $\pi/2$, the above Hamiltonian can be put into the more familiar Kitaev form:
\begin{align}
H_{2D;\text{eff}} = \sum_{\langle i j \rangle = \text{z-link}} J_z S^z_i S^z_j + \sum_{\langle i j \rangle = \text{x-link}}J_{yx} S^x_i S^x_j
- \sum_{\langle i j \rangle = \text{y-link}} J_{yx} S^y_i S^y_j,
\end{align}

\section{Realizing the Ising topological order}

As we have shown, the physical architectures described above can give rise to an effective
realization of the 2D Kitaev honeycomb spin model:
\begin{align}
\label{kitaevH}
H_{\text{K}} = J_x \sum_{\text{x-links}} S_i^x S_j^x + J_y \sum_{\text{y-links}} S_i^y S_j^y + J_z \sum_{\text{z-links}}  S_i^z S_j^z.
\end{align}
It is well-known that this model is proximate to a non-Abelian topological state with Ising topological order.
There are a number of known ways to access the non-Abelian state. One way to access the non-Abelian state is
to apply a small effective Zeeman field:
\begin{align}
\delta H = \sum_{\vec{r}} h_x S^x_{\vec{r}} + h_y S^y_{\vec{r}} + h_z S^z_{\vec{r}}.
\end{align}
As explained in Sec. \ref{chargingSecSS}, The $h_z$ term above can be generated by tuning the gate voltage on each of the superconducting islands.
The $h_x$ and $h_y$ terms are more difficult, but possible, to generate as well. They require connecting the Majorana zero modes with
additional semiconductor wires, as shown in Figs. \ref{capSpin}c, to allow for electron tunneling as shown.

\begin{figure}
	\centering
	\includegraphics[width=4.0in]{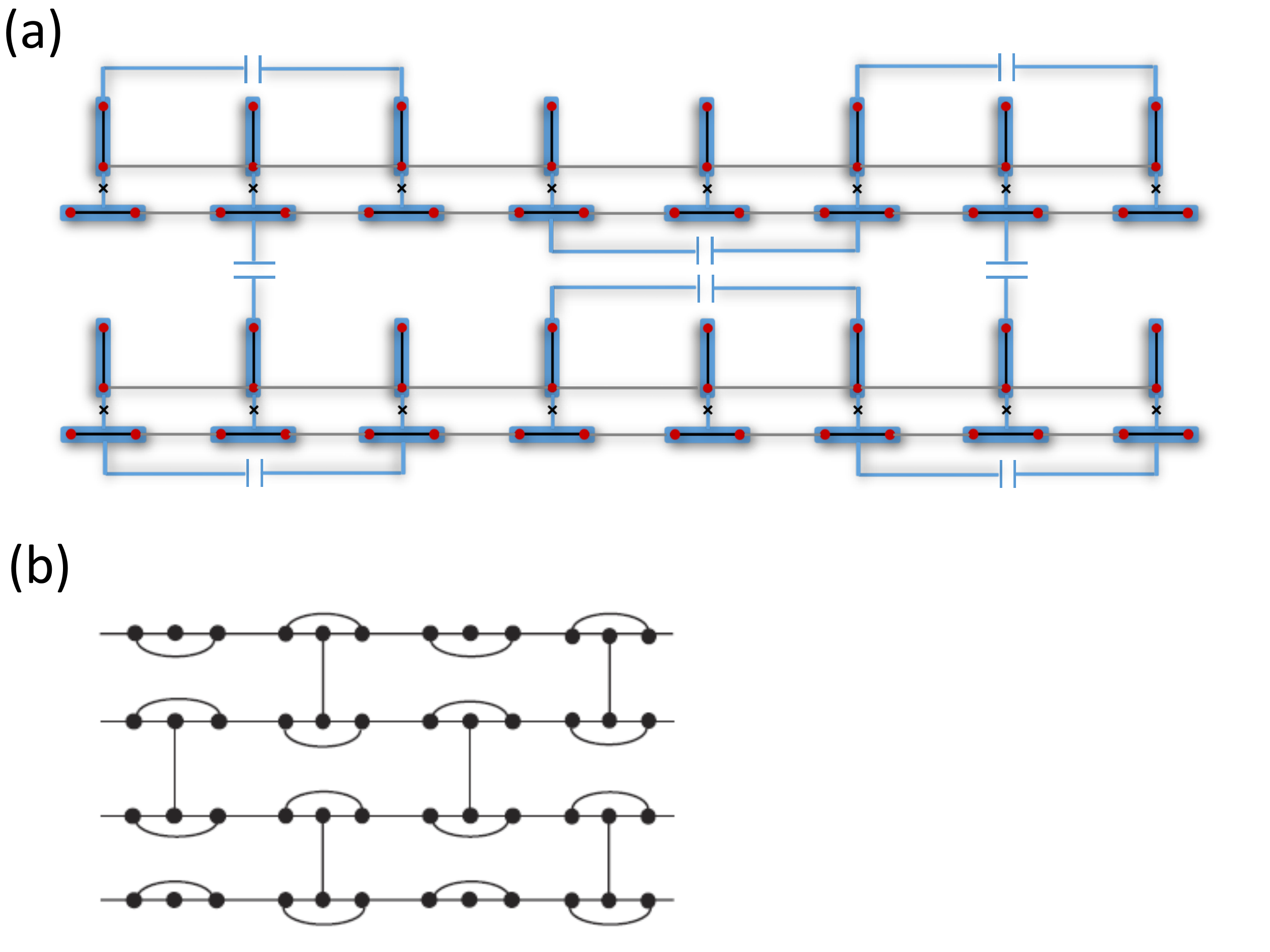}
	\caption{\label{microArcYK} Micro-Architecture for the Yao-Kivelson realization of the model.
}
\end{figure}
A second way of realizing the Ising topological order is to consider effectively the same model, but on a different lattice (see Fig. \ref{microArcYK}),
as proposed by Yao and Kivelson.
As a spin model, the Kitaev Hamiltonian, Eq. (\ref{kitaevH}) is time-reversal invariant. On the lattice structure proposed by Yao-Kivelson
the ground state spontaneously breaks time reversal symmetry, yielding a ground state with Ising (or its time-reversed partner, $\overline{\text{Ising}}$)
topological order. In the completely isotropic limit where all couplings are equal to $J$, the energy gap of the Ising state is also
equal to $J$. Interestingly, disorder in the spin couplings can actually be beneficial and can enlarge the region of stability of the
Ising phase.\cite{chua2011}

By adding a small effective time-reversal symmetry breaking perturbation to the spin model, we can tune whether the
topological order is Ising or $\overline{\text{Ising}}$, and avoid having domains of either, as would be realistically expected in
the case where the effective time-reversal symmetry is broken spontaneously. A Zeeman term $\sum_{\vec{r}} h_z S^z_{\vec{r}}$
by itself is insufficient. We can consider either a Zeeman term that includes both $h_z$ and $h_y$. Or, in order to avoid
requiring $h_y$ or $h_x$ terms, which are more difficult to generate, we can make use of the smaller perturbations
$S^y_{\vec{R}} S^x_{\vec{R}+\hat{x}}$, $S^x_{\vec{r}} S^y_{\vec{r}+\hat{x}}$, that are naturally generated in the
first charging energy based implementation presented above.

\section{Genons}

\subsection{Ising $\times$ Ising topological order}

\begin{figure}
	\centering
	\includegraphics[width=6in]{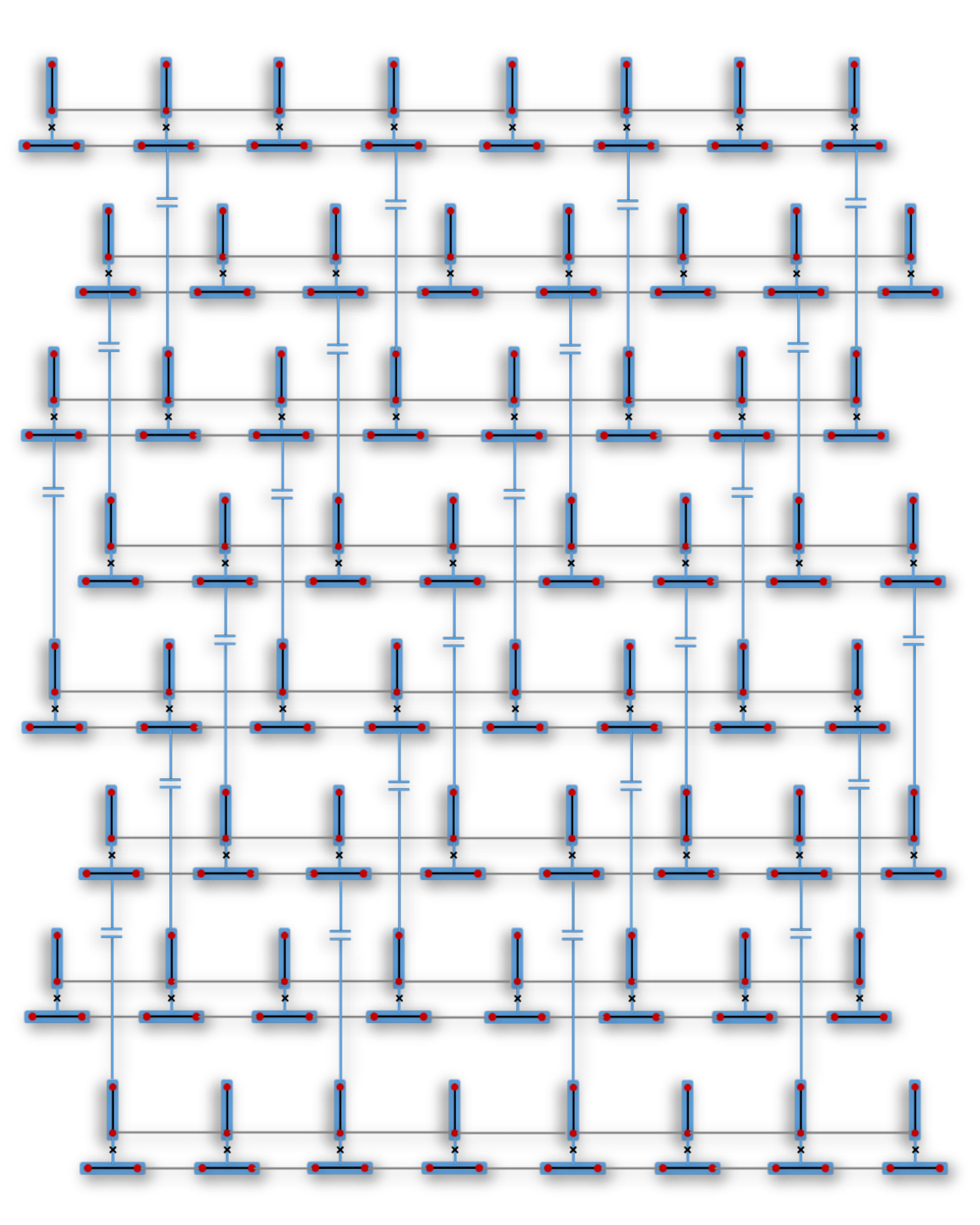}
	\caption{\label{bilayerFig} Two essentially decoupled copies of the capacitance-based model can be created by using short overpasses to couple
next nearest neighbor chains.
}
\end{figure}

Given a microscopic architecture to realize a quantum state with Ising topological order, one can then consider designing two independent copies
of such a state (referred to as the Ising $\times$ Ising state)  by utilizing present-day nanofabrication technology to create short overpasses among different
superconducting wires, as shown in Fig. \ref{bilayerFig}.

\subsection{Creating genons}

A genon in an Ising $\times$ Ising state can then be realized by modifying the overpass connections to create a segment along which the connections among the
horizontal chains is twisted, as shown in Fig. \ref{microArcGenon}. These segments effectively create branch lines that connect one layer to the
other, and vice versa. The end-points of the segments realize exotic non-Abelian twist defects, which have been referred to as genons. The topological
degeneracy of the system in the presence of the genons mimics that of a single copy Ising system on a surface of non-trivial topology.\cite{barkeshli2010,barkeshli2012a,barkeshli2013genon}

\begin{figure}
	\centering
	\includegraphics[width=4.0in]{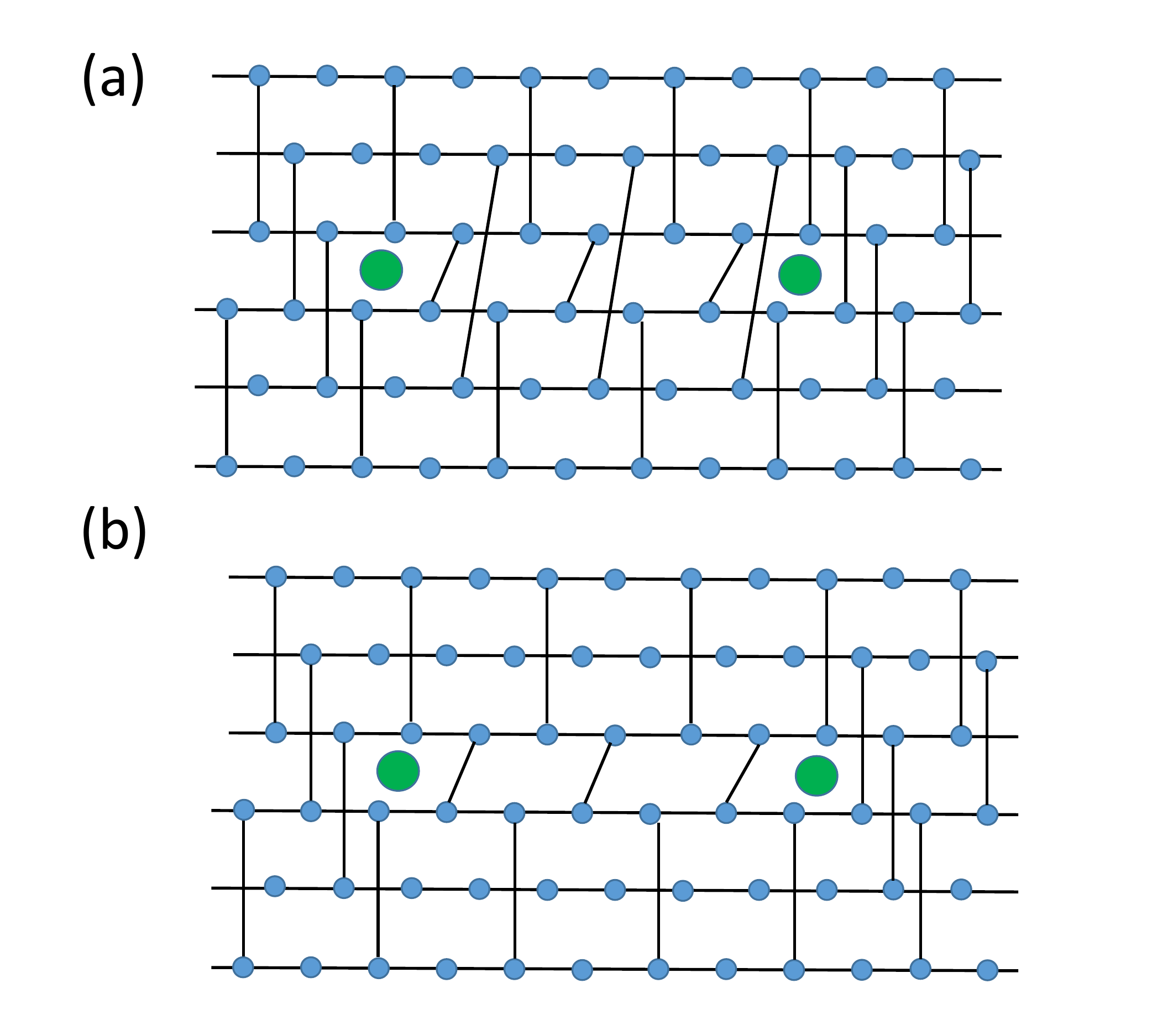}
	\caption{\label{microArcGenon} Architecture for creating genons in the effective spin model. The location
of the genons corresponds to the end point of the branch lines, and are marked by green circles. (a) Full lattice dislocation.
(b) The vertical bonds that skip two chains along the branch cut are removed. The fact that these vertical bonds can be removed
follows the analysis of Ref. \onlinecite{barkeshli2014exp}, which showed in a different context that half of the branch cut is sufficient.
}
\end{figure}

Technically, there are three topologically distinct types of genons, which we label as $X_\mathbb{I}$, $X_\psi$, and $X_\sigma$.\cite{barkeshli2014sdg}
Physically, they correspond to whether a $\mathbb{I}$, $\psi$, $\sigma$ particle, from either layer, is bound to the genon.
The genons have the following fusion rules:

\begin{align}
X_\mathbb{I} \times X_\mathbb{I} = (\mathbb{I},\mathbb{I}) + (\psi, \psi) + (\sigma, \sigma)
\nonumber \\
X_\mathbb{I} \times (\mathbb{I},\psi) = X_\mathbb{I} \times (\psi,\mathbb{I}) = X_\psi
\nonumber \\
X_\mathbb{I} \times (\mathbb{I},\sigma) = X_\mathbb{I} \times (\sigma,\mathbb{I}) = X_\sigma
\end{align}
It follows from the above that $X_\mathbb{I}$, $X_\psi$ have quantum dimension $2$, while $X_\sigma$ has quantum dimension $2\sqrt{2}$.

\subsection{Effective braiding of genons and the topological $\pi/8$ phase gate}

In Ref. \onlinecite{barkeshli2013genon}, it was shown that the braiding of genons can be used to realize a topologically protected
$\pi/8$ phase gate. In the present system, the braiding of genons is complicated by the fact that
it is difficult to continuously modify the physical location of the genons to execute a braid loop
in real space. Fortunately, this is not necessary, as the braiding of the genons can be implemented through a
measurement-based approach. To do this, we require that it be possible to measure the joint fusion channel of
any pair of genons and project it into either the $(\mathbb{I},\mathbb{I})$ channel or the $(\psi,\psi)$ channel.

\subsection{Measurement-based braiding of genons}

Importantly, the braiding of the genons can be achieved without moving them continuously around each other in space, but rather through tuning the effective interactions between them. Specifically, what is required to braid two genons is the ability to project the fusion channel of pairs of genons onto an Abelian charge sector.

In order to implement the $\pi/8$ phase gate, we wish to start with two pairs of genons, labelled $1,..., 4$, and have the ability to braid genons $2$ and $3$. In order to do this, we use an ancillary pair of genons, labelled $5$ and $6$. The braiding process is then established by projecting the genons $5$ and $6$ onto the fusion channel $b_{56}$, then the genons $5$ and $3$ onto the fusion channel $b_{35}$, the genons $5$ and $2$ onto the fusion channel $b_{25}$, and finally again the genons $5$ and $6$ onto the fusion channel $b_{56}'$.

\begin{figure}
	\centering
	\includegraphics[width=4.0in]{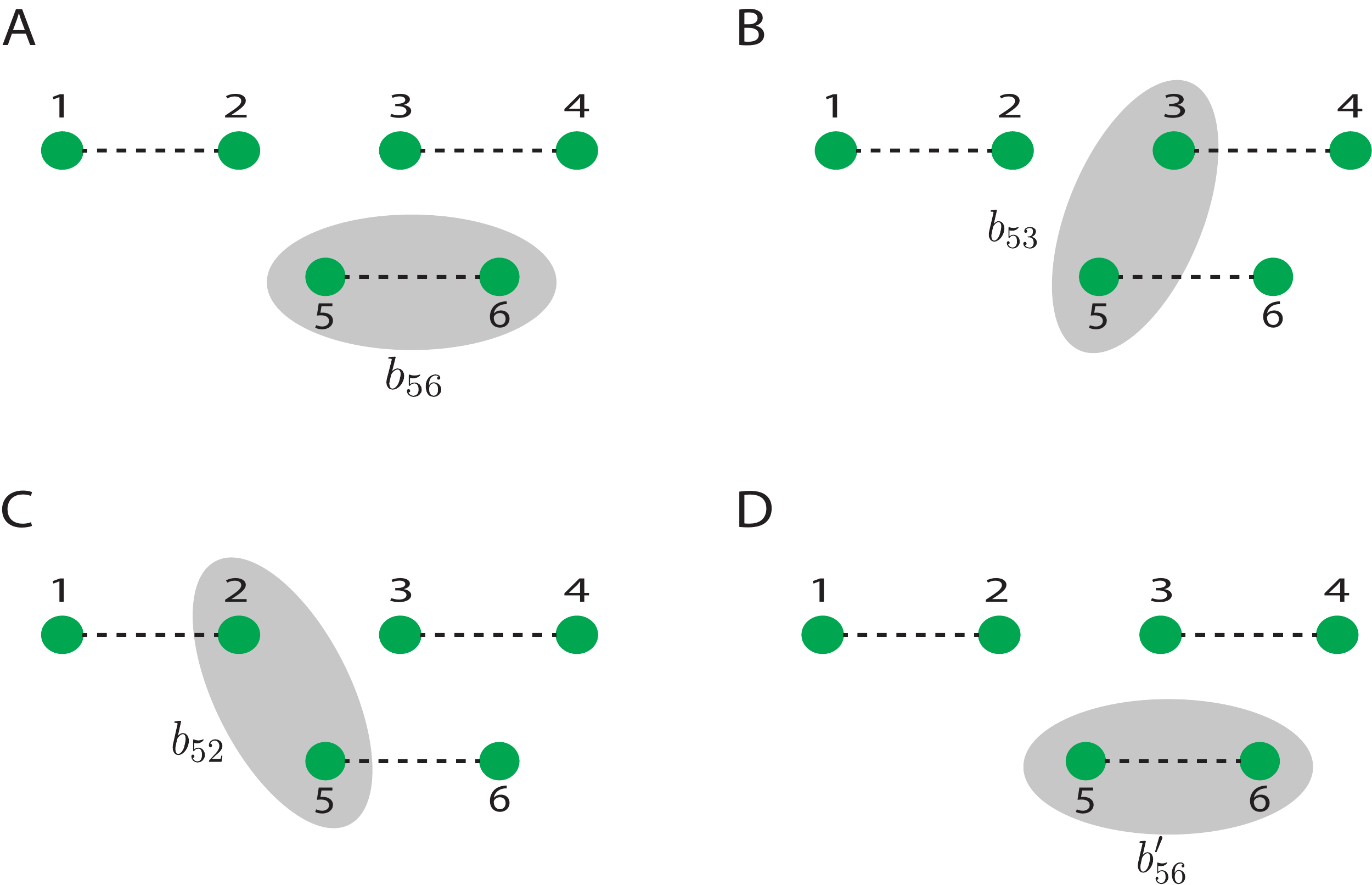}
	\caption{\label{genonBraiding} Schematic for measurement based braiding of genons. In order to effectively braid genons 2 and 3, the following protocol is performed. (a) The fusion channel of genons 5 and 6 is projected onto the anyon $b_{56}$. (b) The fusion channel of genons 5 and 3 is projected onto $b_{53}$. (c) The fusion channel of genons 5 and 2 is projected onto $b_{52}$. (d) The fusion channel of genons 5 and 6 is projected onto $b_{56}'$.
}
\end{figure}

We will asssume here that $b_{56} = b_{56}'$. If the genons $5$ and $6$ are created out of the vacuum, then it will in fact be natural to have $b_{56} = b_{56}' = (\mathbb{I}, \mathbb{I})$. In this situation, we can derive the resulting braid matrix for the genons, following the results of Ref. \onlinecite{bonderson2013braiding}.
When $b_{25} = b_{35}$, the braid matrix for genons $2$ and $3$ is given by
\begin{align}
R_{23} = e^{i\phi} \left(\begin{matrix} 1 & 0 & 0 \\ 0 & -1 & 0 \\ 0 & 0 & e^{i\pi/8}\end{matrix} \right),
\end{align}
where $e^{i\phi}$ is an undetermined, non-topological phase.
In other words, the state obtains a phase of $\pm 1$ or $e^{i\pi/16}$, depending on whether the fusion channel of genons $2$ and $3$ is
$(\mathbb{I}, \mathbb{I})$, $(\psi, \psi)$, or $(\sigma, \sigma)$.
If instead $b_{25} \neq b_{35}$, then
\begin{align}
R_{23} = e^{i\phi} \left(\begin{matrix} -1 & 0 & 0 \\ 0 & 1 & 0 \\ 0 & 0 & e^{i\pi/8}\end{matrix} \right)
\end{align}

\subsection{Physical implementation of projection of pairs of genons onto $(\mathbb{I},\mathbb{I})$ or $(\psi,\psi)$ fusion channels }

When two genons are separated by a distance $L$, the effective Hamiltonian in the degenerate subspace spanned by the genons obtains non-local Wilson loop operators:
\begin{align}
H_{\text{genon}} = t_{(\psi,\psi)} W_{(\psi,\psi)}(C) + t_{(\sigma,\sigma)} W_{(\sigma,\sigma)}(C) + H.c.
\end{align}
$W_{(\psi,\psi)}(C)$ describes the exchange of a $(\psi,\psi)$ particle between the two genons, which equivalently corresponds to a $(\mathbb{I},\psi)$ or $(\psi, \mathbb{I})$ particle
encircling the pair of genons. Similarly, $W_{(\sigma,\sigma)}(C)$ describes the exchange of a $(\sigma,\sigma)$ particle between the two genons, which equivalently corresponds to a $(\mathbb{I},\sigma)$ or $(\sigma, \mathbb{I})$ particle encircling the pair of genons. $t_\psi \propto e^{-L \epsilon_\psi/v_\psi}$ and $t_\psi \propto e^{-L \epsilon_\sigma/v_\sigma}$ are the tunneling amplitudes, with $\epsilon_\psi$ and $\epsilon_\sigma$ being the energy gaps for the $\psi$ and $\sigma$ particles, and $v_\psi$, $v_\sigma$ some appropriate velocity scales.

Next, let us suppose that the pair of genons shown in the figure fuse to the quasiparticle $(b,b)$. The outcome of the process where a $(1,a)$
quasiparticle encircles a topological charge $(b,b)$ is determined by the topological $S$ matrix of the Ising phase, and is given by $S_{ab}/S_{b\mathbb{I}}$,
where
\begin{align}
S =  \left(\begin{matrix}
1/2 & 1/\sqrt{2} & 1/2 \\
1/\sqrt{2} & 0 & -1/\sqrt{2} \\
1/2 & -1/\sqrt{2} & 1/2
\end{matrix} \right),
\end{align}
and where the entries are ordered $\mathbb{I}, \sigma, \psi$. In other words, the eigenvalues of $W_{(a,a)}(C)$ are given by $S_{ab}/S_{b\mathbb{I}}$,
where $(b,b)$ is the fusion channel of the two genons connected by the path $C$.

The ground state of $H_{\text{genon}}$, which depends on $t_\sigma, t_\psi$
therefore corresponds to a definite fusion channel for the pair of genons involved. We can distinguish the following possibilities:
\begin{enumerate}
\item $|t_\sigma| > |t_\psi|$ and $t_\sigma > 0$. The ground state subspace of $H_{\text{genon}}$ corresponds to the case where the two genons have fused to the $(\psi,\psi)$ channel.
\item $|t_\sigma| > |t_\psi|$ and $t_\sigma < 0$. The ground state subspace of $H_{\text{genon}}$ corresponds to the case where the two genons have fused to the $(\mathbb{I},\mathbb{I})$ channel.
\item $|t_\sigma| < |t_\psi|$, $t_\psi < 0$, $t_\sigma > 0$. The ground state subspace of $H_{\text{genon}}$ corresponds to the case where the two genons have fused to the $(\psi, \psi)$ channel.
\item $|t_\sigma| < |t_\psi|$, $t_\psi < 0$, $t_\sigma < 0$. The ground state subspace of $H_{\text{genon}}$ corresponds to the case where the two genons have fused to the $(\mathbb{I}, \mathbb{I})$ channel.
\item $|t_\sigma| < |t_\psi|$, $t_\psi > 0$. The ground state subspace of $H_{\text{genon}}$ corresponds to the case where the two genons have fused to the $(\sigma, \sigma)$ fusion channel.
\end{enumerate}
We see that there is only one possibility that needs to be prevented, which is the last one, where $|t_\sigma| < |t_\psi|$, $t_\psi > 0$. To do this, we flip the sign of $t_\psi$  by flipping the sign of the coupling along a single link of the shortest path that connects the genons. Moreover, note that one can also pick the precise path $C$ along which the quasiparticles tunnel by depressing the gap along that path, which can be done by decreasing the couplings along that path.

\subsection{Ising $\times \overline{\text{Ising}}$ topological order and gapped boundaries}

In the above, we have suggested creating genons in the Ising $\times$ Ising topological state. The braiding of the genons, which can be performed in a measurement-based
fashion, can be used for the topologically protected $\pi/8$ gate. Here we note that the same topologically robust transformations can also be achieved with
the Ising $\times\overline{\text{Ising}}$ topological state in the presence of multiple disconnected gapped boundaries, where
$\overline{\text{Ising}}$ refers to the time-reversed conjugate of the Ising state.

\begin{figure}
	\centering
	\includegraphics[width=4.0in]{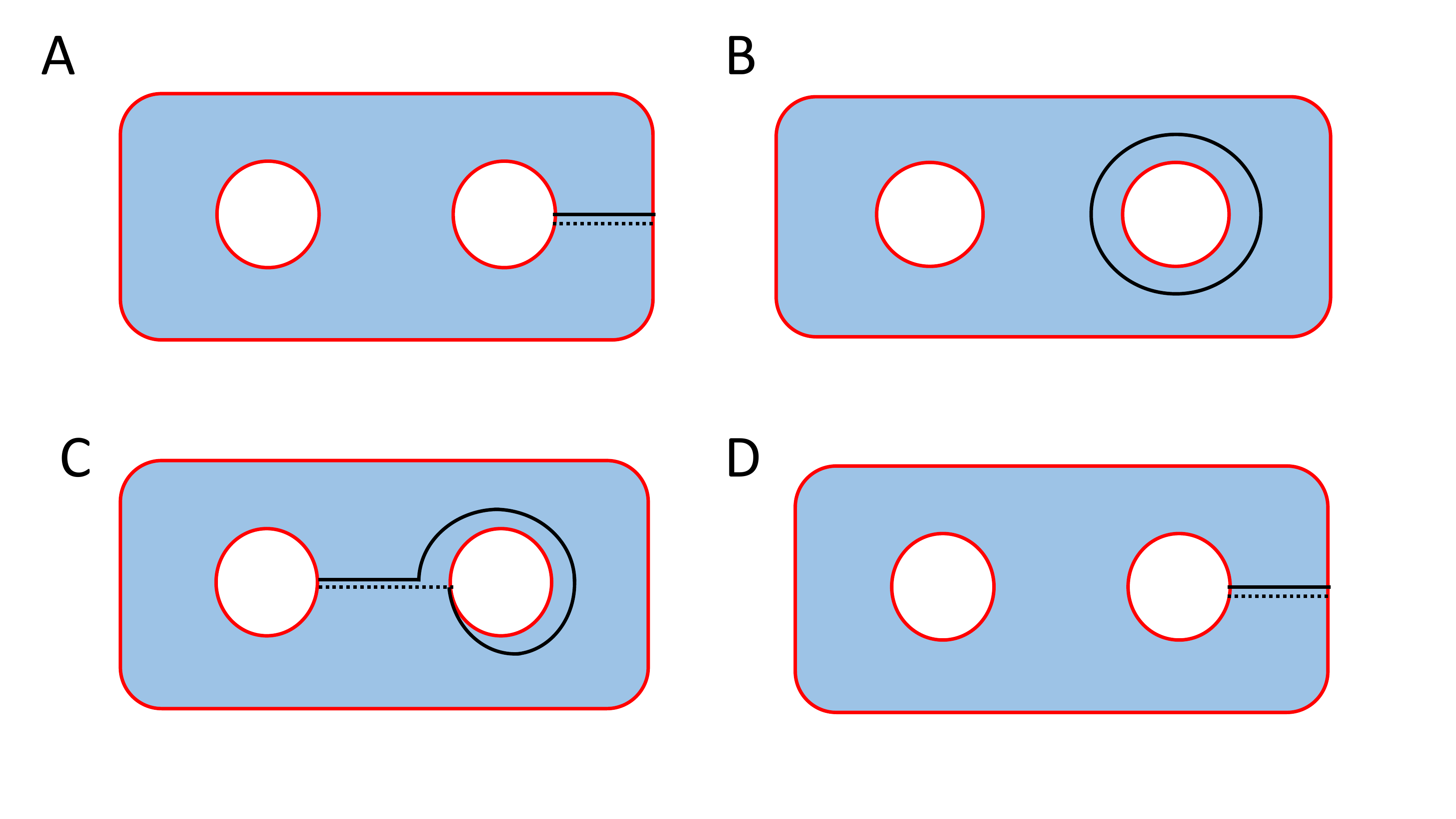}
	\caption{\label{holeBraiding} Ising $\times\overline{\text{Ising}}$ system in the presence of two holes (three gapped boundaries). This is effectively equivalent to the Ising $\times$ Ising system with 6 genons. The topologically robust operation described in Fig. \ref{genonBraiding} can be adapted to this case, by projecting the topological
charge through the loops shown to be equal to $b_1, ..., b_4$. The solid line indicates that it is in the "top" layer, while the dashed line indicates that it is in the "bottom" layer.
}
\end{figure}

The main observation is that the Ising $\times\overline{\text{Ising}}$ state in the presence of $n$ disconnected gapped boundaries can be viewed as effectively a flattened
version of a single Ising state on a genus $g = n - 1$ surface. This is similar to the fact that an Ising $\times$ Ising state, in the presence of $n$ pairs of genons,
can also be effectively mapped onto a single Ising state on a genus $g = n - 1$ surface.
The measurement-based braiding protocol for the genons of the Ising $\times$ Ising state can be readily adapted to the case of the Ising $\times\overline{\text{Ising}}$
with gapped boundaries. Below we will briefly sketch this adapted protocol, leaving a more detailed discussion for future work.

Let us consider the Ising $\times\overline{\text{Ising}}$ state in the presence of two disconnected gapped boundaries, i.e. on an annulus. This is equivalent
to a single Ising state on a torus, similar to the case of the Ising $\times$ Ising state with four genons. In order to carry out an effective Dehn twist in this effective
torus, we use an ancillary gapped boundary, as shown in Fig. \ref{holeBraiding}. This system is now equivalent to a genus 2 surface. In order to effectively carry out
the Dehn twist, we perform a series of projections along various loops in the system. We consider the loops shown in Fig. \ref{holeBraiding}A-D, and sequentially
project the topological charge through those loops to be $b_1, \cdots, b_4$. This is precisely analogous to the genon braiding case described in Fig. \ref{genonBraiding},
where the fusion channel of genons was projected onto $b_{56}$, $b_{35}$, $b_{25}$, and $b_{56}'$. Thus if we take $b_1 = b_4$ and $b_1,\cdots, b_4$
to be all Abelian topological charges, the equivalence between the Ising $\times$ Ising system with genons and the Ising $\times\overline{\text{Ising}}$ with
gapped boundaries implies that we will have effectively carried out the desired operation.

As in the case of the genons in the Ising $\times$ Ising state described in the previous sections, these projections can effectively be implemented by reducing the gap
for quasiparticle tunneling along the various loops as required.

\end{widetext}

\end{document}